\documentclass[prx, twocolumn, longbibliography]{revtex4-1}
\usepackage{graphicx}
\usepackage{amsmath}
\usepackage{amssymb}
\usepackage{color}
\usepackage{appendix}
\usepackage{braket}
\usepackage{bbm}
\usepackage{subfigure}

\begin{document}
	\title{Non-Hermitian second-order skin and topological modes}
	\author{Yongxu Fu}
	  \email{yancyfoy@mail.ustc.edu.cn}
	\author{Jihan Hu}
	\author{Shaolong Wan}
	  \email{slwan@ustc.edu.cn}
	\affiliation{Department of Modern Physics, University of Science and Technology of China, Hefei, 230026, China}
\begin{abstract}
The skin effect and topological edge states in non-Hermitian system have been well-studied, and the second-order skin effect and corner modes have also been proposed in non-Hermitian system recently. In this paper, we construct the nested tight-binding formalism to research the second-order corner modes analytically, which is a direct description of the generic non-Hermitian tight-binding model without other assumptions. Within this formalism, we obtain the exact solutions of second-order topological zero-energy corner modes for the non-Hermitian four-band model. We validate the nested tight-binding formalism in the hybrid skin-topological corner modes for the four-band model and a non-Hermitian two-dimensional (2D) extrinsic model. In addition, we exactly illustrate the corner modes induced by second-order skin effect for a simplest 2D non-Hermitian model by the nested tight-binding formalism.
\end{abstract}
\maketitle
\section{Introduction}
Beyond the conventional hotspot for topological insulators and superconductors~\cite{hasan2010, qi2010, fu2007, qi2008, liu2010, qi2010, teo2010, mong2011, hug2011} and their classification~\cite{ryu2008, mor2013, chiu2013, sato2014, chiu2014, qiu2015, sato2016,Kruthoff2017,sato2017, con2019} in condensed physics past decades, it rapidly ramifies into two patulous fields which involve higher-order topological phases~\cite{Slager2015,schindler2018, dipole2017, huang2017, sha2018, ezawa2018, geier2018, khalaf2018, flore2018, akishi2018, li2018, lin2018, ryo2019, tanaka2020, reflection, fu2011, song2017, ezawa2018, park2019, yan2018, yan22018} and non-Hermitian topological systems~\cite{sato2011, kawabata2018, Alvarez2018,lee2016, xiong2017, leykam2018, shen2018, kunst2018, yao2018, yao20182,Thomale2019,londhi2019, yokomizo2019, kawabata2019, song2019, imura2019, okuma2019,Borgnia2020,origin2020, xue2020, gong2018, kawabataprx} in recent years. An $n$th-order topological insulator, which originates from the topological crystalline insulators~\cite{fu2011}, has topologically protected gapless states at a boundary of the system of co-dimension n~\cite{schindler2018, reflection}, but is gapped otherwise. For example, a two-dimensional second-order topological insulator has topological corner states but a gapped bulk and no gapless edge states. The non-Hermitian Hamiltonians are widely used in describing open systems~\cite{malzard2015, open1, open2, open3, open4, open5} and wave systems with gain and loss~\cite{gain1, gain2, gain3, gain4, gain5, gain6, gain7, gain8, gain9, gain10, gain11, gain12}~(e.g., photonic and acoustic), etc. Of all properties in non-Hermitian systems, the existence of exceptional points~\cite{heiss2012, shen2018, kawabata2019} and the skin effect~\cite{yao2018, yao20182, song2019,Thomale2019,londhi2019, origin2020} are the most intriguing. The exceptional points are the points where complex energy bands coalesce, while the skin effect describes the localized bulk states in non-Hermitian systems. We call the localized bulk states in non-Hermitian systems the skin bulk states in this paper. Recently, the higher-order states of the non-Hermitian systems have been studied~\cite{Edvardsson2019,Ezawa2019,kawabaras, zhang2019, lee2019, denner2020, okugawa} and two novel states, the second-order skin ($SS$) and skin-topological ($ST$) states~\cite{lee2019}, have been proposed.
\setlength{\parskip}{0.1em} 

The abundant localized states in first-order non-Hermitian systems exploit more possible second-order localized states. The contribution from two directions with topological edge ($T$) states or skin bulk ($S$) states induces three possible types of second-order localized corner modes: second-order topological ($TT$), skin-topological ($ST$), and second-order skin ($SS$) modes, which have been numerically calculated in Ref.~\cite{lee2019}. However, the analytical forms of these corner modes are still not obtained. The meaning and configuration of these corner modes are also not clear enough in Ref.~\cite{lee2019}. In this paper, we investigate the three types of corner modes and deduce their localization behavior analytically in non-Hermitian systems. Based on the nested tight-binding formalism constructed in Sec.~\ref{NTB}, we exactly research the $TT$, $ST$, and $SS$ corner modes, and clarify the meaning and configuration of these corner modes. By this formalism, we can analytically study the generic tight-binding model without any other assumptions. We obtain the analytical solutions of $TT$ corner modes from the effective Hamiltonian in the subspace of the edge-states, generated from the generic two-dimensional (2D) tight-binding Hamiltonian. Although not protected by bulk-energy band topology, the nonzero-energy edge states still contribute to the second-order corner modes. Actually, the gapped edge-localized states are protected by Wannier band topology in Hermitian systems with higher-order topological phases~\cite{dipole2017}, and the gapless edge-localized states are protected by bulk-energy band topology. Hence we do not distinguish the zero- and nonzero-energy edge states when we research the second-order corner modes. In this sense, the definitions of $ST$ and $TT$~\cite{lee2019} modes are reasonable. In principle, the possible higher-order cases can be obtained from the first- and second-order cases. Hence we mainly concentrate on the second-order corner modes in this paper.
\setlength{\parskip}{0.1em} 

This paper is organized as follows. In Sec.~\ref{WNSS}, inspired by the topological origin of the skin
effect~\cite{origin2020}, we study two typical one-dimensional (1D) non-Hermitian models with first-order skin effect. Then we illustrate the second-order skin effect for the simplest 2D non-Hermitian model~\cite{lee2019}. In Sec.~\ref{NTBF}, we construct the nested tight-binding formalism and investigate the $TT$ and $ST$ corner modes. Utilizing this formalism, we study the four-band model~\cite{lee2019} with $TT$ and $ST$ corner modes and the 2D model~\cite{okugawa} with extrinsic corner modes. Finally, the conclusion and discussion are given in Sec.~\ref{CD}.
\section{Winding Number and Second-Order Skin Effect}
\label{WNSS}
The skin effect, which is a remarkable difference between complex energy spectra under periodic boundary condition (PBC) and those under open boundary condition (OBC), is the most charming property in non-Hermitian systems. There are extensive number of skin bulk modes localized at arbitrary boundaries. In Sec.~\ref{seo}, after a brief review of the topological origin of first-order skin effect~\cite{origin2020}, we emphasize the difference between winding number protecting first-order topological edge states and that protecting skin effect, and study two typical 1D non-Hermitian models with first-order skin effect. In addition, the second-order skin effect is investigated for the simplest 2D non-Hermitian model in Sec.~\ref{2dss}.
\subsection{Winding number and first-order non-Hermitian skin effect}
\label{seo}
The first-order skin effect, which originates from intrinsic point-gap topology of non-Hermitian systems~\cite{origin2020}, is determined by the winding number of the complex energy contour for a 1D Hamiltonian. For simplicity, we refer the skin effect and edge states to the first-order cases and specify the order for higher-order cases hereafter. The topological invariant for point-gap is the winding number of complex spectra under PBC around the reference skin mode point $E$
\begin{equation}
\label{winding}
W(E)=\frac{1}{2\pi i}\int_{0}^{2\pi}dk\frac{d}{dk}\log \det[H(k)-E].
\end{equation}
We should distinguish the meaning of the winding number protecting first-order topological edge states from that protecting skin effect. The conventional winding number of a $(2n+1)$-dimensional Hermitian Hamiltonian $H(k)$ with chiral symmetry $S$, which protects topological edge states at the $2n$-dimensional surface, comes from a homotopy map: $BZ^{2n+1}\rightarrow U(N)$,
\begin{equation}
W_{2n+1}=\frac{n!}{2(2\pi i)^{n+1}(2n+1)!}\int_{BZ^{2n+1}}tr(SH^{-1}dH)^{2n+1}.
\end{equation}
\begin{figure}
	\centering
	\subfigure[]{\includegraphics[width=0.301\textwidth]{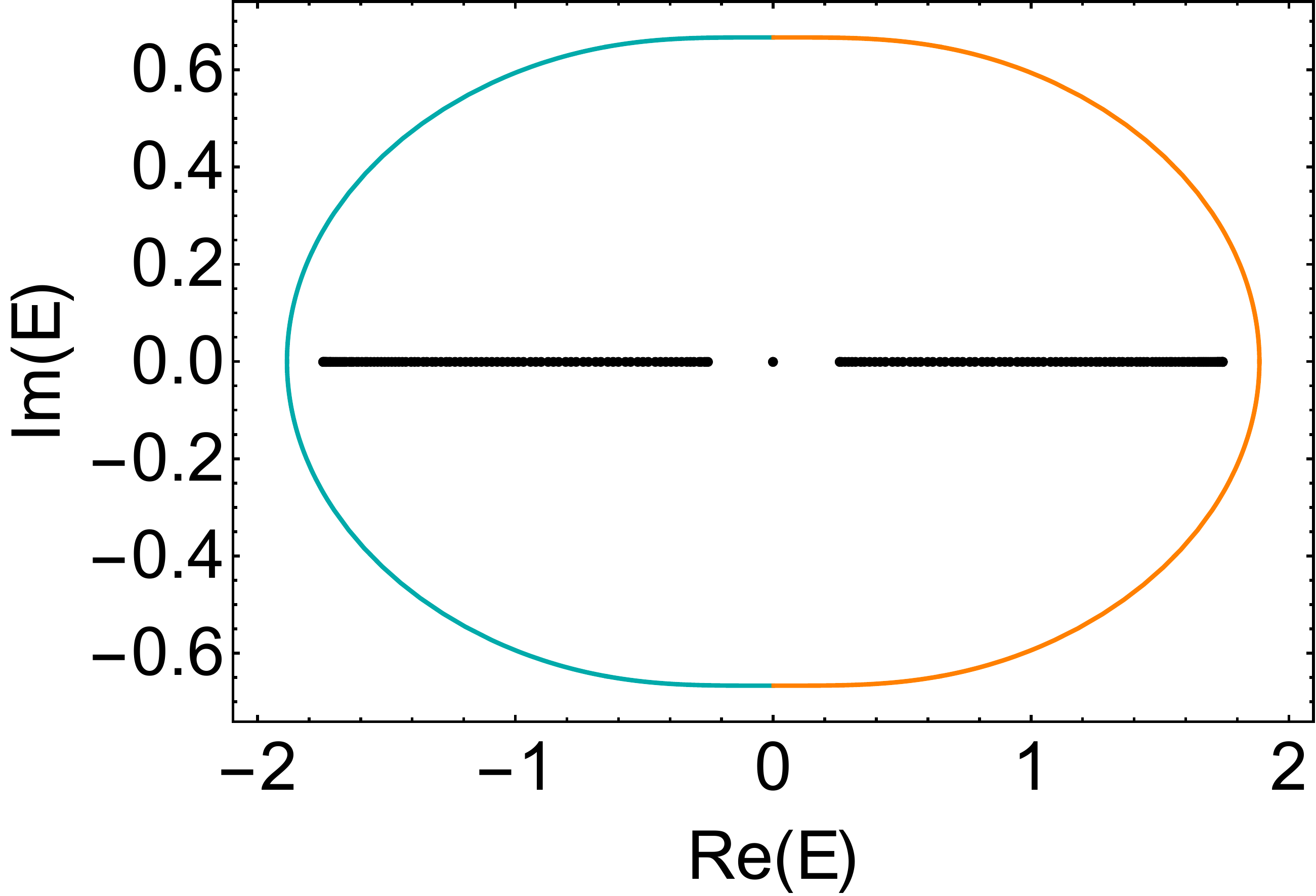}}
	\label{fig: 1a}
	\subfigure[]{\includegraphics[width=0.3\textwidth]{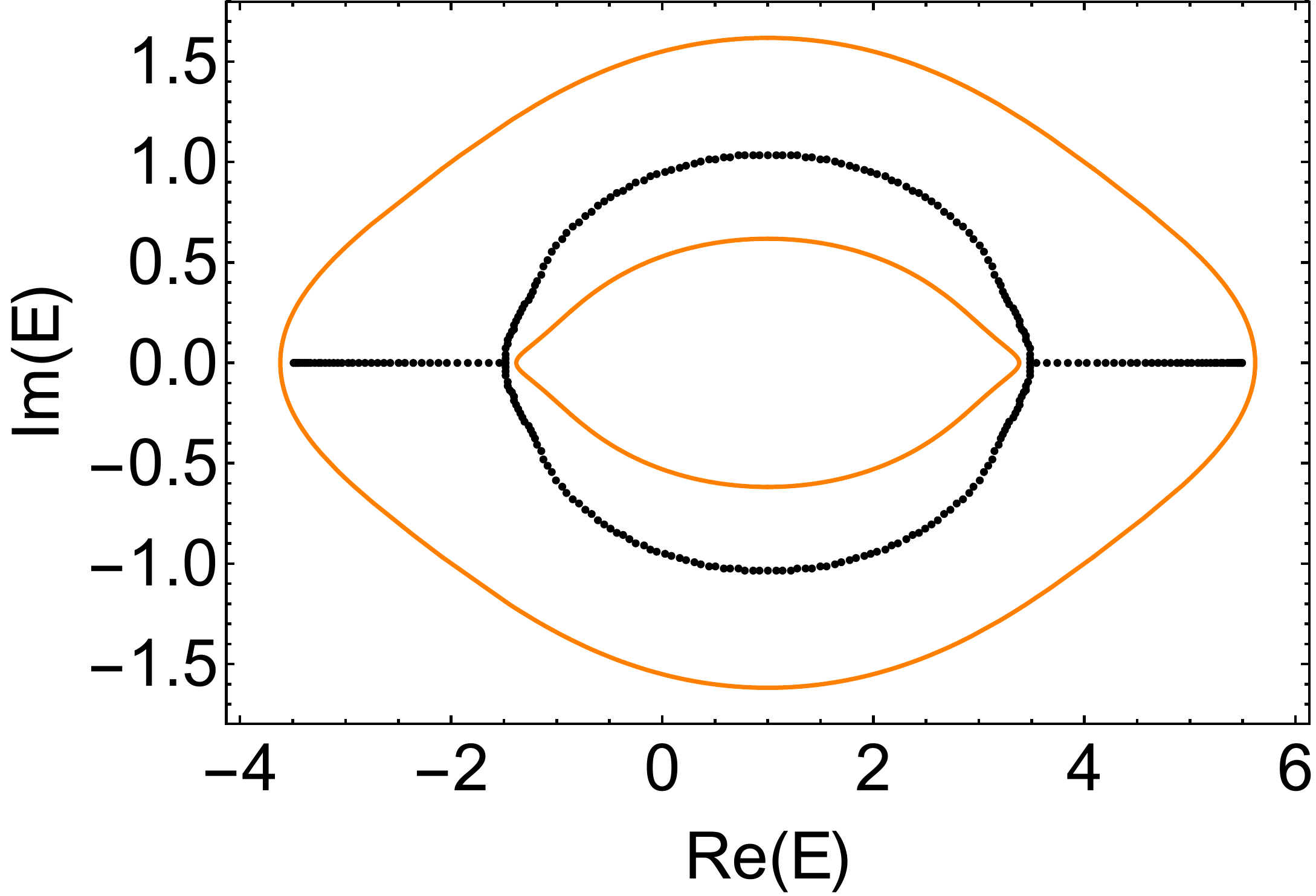}}
	\label{fig: 1b}
	\subfigure[]{\includegraphics[width=0.301\textwidth]{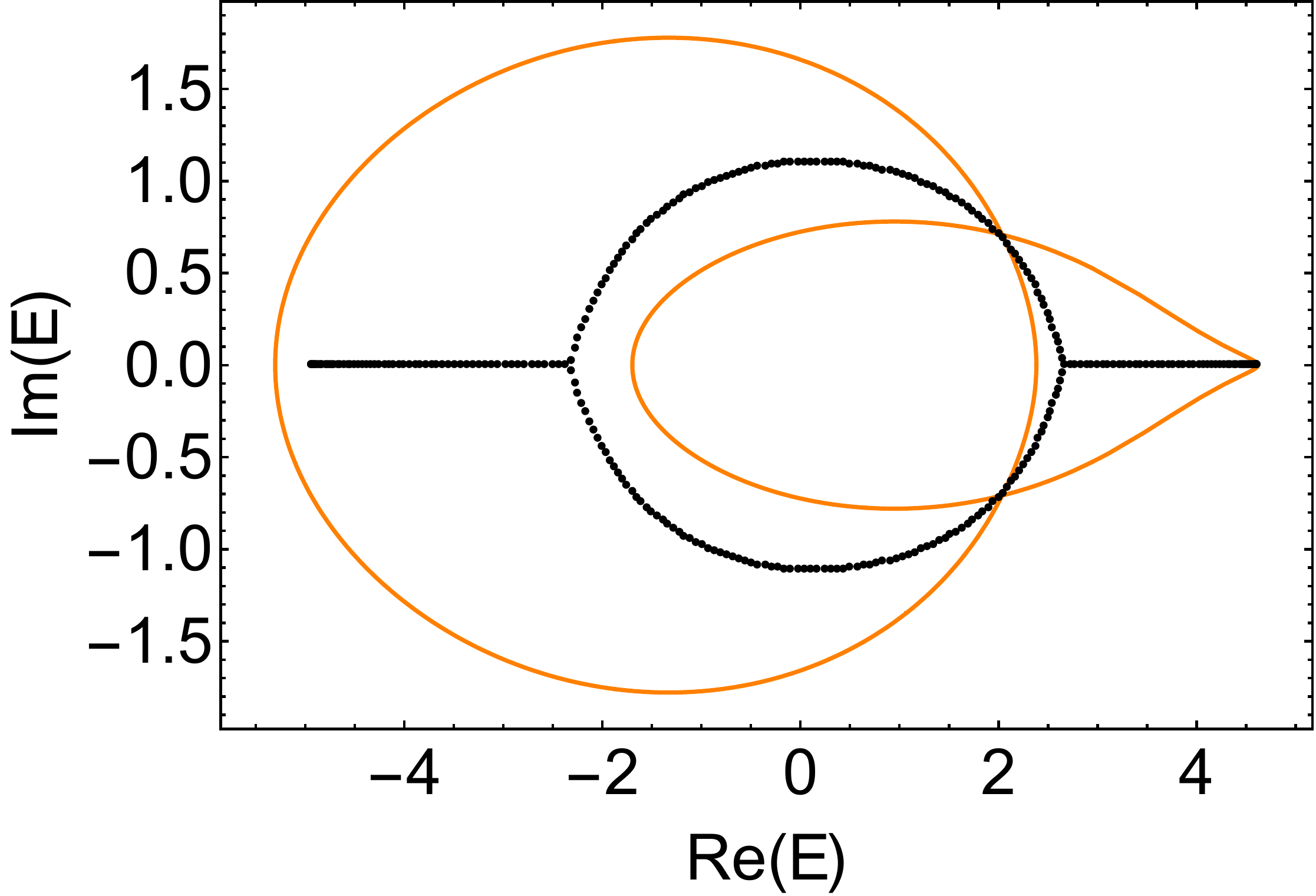}}
	\label{fig: 1c}
	\caption{(a) The complex energy spectra for the non-Hermitian SSH model with $t_{1}=1, t_{2}=1, \gamma=4/3$. There are two degenerate topological edge modes located exactly at zero energy. The complex energy spectra for the two-band model, Eq.~(\ref{1ds}) with $t_{0}=1, t_{-}=2, t_{+}=1, w_{0}=1, w_{-}=1, w_{+}=3, c=1$ are plotted in (b) and those with $t_{0}=1, t_{-}=2, t_{+}=1, w_{0}=-1, w_{-}=1, w_{+}=3, c=1$ are plotted in (c). The spectra under PBC are plotted as orange or cyan loops, while the spectra under OBC are plotted as black parts.}
	\label{fig: 1}
\end{figure}
In addition, the conventional winding number has been generalized to the winding number of non-Bloch Hamiltonian $H(\beta)$ in 1D non-Hermitian systems recently~\cite{yao2018,yokomizo2019}, where $\beta$ is in the generalized Brillouin zone~(see Appendix~\ref{appendix: EEOTB}).
However, the winding number protecting the skin bulk part of the spectra under OBC~[Eq.(\ref{winding})] is calculated from the complex energy spectra under PBC for a system with point-gap. The winding number of the skin effect vanishes for the Hermitian Hamiltonian since the energy spectra are always real. Note that, the conventional winding number, which protects the edge states of a 1D Hermitian Hamiltonian $H_{h}$ with chiral symmetry, is actually the winding number of the chiral non-Hermitian block Hamiltonian
\begin{equation}
W_{1}^{h}=\frac{1}{2\pi i}\int_{0}^{2\pi} dk \frac{d}{dk}\log\det[h(k)],\, \, \, \, \, \,
H_{h}=
\begin{bmatrix}
0&h(k)\\
h^{\dagger}(k)&0
\end{bmatrix}.
\end{equation}
Moreover, the value of winding number $W(E)$ counts the degenerate skin modes at reference energy $E$~\cite{gong2018}.
When we study the skin effect for a generic 1D multiple-band system, we should sum over all the winding numbers for each band $E^{\mu}(k)$
\begin{equation}
W(E)=\frac{1}{2\pi i}\sum_{\mu=1}^{q}\int_{0}^{2\pi}dk\frac{d}{dk}\log[E^{\mu}(k)-E].
\end{equation}

Firstly, we consider the typical non-Hermitian Su-Schrieffer-Heeger (SSH) model $H_{nSSH}(k)=(t_{1}+t_{2}\cos k)\sigma_{x}+(t_{2}\sin k+i\gamma/2)\sigma_{y}$~\cite{yao2018}. The energy spectra of this model under PBC form two energy bands  $E_{\pm}(k)=\pm\sqrt{(t_{1}+t_{2}\cos k)^{2}+(t_{2}\sin k+i\gamma/2)^{2}}$. Each band forms a semicircle  [cyan and orange semicircles in Fig.~\ref{fig: 1}(a)] in the complex plane. The winding number for each skin mode $E_{s}$ under OBC [point on the black lines in Fig.~\ref{fig: 1}(a)] is
\begin{eqnarray}
W(E_{s})=W^{+}(E_{s})+W^{-}(E_{s})=1.\nonumber
\end{eqnarray}
Therefore each point on the black lines in Fig.~\ref{fig: 1}(a) is an eigenenergy of one skin mode localized at one boundary for the Hamiltonian under OBC. However, the modes at origin in Fig.~\ref{fig: 1}(a) are not skin modes, which contain two degenerate topological edge states.

Secondly, we consider the model with two energy bands~\cite{yang2019}, and the Hamiltonian reads
\begin{eqnarray}
H_{2}(k)=\begin{bmatrix}
h_{1}(k)&c\\
c&h_{2}(k)
\end{bmatrix},
\label{1ds}
\end{eqnarray}
where $h_{1}(k)=t_{0}+t_{-}e^{-ik}+t_{+}e^{ik}$ and $h_{2}(k)=w_{0}+w_{-}e^{-ik}+w_{+}e^{ik}$. The two energy bands are $E_{\pm}(k)=h_{+}(k)\pm\sqrt{c^{2}+h_{-}^{2}(k)}$, where $h_{\pm}(k)=(h_{1}(k)\pm h_{2}(k))/2$. In Figs.~\ref{fig: 1}(b) and \ref{fig: 1}(c), the complex energy spectra under PBC and OBC are plotted as orange loops and black parts respectively. The skin modes (black parts) only exist in the area with nonvanishing winding number.
\begin{figure}
	\centering
	\subfigure[]{\includegraphics[width=0.17\textwidth]{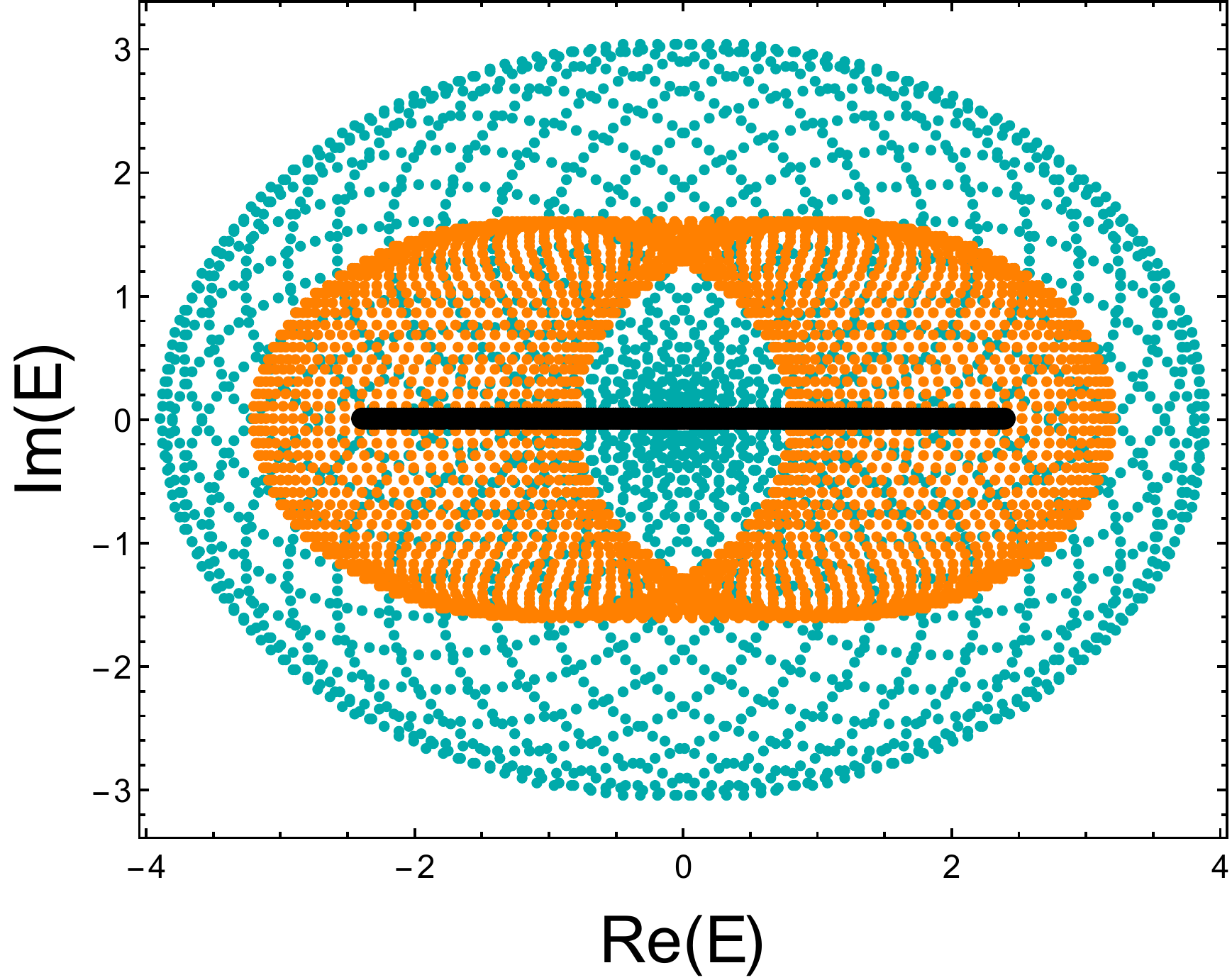}}
	\subfigure[]{\includegraphics[width=0.29\textwidth]{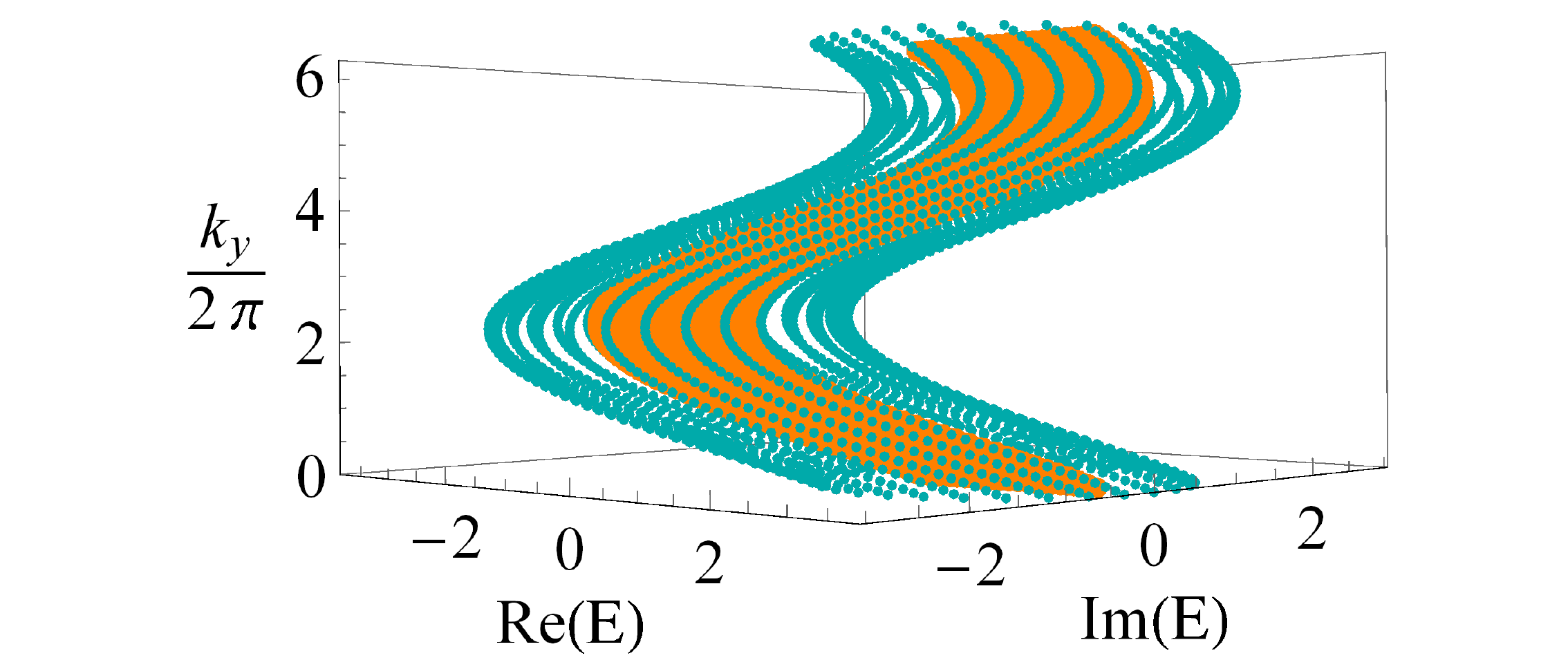}}\\
	\subfigure[]{\includegraphics[width=0.175\textwidth]{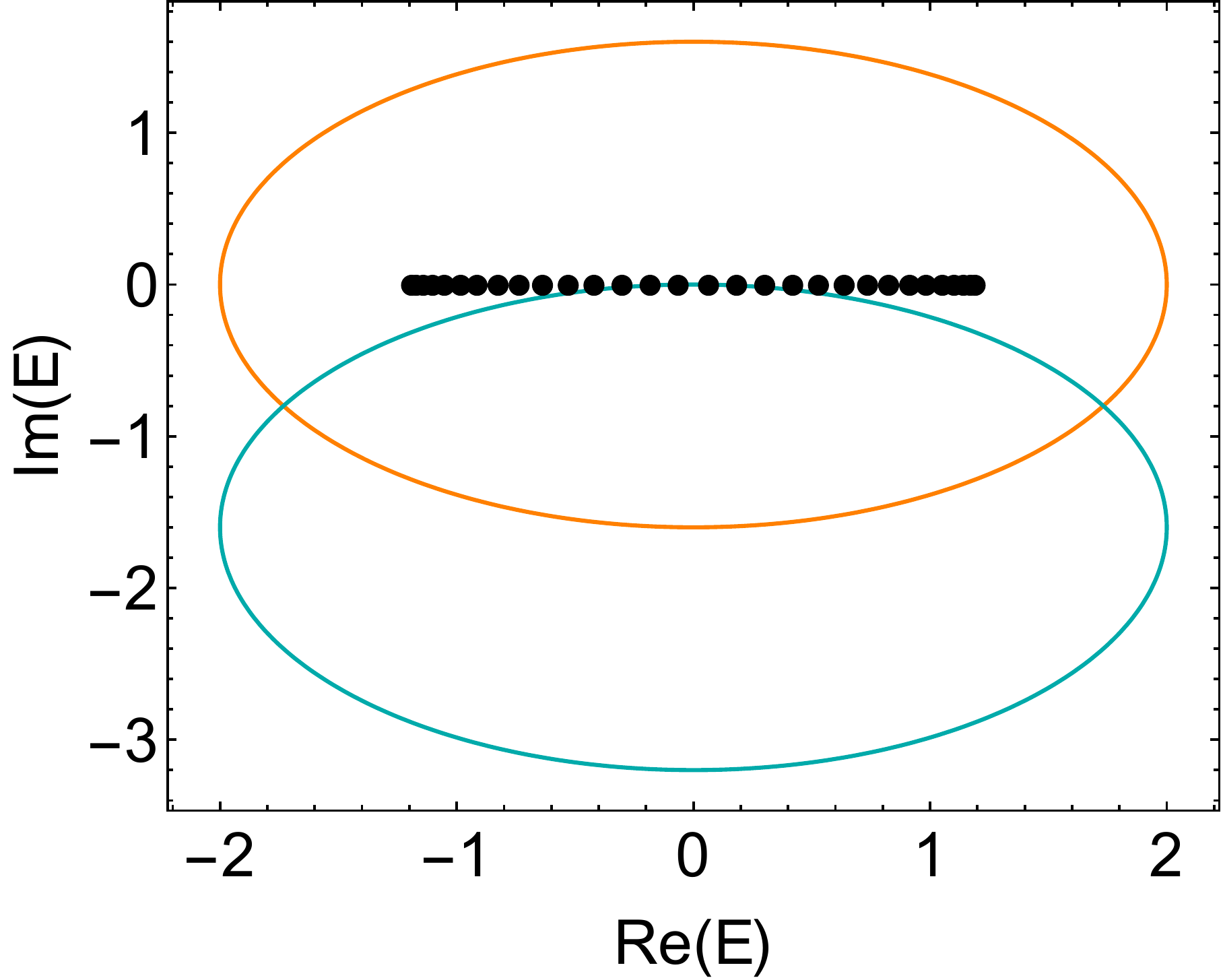}}
	\subfigure[]{\includegraphics[width=0.295\textwidth]{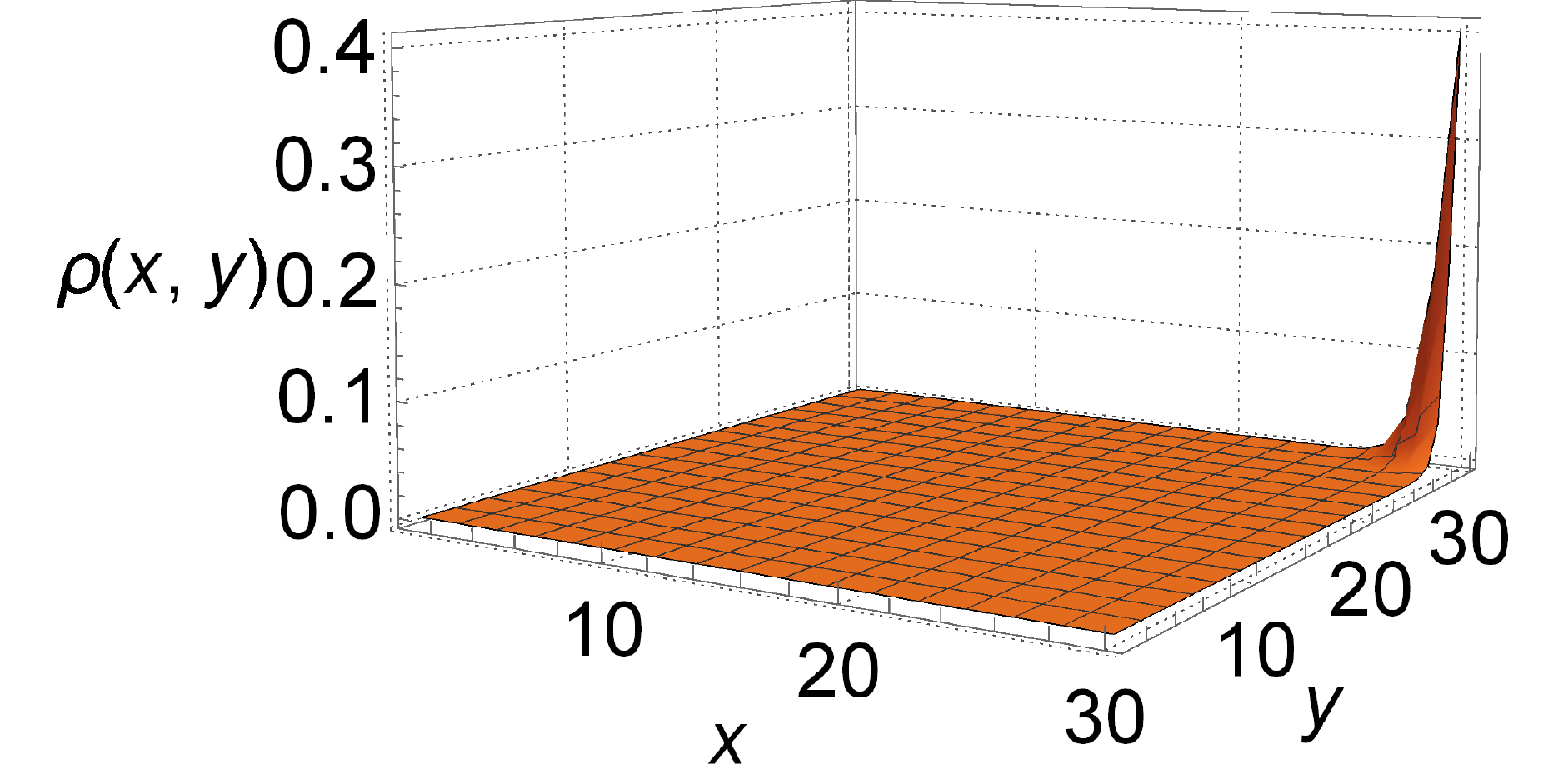}}\\
	\caption{Complex energy spectra of the simplest 2D model in Eq.~(\ref{2ds}). The number of unit cells is $30\times 30$ with parameters $t^{x}=t^{y}=1, \gamma^{x}=\gamma^{y}=0. 8$. (a) Spectra under double-PBC (cyan), $x$ OBC/$y$ PBC (orange) and full OBC (black) respectively. (b) Spectra under double-PBC (cyan) and $x$ OBC/$y$ PBC (orange) are plotted in $E$-$k_{y}$ space. (c) The loops projected from (b) for a fixed energy band under $x$ PBC (cyan, $k_{x}=\pi/2$) and $x$ OBC (orange, $k=\pi/2$) respectively. The orange loop surrounds its corresponding spectra under further taking $y$ OBC (black points). (d) A typical second-order skin mode with $E_{SS}=-2.38769$ localized at one corner.}
	\label{fig: 2}
\end{figure}
\subsection{The second-order skin effect}
\label{2dss}
Consider the simplest 2D non-Hermitian model~\cite{lee2019} possessing second-order skin effect. The Hamiltonian in momentum space is
\begin{equation}
H_{2D} (\vec{k})=t_{+}^{x}e^{-ik_{x}}+t_{-}^{x}e^{ik_{x}}+t_{+}^{y}e^{-ik_{y}}+t_{-}^{y}e^{ik_{y}},
\label{2ds}
\end{equation}
where $t^{x, y}_{\pm}=t^{x, y}\pm \gamma^{x, y}$ are the real nonreciprocal hopping terms inducing non-Hermiticity. This Hamiltonian respects time-reversal symmetry $TH_{2D}(-\vec{k})T^{-1}=H_{2D}(\vec{k})$ and $T$ is the complex conjugation operator. Hence, $H_{2D}$ belongs to class AI with point gap~\cite{ryu2008, qiu2015,kawabataprx}, which is topologically trivial in 2D resulting in the absence of first-order edge states. It follows that the pure first- and second-order skin effect are not protected by the conventional topological invariant but protected by the point-gap topology.

From the simplest 2D model mentioned above, the single $y$-layer Hamiltonian [see Sec.~\ref{NTB}] $H_{s}$, which is the Hatano-Nelson model~\cite{hn}, reads
\begin{eqnarray}
\hat{H}_{s}^{2D}=\sum_{x}[\hat{c}_{x+1, y}^{\dagger}t^{x}_{+}\hat{c}_{x, y}+\hat{c}_{x-1, y}^{\dagger}t^{x}_{-}\hat{c}_{x, y}].
\end{eqnarray}
We can obtain $\beta_{x}=\sqrt{\frac{t^{x}_{+}}{t^{x}_{-}}}e^{ik}$~($k\in[0,2\pi]$) forming the circular generalized Brillouin zone~(Appendix~\ref{appendix: EEOTB}). The energy spectrum under OBC is $\epsilon(k)=2\sqrt{t^{x}_{+}t^{x}_{-}}\cos k$, which is derived in Ref.~\cite{yokomizo2019} by non-Bloch band theory, while that under PBC is $\epsilon_{P}(k_{x})=t_{+}^{x}e^{-ik_{x}}+t_{-}^{x}e^{ik_{x}}$ forming a loop, which is obtained by Fourier transformation of $\hat{H}_{s}^{2D}$. The former lies in the interior of the latter, indicative of skin effect along $x$ direction. Since the internal degree of freedom is $1$ in this model, we obtain the effective Hamiltonian for second-order skin effect~[see Sec.~\ref{NTB}]
\begin{eqnarray}
H_{eff}(k_{y})=\sum_{k}(t_{-}^{y}e^{ik_{y}}+\epsilon(k)+t_{+}^{y}e^{-ik_{y}}).
\label{1deff}
\end{eqnarray}
This effective Hamiltonian, as a function of $k_{y}$, is expressed in one skin-mode subspace along $x$ direction for each $k$ value. Consequently, we obtain the second-order skin modes~($SS$ modes) under further taking $y$ OBC for $H_{eff}(k)$. The meaning and configuration of $SS$ corner modes is that the corner modes under full OBC are contributed from skin modes along $x$ direction~(skin-mode subspace) and $y$ direction~(further taking $y$ OBC).
For each fixed $k$ value, the complex energy spectrum forms a loop $C(k)$, for which $\epsilon(k)$ assigns the loop center varying in $\left[ -2\sqrt{t^{x}_{+}t^{x}_{-}}, 2\sqrt{t^{x}_{+}t^{x}_{-}}\right]$.

We illustrate the second-order skin effect of this model in Fig.~\ref{fig: 2}. The full OBC energy spectra (black) lie within $x$ OBC/$y$ PBC energy spectra (orange), which in turn lie within double-PBC energy spectra (cyan)~[Figs.~\ref{fig: 2}(a) and \ref{fig: 2}(b)]. The loops~[Fig.~\ref{fig: 2}(c)] projected from Fig.~\ref{fig: 2}(b) for a fixed $x$ PBC (cyan, $k_{x}=\pi/2$) and $x$ OBC (orange, $k=\pi/2$) energy band indicate the skin effect along $y$ direction and second-order skin effect, respectively. All the cyan and orange loops, with varying $k_{x}$ and $k$, respectively, form the corresponding cyan and orange energy spectra in Fig.~\ref{fig: 2}(a). As the topological origin of the skin effect clarified by Ref.~\cite{origin2020}, each loop $C(k)$ surrounds its corresponding spectra under further taking $y$ OBC [black points in Fig.~\ref{fig: 2}(c)], which are the $SS$ modes localized at one corner under full OBC~[Fig.\ref{fig: 2}(d)]. Therefore the second-order skin effect indeed originates from the point-gap topology along each of the two directions with first-order skin effect respectively. The conventional winding number does not protect the $SS$ modes and there are no edge states of the simplest 2D non-Hermitian model, Eq.~(\ref{2ds}). The $SS$ modes~[black points in Fig.~\ref{fig: 2}(c)] are protected by the winding number of $C(k)$~[orange loop in Fig.~\ref{fig: 2}(c)] around corresponding $SS$ modes, i.e., the point-gap topology of $H_{eff}(k_{y})$ for fixed $k=\pi/2$.

\section{Nested Tight-Binding Formalism and Second-Order Corner Modes}
\label{NTBF}
\subsection{The nested tight-binding formalism}
\label{NTB}
One of the simplest perspectives to give the second-order corner modes is working out the localized states in turn along two related directions. It means that we put the localized information of one direction into the other directions, for which we call the nested process. For the lattice tight-binding model, our general formalism for second-order phase is called the nested tight-binding formalism.

A generic tight-binding 2D Hamiltonian, with $L_{x}, L_{y}$ lattice sites, $R_{x}, R_{y}$ the hopping range along $x, y$ directions, respectively, and $q$ the internal degrees of freedom per unit cell, is
\begin{eqnarray}
\hat{H}=\sum_{x=1}^{L_{x}}\sum_{y=1}^{L_{y}}\sum_{\mu,\nu=1}^{q}\bigg[\sum_{i=-R_{x}}^{R_{x}}\hat{c}_{x+i, y}^{\mu\dagger}t_{i, \mu\nu}^{x}\hat{c}_{x, y}^{\nu}
\nonumber \\
+\sum_{j=-R_{y}}^{R_{y}}\hat{c}_{x, y+j}^{\mu\dagger}t_{j, \mu\nu}^{y}\hat{c}_{x, y}^{\nu}\bigg].
\label{2dgf}
\end{eqnarray}
We first deal with a fixed single $y$ layer
\begin{equation}
\label{layer}
\hat{H}_{y}=\sum_{x=1}^{L_{x}}\sum_{\mu,\nu=1}^{q}\sum_{i=-R_{x}}^{R_{x}}\hat{c}_{x+i, y}^{\mu\dagger}T_{i, \mu\nu}^{x}\hat{c}_{x, y}^{\nu},
\end{equation}
where $T_{0, \mu\nu}^{x}=t_{0, \mu\nu}^{x}+t_{0, \mu\nu}^{y}$ and $T_{i, \mu\nu}^{x}=t_{i, \mu\nu}^{x}$~($i\neq0$).
We can formally give $qL_{x}$ right eigenstates with eigenenergies $\epsilon^{\mu}(\beta_{\alpha})$ for the above Hamiltonian,
\begin{equation}
\ket{\Phi_{\alpha, y}^{R,\mu}}=\sum_{x=1}^{L_{x}}\sum_{j=1}^{N}\beta_{\alpha, j}^{x}\ket{\phi_{\alpha, y}^{Rj, \mu}}\ket{x}: =\sum_{x=1}^{L_{x}}\sum_{\nu=1}^{q}\tilde{\phi}^{R,\mu\nu}_{\alpha x}\ket{\nu}\ket{x},
\end{equation}
where $\alpha=1, 2, \ldots, L_{x}$ and $\mu=1, 2, \ldots, q$. We denote that $\tilde{\phi}^{R,\mu\nu}_{\alpha x}$ contains all the contributions from solutions $\beta_{j}$ with its multipliers $s_{j}$, of which the detail is given in Ref.~\cite{alase2017}. Focusing on the general forms of the solutions, we do not elaborate $\beta_{j}$ with its multiplier $s_{j}$ here. If we impose PBC along $x$ direction, we consider the standard Bloch theorem with $k_{x}: =-i\log\beta_{\alpha}=\frac{2\pi}{L_{x}}\alpha$~[$\alpha=0, 1, \ldots, (L_{x}-1)$], while if imposing OBC we extend that to the generalized Bloch theorem~\cite{alase2017}. In non-Hermitian systems, $|\beta_{\alpha}|\neq1$ does indicate the skin effect of the continuous bulk bands.

Using biorthogonal relation of the eigenstates,  we can diagonalize the single-particle Hamiltonian of $\hat{H}_{y}$ to diagonal eigenenergy matrix $\left\{\epsilon^{\mu}(\beta_{\alpha})\right\}$ in the right eigenstate basis $\left\{\ket{\Phi^{R,\mu}_{\alpha, y}}\right\}$~(see Appendix~\ref{appendix: BD} for details)
\begin{equation}
\epsilon=U_{L}^{\dagger}\cdot H_{y}\cdot U_{R}.
\end{equation}
The remaining inter-layer hopping terms along the $y$ direction of the total Hamiltonian are thus similarly transformed by
\begin{equation}
\mathbb{T}^{y}_{j}=U_{L}^{\dagger} T^{y}_{j} U_{R},
\end{equation}
where $(\mathbb{T}^{y}_{j})_{\alpha\mu, \beta\nu}=\sum_{i=1}^{L_{x}}\sum_{\rho, \sigma=1}^{q}\tilde{\phi}^{L,\mu\rho*}_{\alpha i}(t^{y}_{j})_{\rho\sigma}\tilde{\phi}^{R,\nu\sigma}_{\beta i}$ with $\alpha=1, \ldots, L_{x}$ and $j=-R_{y}, \ldots, \hat{0}, \ldots, R_{y}$~(see Appendix~\ref{appendix: BD}),
\begin{displaymath}
T^{y}_{j}=\begin{bmatrix}
t_{j}^{y}&\ldots&0\\
\vdots&\ddots&\vdots\\
0&\ldots&t_{j}^{y}
\end{bmatrix}_{L_{x}\times L_{x}},
\, \, \, \, \,
t^{y}_{j}=\begin{bmatrix}
t_{j, 11}^{y}&\ldots&t_{j, 1q}^{y}\\
\vdots&\ddots&\vdots\\
t_{j, q1}^{y}&\ldots&t_{j, qq}^{y}
\end{bmatrix}_{q\times q}.
\end{displaymath}
The entry below \,\,$\hat{ }$\,\, means excluded.
We finally obtain a 1D effective Hamiltonian along $y$ direction, in the biorthogonal basis along $x$ direction, which is given as
\begin{equation}
\label{eff}
\hat{H}_{eff}=\sum_{y=1}^{L_{y}}\sum_{j=-R_{y}}^{R_{y}}\sum_{\alpha,\beta=1}^{L_{x}}\sum_{\mu,\nu=1}^{q}\hat{\Phi}_{\alpha,y+j}^{R, \mu\dagger}\cdot(\mathbb{T}^{y}_{j})_{\alpha\mu, \beta\nu}\cdot\hat{\Phi}_{\beta,y}^{L, \nu},
\end{equation}
In this Hamiltonian, $\hat{\Phi}^{R, \mu\dagger}_{\alpha, y}=\sum_{x=1}^{L_{x}}\sum_{\nu=1}^{q}\tilde{\phi}^{R, \mu\nu}_{\alpha x}\hat{c}^{\nu\dagger}_{x, y}$ and $\hat{\Phi}_{\beta,y}^{L, \nu}$ is the annihilation operator of the corresponding biorthogonal left eigenstate~[see Appendix~\ref{appendix: BD}].

Here we give the difference between the construction of topological phases via coupled layers and our nested tight-binding formalism. For the former, people couple the fermionic operators ($\hat{c}$) between different layers; for the latter, we couple the operators ($\hat{\Phi}^{R},\hat{\Phi}^{L}$)~(corresponding biorthogonal eigenstates of a single layer under OBC) between different layers. In other words, we first solve the eigenstates for a single-layer Hamiltonian~(such as $y$-layer Hamiltonian $H_{y}(k_{x})$) under OBC~($x$ OBC), and then couple the fermionic operators of these eigenstates between different layers in the perpendicular direction~($y$ direction), by which we obtain the 1D effective Hamiltonian $\hat{H}_{eff}$. The internal degrees of freedom of this effective Hamiltonian are exactly the eigenstates of the single-layer Hamiltonian~($H_{y}(k_{x})$) under OBC~($x$ OBC). After further solving this effective Hamiltonian under OBC~($y$ OBC), we arrive at the results under full OBC. By our formalism, we can clarify the meaning and configuration of $TT$, $ST$, and $SS$ corner modes. Actually, our formalism also validates more general 2D tight-binding Hamiltonian, which contains the hopping $t_{i,j}$ between $\hat{c}_{x+i,y+j}^{\dagger}$ and $\hat{c}_{x,y}$ for any integer values $i,j$, i.e., the coupling between $k_{x}$ and $k_{y}$ in momentum space. Subsequently, $T_{j}^{y}(j=1,2,\ldots)$ is no longer block diagonal~(banded block) and $\hat{H}_{eff}$ is more intricate. To elucidate the nested tight-binding formalism, we research the models with nearest-neighbor hopping. The more general models will be studied in future work.

The nested tight-binding formalism is valid to investigate $TT$ and $ST$ modes when the edge-state subspace part of $\hat{H}_{eff}$ is independent from the bulk part, in other words, the degrees of freedom of topological edge eigenstates along $x$ direction are not coupled with that of bulk eigenstates in $\hat{H}_{eff}$. In next part, this formalism will be further confirmed for the four-band model (complete block diagonalization of $\hat{H}_{eff}$ for typical parameter choices) to obtain $TT$ and $ST$ corner modes analytically, and the meaning and configuration of these corner modes will be clarified. Moreover, the block diagonal result also applies to the 2D model with extrinsic second-order corner modes~\cite{okugawa}. When the skin bulk block part of $\hat{H}_{eff}$ is independent from the edge-state subspace part, the $\hat{H}_{eff}$ induces the pure second-order skin effect, which is the combination of skin bulk eigenstates along the $x$ direction~(internal degrees of freedom of $\hat{H}_{eff}$) and the skin effect of $\hat{H}_{eff}$ along the $y$ direction. In other words, skin bulk block of $\hat{H}_{eff}$ also has nontrivial point-gap topology indicating the existence of the skin effect along the $y$ direction. The simplest 2D model~[Eq.~(\ref{2ds})] with pure $SS$ modes has already been given in Sec.~\ref{WNSS}, of which the effective Hamiltonian is easily obtained as Eq.~(\ref{1deff}). Although it is cumbersome to analyze the $SS$ modes for a more complicated model due to the complexity of skin bulk states, the numerical result also can indicate the $SS$ modes. Hence, we focus on the generally analyzable $ST$ and $TT$ modes hereinafter.

\subsection{The four-band model}
\label{4bandmodel}

Consider a 2D non-Hermitian four-band model~\cite{kawabaras, lee2019}
\begin{equation}
H(\vec{k})=\begin{bmatrix}
0&0&H_{1, -}&-H_{4, -}\\
0&0&H_{3, -}^{*}&H_{2, -}^{*}\\
H_{1, +}^{*}&H_{3, +}&0&0\\
-H_{4, +}^{*}&H_{2, +}&0&0
\end{bmatrix},
\end{equation}
where $H_{j, \pm}=t_{x}\pm\delta_{j}+\lambda e^{i k_{x}}$ for $j=1, 2$ and $H_{j, \pm}=t_{y}\pm\delta_{j}+\lambda e^{i k_{y}}$ for $j=3, 4$, setting $t_{x}=t_{y}=t$ for simplicity. The Hermitian counterpart of this model $(\delta_{j}=0, j=1, 2, 3, 4)$ has already been investigated in Refs.~\cite{dipole2017, li2018}. Without any other parameter assignments, the Hamiltonian of this model only preserves sublattice symmetry $S^{-1}H(k)S=-H(k)$ with $S=\tau_{z}$. We set $\delta_{1}=-\delta_{2}=-\delta_{3}=\delta_{4}=\gamma$ for simplicity, from which we consider the model investigated in Ref.~\cite{kawabaras} with net nonreciprocities for both $x$ and $y$ directions, i.e.
\begin{eqnarray}
&&H(\vec{k})=\left(t+\lambda\cos k_{x}\right)\tau_{x}-\left(\lambda\sin k_{x}+i \gamma\right)\tau_{y}\sigma_{z}
\nonumber \\
&&\qquad +\left(t+\lambda\cos k_{y}\right)\tau_{y}\sigma_{y}+\left(\lambda\sin k_{y}+i \gamma\right)\tau_{y}\sigma_{x}.
\label{2d4b}
\end{eqnarray}
Besides sublattice symmetry, this Hamiltonian also preserves mirror-rotation symmetry $M_{xy}^{-1}H(k_{x}, k_{y})M_{xy}=H(k_{y}, k_{x})$ with $M_{xy}=C_{4}M_{y}$, while its Hermitian counterpart preserves both mirror symmetries $M_{x}=\tau_{x}\sigma_{z}, M_{y}=\tau_{x}\sigma_{x}$ and four-fold rotational symmetry $C_{4}=[(\tau_{x}-i\tau_{y})\sigma_{0}-(\tau_{x}+i\tau_{y})(i\sigma_{y})]/2$.

Applying our nested tight-binding formalism, we study a single $x$-layer Hamiltonian
\begin{equation}
\hat{H}_{s}=\sum_{y}(\hat{c}^{\dagger}_{y}m_{0}\hat{c}_{y}+\hat{c}^{\dagger}_{y}t_{y}^{+}\hat{c}_{y+1}+\hat{c}^{\dagger}_{y+1}t_{y}^{-}\hat{c}_{y}),
\end{equation}
where
\begin{eqnarray}
&& m_{0}=t(\tau_{x}+\tau_{y}\sigma_{y})+i\gamma(\tau_{y}\sigma_{x}-\tau_{y}\sigma_{z}),
\nonumber\\
&& t_{y}^{+}=\frac{\lambda}{2}(\tau_{y}\sigma_{y}-i\tau_{y}\sigma_{x}),
\nonumber\\
&& t_{y}^{-}=\frac{\lambda}{2}(\tau_{y}\sigma_{y}+i\tau_{y}\sigma_{x}).
\end{eqnarray}
As usual, we assume the eigenstate of the Hamiltonian under OBC is
\begin{eqnarray}
\ket{\psi}=\sum_{y=1}^{L_{y}}\beta^{y}\ket{y}\ket{\phi},\nonumber
\end{eqnarray}
where $\ket{\phi}$ is a four-component column vector representing the internal degrees of freedom. From the eigen-equation $\hat{H}_{s}\ket{\psi}=\epsilon\ket{\psi}$, the secular equation of the bulk equation reads
\begin{eqnarray}
\det(t_{y}^{-}\beta^{-1}+m_{0}+t_{y}^{+}\beta-\epsilon)=0,\nonumber
\end{eqnarray}
which gives
\begin{eqnarray}
\frac{1}{\beta^{2}}[\lambda(t+\gamma)\beta^{2}+(2t^{2}-2\gamma^{2}+\lambda^{2}-\epsilon^{2})\beta+\lambda(t-\gamma)]^{2}=0.
\nonumber
\end{eqnarray}
The four nonzero finite bulk solutions satisfy the relation
\begin{eqnarray}
\beta_{1}^{b}\beta_{2}^{b}=\beta_{3}^{b}\beta_{4}^{b}=\frac{t-\gamma}{t+\gamma}.\nonumber
\end{eqnarray}
As derived in Refs.~\cite{yao2018,yokomizo2019}, the continuous condition gives
\begin{eqnarray}
|\beta_{1}^{b}|=|\beta_{2}^{b}|=|\beta_{3}^{b}|=|\beta_{4}^{b}|=\sqrt{\bigg|\frac{t-\gamma}{t+\gamma}\bigg|},\nonumber
\end{eqnarray}
which we call the skin effect indicator~(left-localized when $|t|>|\gamma|$) along the $y$ direction (same for the $x$ direction).
In momentum space, the Hamiltonian of this model is
\begin{eqnarray}
H_{s}(k_{y})=t(\tau_{x}+\tau_{y}\sigma_{y})+i\gamma(\tau_{y}\sigma_{x}-\tau_{y}\sigma_{z})
\nonumber\\
+\lambda\cos k_{y}\tau_{y}\sigma_{y}+\lambda\sin k_{y}\tau_{y}\sigma_{x}.
\end{eqnarray}
The above Hamiltonian possesses four nonzero-energy edge states under OBC, which contribute to the second-order corner-localized modes. We emphasize that the gapped edge states of Eq.~(\ref{2d4b}) under $y$ OBC/$x$ PBC~(or $x$ OBC/$y$ PBC) are not protected by bulk-energy band topology due to the vanishing Chern number~\cite{kawabaras}. We need to research the second-order topological modes.

Firstly, we solve the left-localized edge states of the Hamiltonian $H_{s}$ under OBC~\cite{kawabata2018,alase2017}. The bulk and boundary equations are
\begin{eqnarray}
\label{bulkequ}
(t_{y}^{-}\beta^{-1}+m_{0}+t_{y}^{+}\beta)\ket{\phi}=\epsilon\ket{\phi},
\\
(m_{0}+t_{y}^{+}\beta)\ket{\phi}=\epsilon\ket{\phi}.
\end{eqnarray}
We can obtain $\ket{\phi}$ the kernels of $t_{y}^{-}$, which are
\begin{eqnarray}
&&\ket{u_{1}}=u_{1}\ket{\sigma},
\nonumber\\
&&\ket{u_{2}}=u_{2}t\ket{\sigma},
\end{eqnarray}
where $u_{1}=(0, 0, 0, 1),u_{2}=(0, 1, 0, 0)$. We denote $\ket{\sigma}=(\ket{1}, \ket{2}, \ket{3}, \ket{4})^{T}$ as the internal degrees of freedom. Substituting the linear combination of $\ket{u_{1, 2}}$ into the bulk equation, we obtain two solutions as
\begin{equation}
\ket{\phi_{L}^{\pm}}=\ket{u_{1}}\pm r\ket{u_{2}}: =\phi_{L}^{\pm}\ket{\sigma},
\end{equation}
where $r=\sqrt{\frac{t+\gamma}{t-\gamma}}$. Accordingly, the two left-localized solutions with energies $\epsilon_{\pm}=\pm\sqrt{(t+\gamma)(t-\gamma)}$, are
\begin{equation}
\label{edgesolution}
\ket{\psi_{L}^{\pm}}=\sum_{y=1}^{L_{y}}\beta_{1}^{y}\ket{y}\ket{\phi_{L}^{\pm}},
\end{equation}
where $\beta_{1}=-\frac{t-\gamma}{\lambda}$.
Additionally, the left-localized condition $|\beta_{1}|<1$ guarantees the above solutions automatically satisfying the right boundary equation for large enough $L_{y}$.

Secondly, the right-localized edge states are given as
\begin{equation}
\ket{\psi_{R}^{\pm}}=\sum_{y=1}^{L_{y}}\beta_{2}^{-L_{y}+y}\ket{y}\ket{\phi_{R}^{\pm}},
\end{equation}
with respective energies $\epsilon_{\pm}$, where $\beta_{2}=-\frac{\lambda}{t+\gamma}$ and
\begin{equation}
\ket{\phi_{R}^{\pm}}=\ket{v_{1}}\pm r^{-1}\ket{v_{2}}: =\phi_{R}^{\pm}\ket{\sigma},
\end{equation}
with
\begin{eqnarray}
\label{procedure}
&&\ket{v_{1}}=v_{1}\ket{\sigma},
\nonumber\\
&&\ket{v_{2}}=v_{2}\ket{\sigma},
\end{eqnarray}
$v_{1}=(0, 0, 1, 0), v_{2}=(1, 0, 0, 0)$.

We find that the numerical results of $U_{L}^{\dagger} T_{x} U_{R}$ and $U_{L}^{\dagger}T_{x}^{\dagger} U_{R}$ are both block-diagonal, of which each block is a $4\times 4$ matrix in this model. Therefore we can deal with the edge-state subspace independently. However, we have to find the corresponding left eigenstates of the right eigenstates $\ket{\psi_{L, R}^{\pm}}$ due to the biorthogonal relation of the non-Hermitian Hamiltonian. So we solve the edge states for eigen-equation $\hat{H}_{s}^{T}\ket{\psi^{'}}^{*}=\epsilon\ket{\psi^{'}}^{*}$. With the same procedure solving right eigenstates, we have the left eigenstates
\begin{eqnarray}
&&\ket{\psi_{L}^{'\pm}}^{*}=\sum_{y=1}^{L_{y}}\beta_{2}^{-y}\ket{y}\ket{\phi_{L}^{'\pm}},
\nonumber\\
&&\ket{\psi_{R}^{'\pm}}^{*}=\sum_{y=1}^{L_{y}}\beta_{1}^{L_{y}-y}\ket{y}\ket{\phi_{R}^{'\pm}},
\end{eqnarray}
where
\begin{eqnarray}
&&\ket{\phi_{L}^{'\pm}}=\ket{u_{1}}\pm r^{-1}\ket{u_{2}}: =\phi_{L}^{'\pm}\cdot\ket{\sigma},
\nonumber\\
&&\ket{\phi_{R}^{'\pm}}=\ket{v_{1}}\pm r\ket{v_{2}}: =\phi_{R}^{'\pm}\cdot\ket{\sigma}.
\end{eqnarray}
We construct the biorthogonal diagonalized matrices in the edge-state subspace as
\begin{eqnarray}
&&U_{R}^{edge}=\bigg((\phi_{L}^{+})^{T}, (\phi_{L}^{-})^{T}, (\phi_{R}^{+})^{T}, (\phi_{R}^{-})^{T}\bigg),
\nonumber\\
&&U_{L}^{edge\dagger}=\bigg((\phi_{L}^{'+})^{T},(\phi_{L}^{'-})^{T}, (\phi_{R}^{'+})^{T}, (\phi_{R}^{'-})^{T}\bigg)^{T}.
\end{eqnarray}
After biorthogonally normalizing of the right and left eigenstates, we finally arrive at the effective Hamiltonian in edge-state subspace,
\begin{equation}
\label{4bandeff}
\hat{H}_{j}^{edge}=\sum_{x=1}^{L_{x}}(\hat{\phi}^{j\dagger}_{x}\epsilon_{0}\hat{\phi}_{x}^{'j}+\hat{\phi}^{j\dagger}_{x}t_{j}^{+}\hat{\phi}_{x+1}^{'j}+\hat{\phi}^{j\dagger}_{x+1}t_{j}^{-}\hat{\phi}_{x}^{'j}),
\end{equation}
where $j=L, R$ corresponding to left- or right-localized edge-state subspace and
\begin{eqnarray}
 &&\hat{\phi}^{j\dagger}_{x}=(\hat{\phi}^{j+\dagger}_{x}, \hat{\phi}^{j-\dagger}_{x}),
\nonumber\\
 &&\hat{\phi}^{'j}_{x}=(\hat{\phi}^{'j+}_{x}, \hat{\phi}^{'j-}_{x})^{T}.
\end{eqnarray}
The fermionic operators in above equations are
\begin{eqnarray}
&&\hat{\phi}^{L\pm\dagger}_{x}=\sum_{y=1}^{L_{y}} \mathcal{N}_{L}^{y}\beta_{1}^{y}(\hat{c}_{x, y}^{1\dagger}, \hat{c}_{x, y}^{2\dagger}, \hat{c}_{x, y}^{3\dagger}, \hat{c}_{x, y}^{4\dagger})\cdot(\phi_{L}^{\pm})^{T},
\nonumber\\
&&\hat{\phi}^{R\pm\dagger}_{x}=\sum_{y=1}^{L_{y}} \mathcal{N}_{R}^{y}\beta_{2}^{y-L_{y}}(\hat{c}_{x, y}^{1\dagger}, \hat{c}_{x, y}^{2\dagger}, \hat{c}_{x, y}^{3\dagger}, \hat{c}_{x, y}^{4\dagger})\cdot(\phi_{R}^{\pm})^{T},
\nonumber\\
&&\hat{\phi}^{'L\pm}_{x}=\sum_{y=1}^{L_{y}}\mathcal{N}_{L}^{y}\beta_{2}^{-y}\phi_{L}^{'\pm}\cdot(\hat{c}_{x, y}^{1}, \hat{c}_{x, y}^{2}, \hat{c}_{x, y}^{3}, \hat{c}_{x, y}^{4})^{T},
\nonumber\\
&&\hat{\phi}^{'R\pm}_{x}=\sum_{y=1}^{L_{y}}\mathcal{N}_{R}^{y}\beta_{1}^{-y+L_{y}}\phi_{R}^{'\pm}\cdot(\hat{c}_{x, y}^{1}, \hat{c}_{x, y}^{2}, \hat{c}_{x, y}^{3}, \hat{c}_{x, y}^{4})^{T},
\end{eqnarray}
where the biorthogonally normalized coefficients $\mathcal{N}_{j}^{y}$ $(j=L,R)$ are given as
\begin{eqnarray}
&&\mathcal{N}_{L}^{y}=[2\sum_{y=1}^{L_{y}}(\beta_{1}\beta_{2}^{-1})^{y}]^{-1/2},
\nonumber\\
&&\mathcal{N}_{R}^{y}=[2\sum_{y=1}^{L_{y}}(\beta_{1}^{-1}\beta_{2})^{-L_{y}+y}]^{-1/2}.
\end{eqnarray}
The hopping matrices are given by
\begin{eqnarray}
\epsilon_{0}=\sqrt{(t+\gamma)(t-\gamma)}\sigma_{z}\nonumber
\end{eqnarray}
and
\begin{eqnarray}
t^{\pm}=\frac{1}{2} U^{edge\dagger}_{L}\cdot t_{x}^{\pm}\cdot U^{edge}_{R}=\begin{bmatrix}
t_{L}^{\pm}&0\\
0&t_{R}^{\pm}
\end{bmatrix},\nonumber
\end{eqnarray}
where
\begin{equation}
t_{L}^{\pm}=\frac{\lambda}{2}r^{\pm}\begin{bmatrix}
1&\mp1\\
\pm1&-1
\end{bmatrix}
\end{equation}
and
\begin{equation}
t_{R}^{\pm}=\frac{\lambda}{2}r^{\pm}\begin{bmatrix}
1&\pm1\\
\mp1&-1
\end{bmatrix}.
\end{equation}
This effective Hamiltonian Eq.~(\ref{4bandeff}) is the main result of applying our nested tight-binding formalism to the four-band model. Here, we clarify the meaning and configuration of $TT$ and $ST$ corner modes. The superscript $edge$ means this Hamiltonian is in the edge-state subspace along the $y$ direction, which contributes the topological edge~($T$) modes. Therefore, combining with the skin bulk~($S$) modes and topological edge~($T$) modes deduced from $H_{j}^{edge}$ under $x$ OBC, we obtain the $ST$ and $TT$ corner modes under full OBC respectively.
\begin{figure}
	\centering
	\subfigure[]{\includegraphics[width=0.265\textwidth]{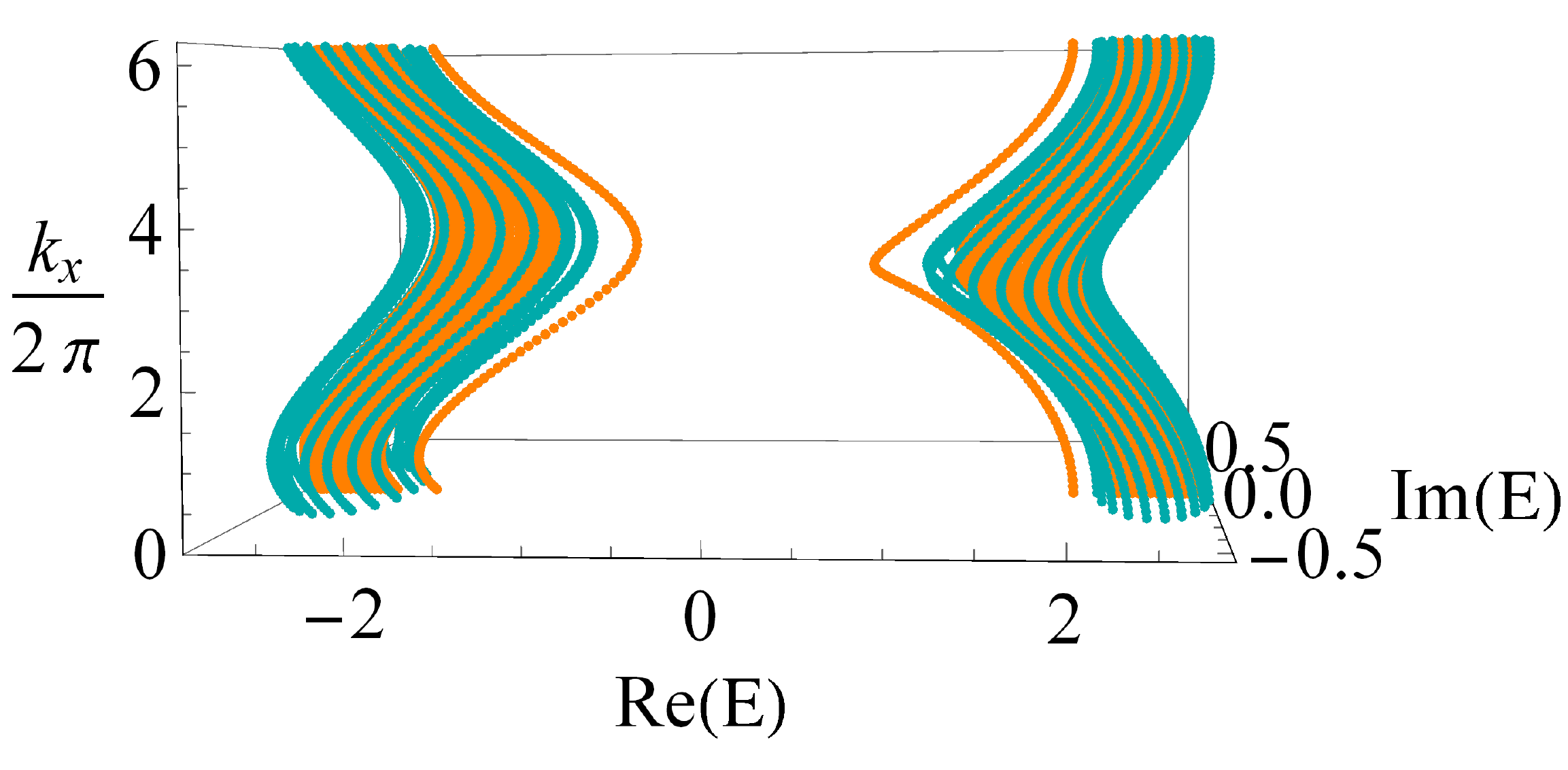}}
	\subfigure[]{\includegraphics[width=0.19\textwidth]{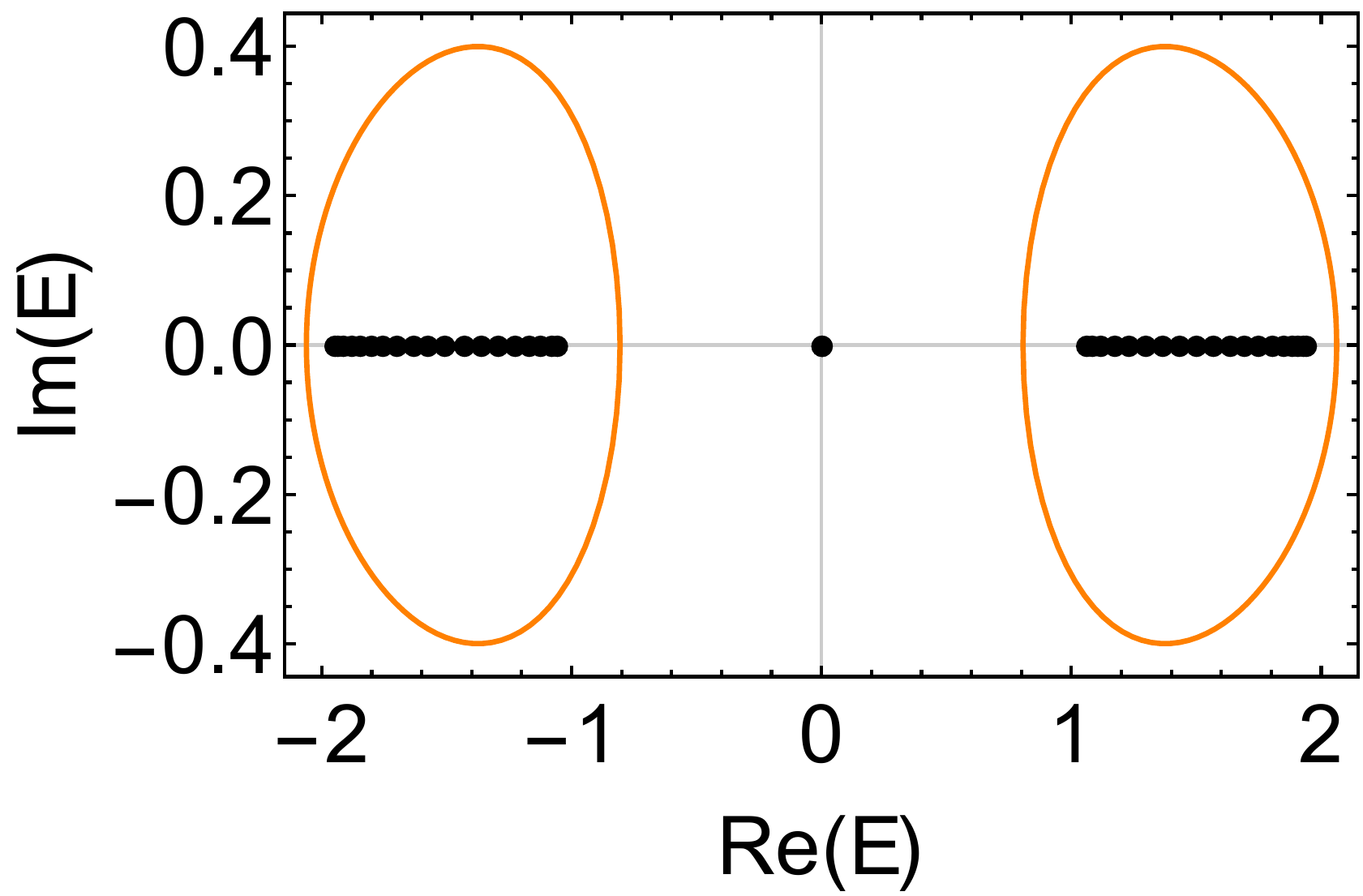}}\\
	\subfigure[]{\includegraphics[width=0.23\textwidth]{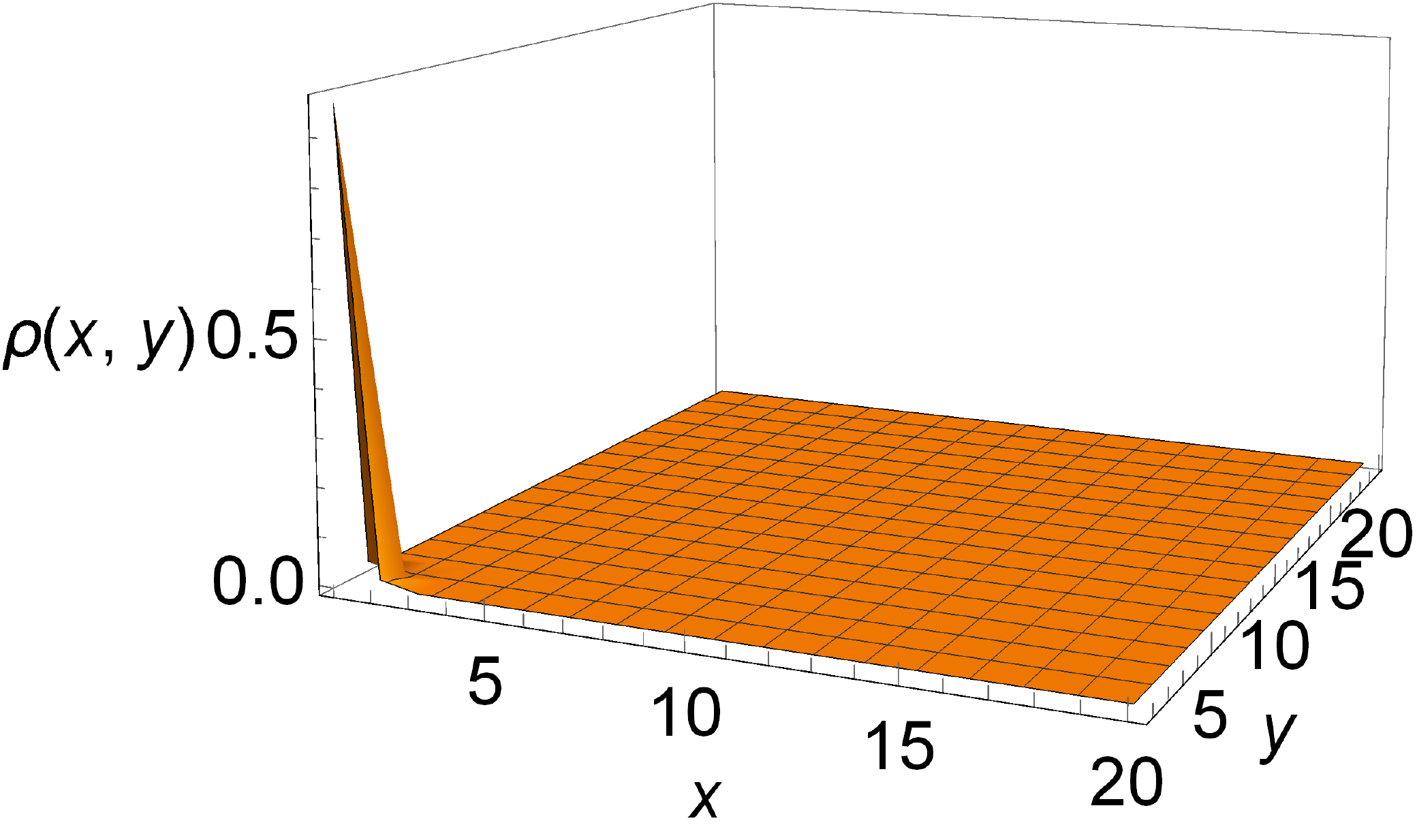}}
	\subfigure[]{\includegraphics[width=0.24\textwidth]{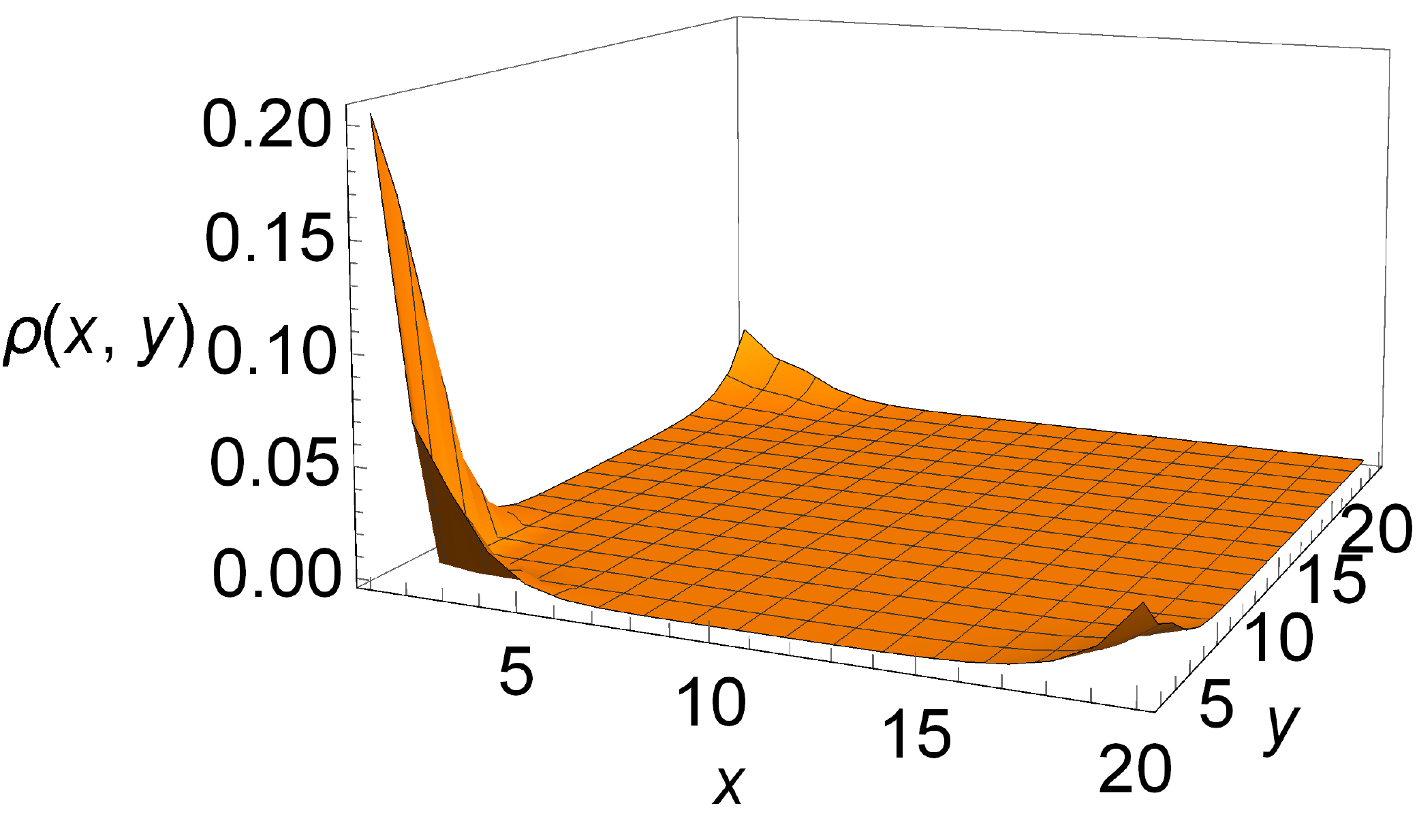}}\\
	\caption{Complex energy spectra of the four band model in Eq.~(\ref{2d4b}) with parameters $t=0. 6, \lambda=1. 5, \gamma=0. 4$. The number of unit cells is $20\times 20$. (a) Spectra under double-PBC (cyan) and $x$ OBC/$y$ PBC (orange) are plotted in the $E$-$k_{y}$ space. (b) The two orange loops, which are projected from the isolated edge spectra in (a), deduce the $ST$ modes (black lines) lying within the orange loops and four degenerate zero-energy $TT$ modes (black point on the origin). The typically localized zero-energy $TT$ mode and nonzero-energy $ST$ mode with energy $E_{ST}=-1.06094$ are plotted in (c) and (d), respectively.}
	\label{fig: 3}
\end{figure}

The edge effective Hamiltonian, Eq.~(\ref{4bandeff}), in momentum space is
\begin{equation}
\label{4deff}
H_{j}^{edge}(k_{x})=t^{-}_{j}e^{-ik_{x}}+\epsilon_{0}+t^{+}_{j}e^{ik_{x}},
\end{equation}
and the energy spectra under PBC read as
\begin{equation}
\epsilon_{j}^{2}(k_{x})=t^{2}-\gamma^{2}+\lambda^{2}+\lambda[(t+\gamma)e^{ik_{x}}+(t-\gamma)e^{-ik_{x}}],\nonumber
\end{equation}
where $j=L, R$. They form two orange loops localized on both sides of the imaginary axis in the complex energy plane~[Fig.~\ref{fig: 3}(b)], which are exactly projected from the $k_{x}$ dependent $x$ PBC/$y$ OBC edge-state subspace spectra [isolated orange lines in Fig.~\ref{fig: 3}(a)]. These two loops depict the skin effect of $H_{j}^{edge}$ under OBC along the $x$ direction leading to the $ST$ modes~\cite{lee2019} under full OBC, which are plotted as black lines lying within the orange loops in Fig.~\ref{fig: 3}(b). The skin effect indicator for $H_{j}^{edge}$ is also $|\rho|=\sqrt{|\frac{t-\gamma}{t+\gamma}|}$, which implies the localization of all bulk states at the left side when $|t|>|\gamma|$. Together with the edge-state subspace along the $y$ direction, the four zero $TT$ modes are localized at the four corners and the $ST$ modes are localized at the low-left and up-left corners when  $|\lambda|>|t-\gamma|,  |t+\gamma|$. The four zero-energy corner modes localized at low-left (LL), low-right (LR), up-left (RL) and up-right (RR) can be written as
\begin{eqnarray}
&&\ket{\Psi_{LL}}=\mathcal{N}_{L}^{x}\mathcal{N}_{L}^{y}\sum_{x=1}^{L_{x}}\sum_{y=1}^{L_{y}}\beta_{1}^{x}\beta_{1}^{y}\big[\ket{\phi_{L}^{+}}-\ket{\phi_{L}^{-}}\big]\ket{x}\ket{y},
\nonumber\\
&&\ket{\Psi_{LR}}=\mathcal{N}_{L}^{x}\mathcal{N}_{R}^{y}\sum_{x=1}^{L_{x}}\sum_{y=1}^{L_{y}}\beta_{1}^{x}\beta_{2}^{y-L_{y}}\big[\ket{\phi_{R}^{+}}+\ket{\phi_{R}^{-}}\big]\ket{x}\ket{y},
\nonumber\\
&&\ket{\Psi_{RL}}=\mathcal{N}_{R}^{x}\mathcal{N}_{L}^{y}\sum_{x=1}^{L_{x}}\sum_{y=1}^{L_{y}}\beta_{2}^{x-L_{x}}\beta_{1}^{y}\big[\ket{\phi_{L}^{+}}+\ket{\phi_{L}^{-}}\big]\ket{x}\ket{y},
\nonumber\\
&&\ket{\Psi_{RR}}=\mathcal{N}_{R}^{x}\mathcal{N}_{R}^{y}\sum_{x=1}^{L_{x}}\sum_{y=1}^{L_{y}}\beta_{2}^{x-L_{x}}\beta_{2}^{y-L_{y}}\big[\ket{\phi_{R}^{+}}-\ket{\phi_{R}^{-}}\big]\ket{x}\ket{y},
\nonumber\\
\end{eqnarray}
where the normalized coefficients read~($\delta=x,y$)
\begin{eqnarray}\nonumber
&&\mathcal{N}_{L}^{\delta}=[2\sum_{\delta=1}^{L_{\delta}}(\beta_{1}\beta_{2}^{-1})^{\delta}]^{-1/2},
\nonumber\\
&&\mathcal{N}_{R}^{\delta}=[2\sum_{\delta=1}^{L_{\delta}}(\beta_{1}^{-1}\beta_{2})^{-L_{\delta}+\delta}]^{-1/2}\nonumber.
\end{eqnarray}
Noteworthily, $\ket{\Psi_{LL}}$ and $\ket{\Psi_{RR}}$ are invariant under mirror-rotation transformation when $L_{x}=L_{y}$, while $\ket{\Psi_{LR}}$ and $\ket{\Psi_{RL}}$ are transformed to each other.

However, the $TT$ modes are all numerically localized at the low-left corner~[Fig.~\ref{fig: 3}(c)], while the $ST$ modes at the low-left corner with larger amplitude, low-right and up-left corners with smaller amplitude~[Fig.~\ref{fig: 3}(d)]. In addition, the pure $SS$ modes are also all localized at the low-left corner by numerical result.
The analytical and numerical results are seemingly inconsistent, but we notice that the linear combinations of energy degenerate states are also the eigenstates of the Hamiltonian with the same energy. Based on this consideration, we can eliminate this inconsistence, on which we will elaborate in the following.

Let us focus on the 1D Hamiltonian, Eq.~(\ref{4bandeff}), to explore the difference between analytical and numerical results. Following the procedure of Eqs.~(\ref{bulkequ})-(\ref{procedure}), we figure out the topological zero edge modes for $H_{j}^{edge}$ under OBC analytically. Writing the two zero modes of $H_{L}^{edge}$ as an example,
\begin{eqnarray}
&&\psi_{0, L}=\mathcal{N}_{L}^{x}\sum_{x=1}^{L_{x}}(-\frac{t-\gamma}{\lambda})^{x}(1,-1)^{T},
\nonumber\\
&&\psi_{0, R}=\mathcal{N}_{R}^{x}\sum_{x=1}^{L_{x}}(-\frac{\lambda}{t+\gamma})^{x-L_{x}}(1, 1)^{T}.
\end{eqnarray}
Requiring $|\lambda|>|t-\gamma|, |t+\gamma|$, the two solutions are localized on the left and right sides along the $x$ direction respectively. However, the two numerical edge states are localized only on left side when we set parameters as $t=0. 6, \gamma=0. 4, \lambda=1. 5$.
After carefully comparing these solutions, we find that the numerical solutions are precisely the linear combination of the two analytical zero modes
\begin{eqnarray}
\psi_{0}=\pm\alpha_{L}\psi_{0, L}-\alpha_{R}\psi_{0, R},\nonumber
\end{eqnarray}
but the coefficient $\alpha_{R}$ is much smaller than $\alpha_{L}$, leading to the two zero modes both localized on the left side. In addition, the two combination solutions are not orthogonal normalization since they satisfy biorthogonal relation in non-Hermitian system.

Motivated by the 1D case, we obtain the four second-order zero modes localized at the low-left corner by linear combination of the analytical four zero-energy corner modes
\begin{equation}
\ket{\Psi_{k}}=\sum_{i, j=L, R}\alpha_{ij}^{k}\ket{\Psi_{ij}},
\end{equation}
where $k=1, 2, 3, 4$ denotes the four zero-energy corner modes. The domination of the coefficient $\alpha_{LL}$ induces the final four zero modes all localized at low-left corner, which are indeed consistent with the numerical result~[Fig.~\ref{fig: 3}(c)].  Although the difference between analytical and numerical results exists, the second-order topological invariant, which is constructed in~Ref.~\cite{kawabaras} by utilizing mirror-rotation symmetry $M_{xy}$, characterizes the number of $TT$ zero-energy corner modes not the localization behavior of those. Additionally, based on our tight-binding formalism, the point-gap topology of the edge-state subspace effective Hamiltonian $H^{edge}_{j}(k_{x})$ protects $ST$ modes, i.e., the winding numbers of orange loops in Fig.~\ref{fig: 3}(b) around corresponding $ST$ modes~[black lines in Fig.~\ref{fig: 3}(b)].

Due to the mirror-rotation symmetry, we can also obtain $ST$ modes analytically by first considering a single $y$-layer tight-binding model along the $x$ direction. Then we obtain low-left and low-right localized $ST$ modes, degenerate with the preceding low-left and up-left localized $ST$ modes~(the result by first considering the single $x$ layer Hamiltonian). By properly combining these $ST$ modes with degenerate energy~(i.e. larger coefficient for low-left localized $ST$ modes and smaller coefficients for up-left and low-right localized $ST$ modes), we can obtain the $ST$ modes consistent with the numerical result~[Fig.~\ref{fig: 3}(d)]. In addition, the $SS$ modes, all localized at the low-left corner, are obviously induced by the left-localized skin effect along both directions.

It is analyzable when we take $|\delta_{1}|=|\delta_{2}|$ and $|\delta_{3}|=|\delta_{4}|$. In general, the coupling terms between neighbor lattices can also be different, i.e., $\lambda_{1}, \lambda_{2}$ for $x$ and $y$ directions respectively. Following our nested tight-binding formalism, we solve the Hamiltonian for a single $y$-layer with net nonreciprocity. It is well known that $\sqrt{|\frac{t_{x}-\delta_{1}}{t_{x}-\delta_{2}}|}<(>)1$ indicates the skin bulk states localized on left (right) side along the $x$ direction. Moreover, the localization behavior of analytical edge states is determined by $\beta_{1}=-\frac{t_{x}-\delta_{1}}{\lambda_{1}}$ and $\beta_{2}=-\frac{\lambda_{1}}{t_{x}-\delta_{2}}$. As derived in Ref.~\cite{yao2018}, the merging-into-bulk condition yields the topological phase-transition points
\begin{equation}
|\beta_{1}|=|\beta_{2}|=\sqrt{|\frac{t_{x}-\delta_{1}}{t_{x}-\delta_{2}}|},
\end{equation}
resulting in $(t_{x}-\delta_{1})(t_{x}-\delta_{2})=\pm\lambda_{1}^{2}$. Noticing the nonreciprocity condition $\delta_{1}=-\delta_{2}=\gamma_{1}$, we obtain the phase-transition edge for the $x$ direction $t_{x}^{2}-\gamma_{1}^{2}=\pm\lambda_{1}^{2}$.
Following the above derivation of phase-transition edge, we obtain the similar result $t_{y}^{2}-\gamma_{2}^{2}=\pm\lambda_{2}^{2}$ for the effective Hamiltonian~[Eq.~(\ref{4bandeff})] in the edge-state subspace. Therefore we recover the phase diagram with boundary $t^{2}-\gamma^{2}=\pm\lambda^{2}$ in Ref.~\cite{kawabaras} taking $t_{x}=t_{y}=t$ and $\gamma_{1}=\gamma_{2}=\gamma$. Moreover, we introduce other parameter choices for the four-band model in Appendix~\ref{appendix: OPC}.

\subsection{The 2D model with extrinsic $ST$ modes}
\label{2dext}
We further consider a 2D model possessing extrinsic second-order corner modes, of which the second-order topological invariant has been given in Ref.~\cite{okugawa}. However, the $ST$ modes and $TT$ modes have not been distinguished, to which we apply our nested tight-binding formalism. The simple Hamiltonian~\cite{okugawa} of this model has two internal degrees of freedom and reads
\begin{eqnarray}
\label{ext}
H_{e}(\vec{k})=2t_{x}\cos k_{x}\tau_{0}-2ig_{x}\sin k_{x}\tau_{z}
\nonumber\\
-2it_{y}\cos k_{y}\tau_{y}-2ig_{y}\sin{k_{y}\tau_{x}},
\end{eqnarray}
where $t_{y}>g_{y}>0$ and $t_{x}>g_{x}>0$ without loss of generality.

The complex energy spectrum of single $y$-layer Hamiltonian  $H_{x}(k_{x})$ forms a loop, indicative of skin effect, while those of single $x$-layer Hamiltonian $H_{y}(k_{y})$ form pure imaginary lines, suppressing skin effect. For simplicity, we start from $H_{y}(k_{y})$ with two localized zero topological states,
\begin{equation}
H_{y}(k_{y})=-2it_{y}\cos k_{y}\tau_{y}-2ig_{y}\sin{k_{y}\tau_{x}}.
\end{equation}
We can easily work out the two localized zero modes taking odd lattice sites~(for even sites and details in Appendix~\ref{appendix:ESEM})
\begin{eqnarray}
\label{extso}
&&\ket{\psi_{L}}=\sum_{y=1}^{(L_{y}+1)/2}\beta^{2y-1}\ket{2y-1}\phi_{L},
\nonumber\\
&&\ket{\psi_{R}}=\sum_{y=1}^{(L_{y}+1)/2}\beta^{L_{y}-2y}\ket{2y-1}\phi_{R},
\end{eqnarray}
where $|\beta|=\sqrt{\frac{t_{y}-g_{y}}{t_{y}+g_{y}}}$ and $\phi_{L}=(0, 1)^{T}, \phi_{R}=(1, 0)^{T}$.
\begin{figure}
	\centering
	\subfigure[]{\includegraphics[width=0.21\textwidth]{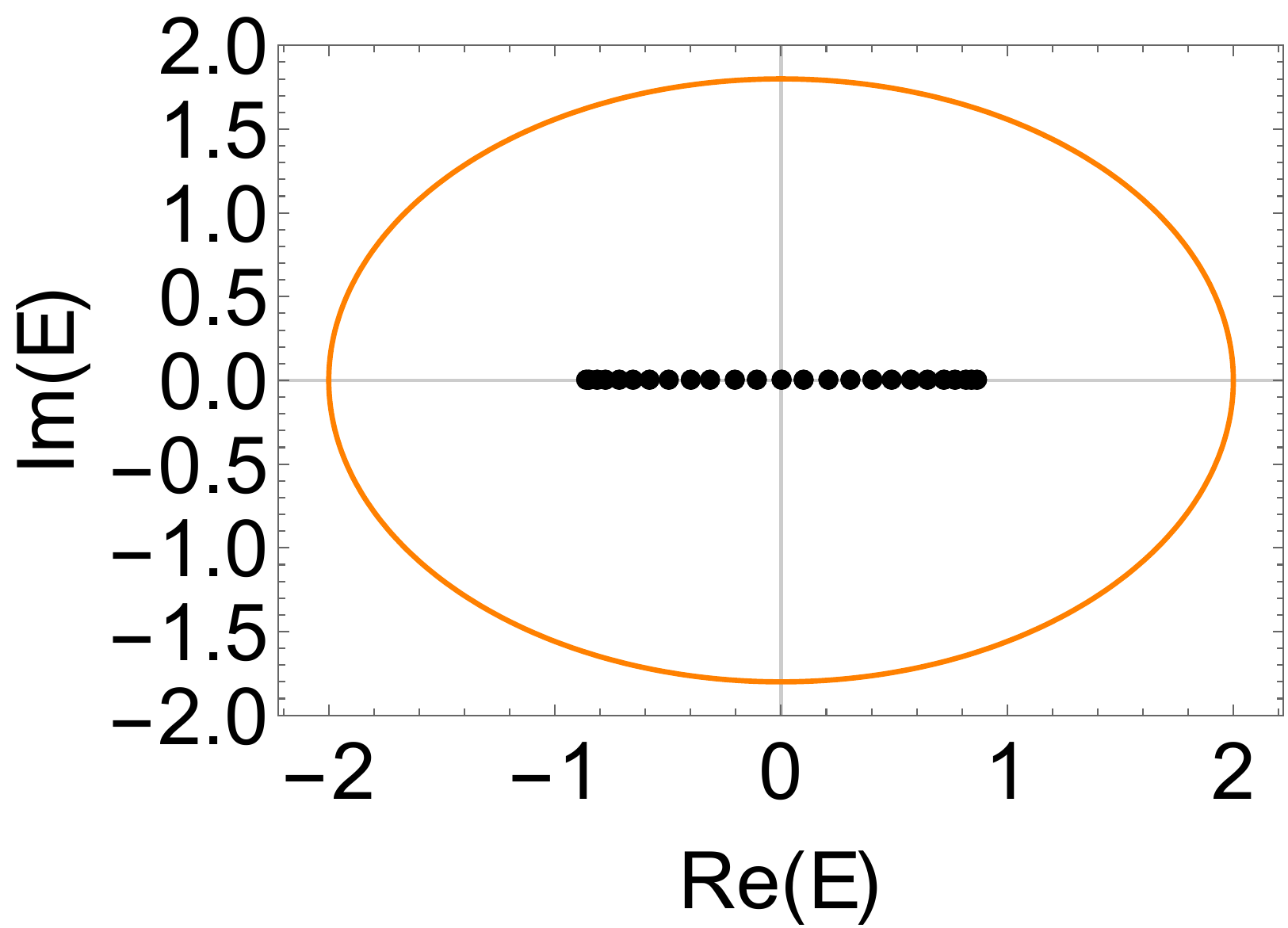}}
	\subfigure[]{\includegraphics[width=0.203\textwidth]{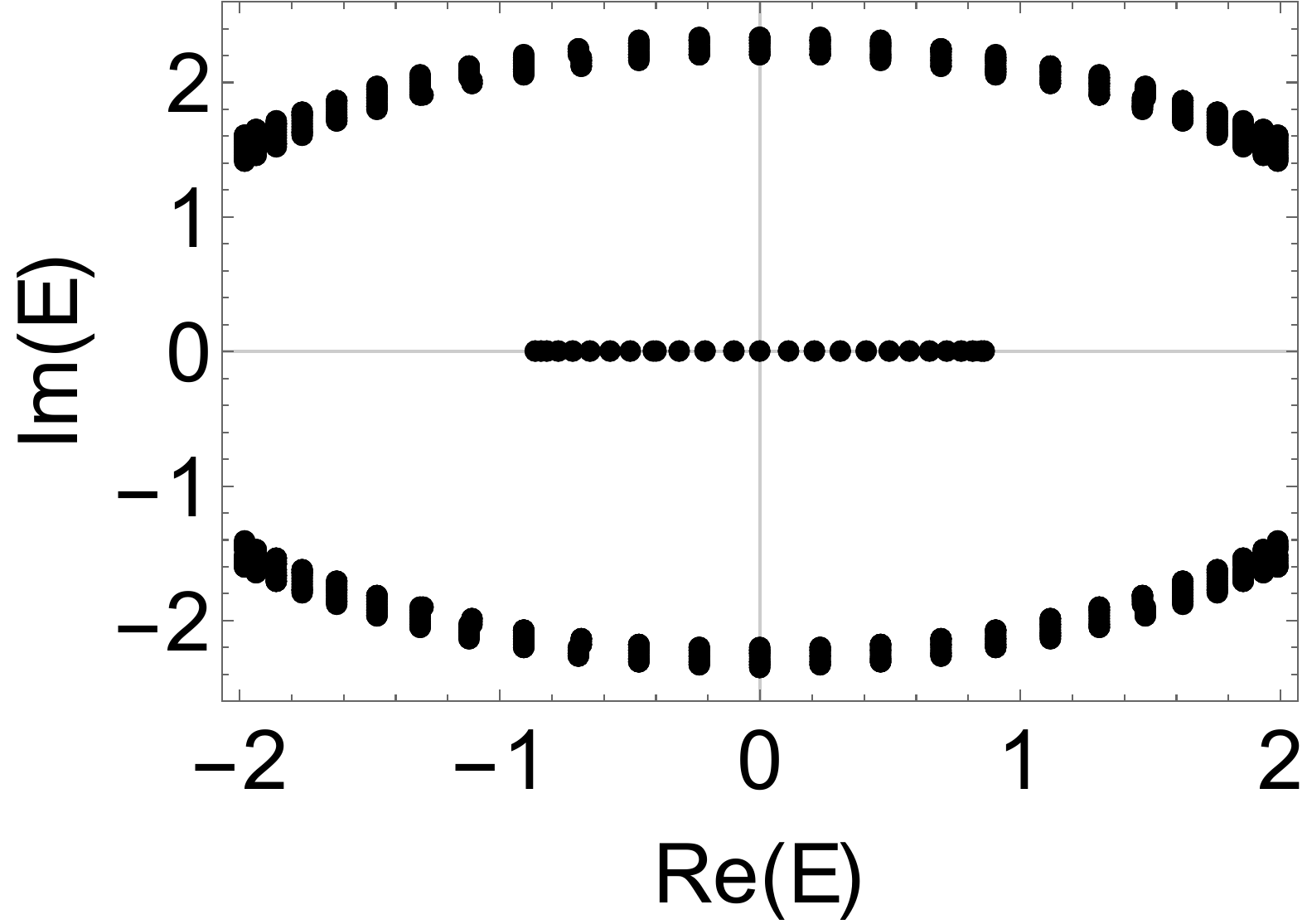}}\\
	\caption{(a) Complex energy spectrum for $H_{eff}(k_{x})$ in Eq.~(\ref{exteff}) under PBC (orange) surrounds that under OBC (black). (b) Complex energy spectra under full OBC for $H_{e}(\vec{k})$ in Eq.~(\ref{ext}). The number of unit cells is $25\times 25$ and the parameters are the same as Ref.~\cite{okugawa}: $t_{x}=1, g_{x}=0. 9, t_{y}=0. 8, g_{y}=0. 7$. There are two degenerate $TT$ corner modes located exactly at zero energy with odd number of lattice sites. }
	\label{fig: 4}
\end{figure}
Hence, we obtain the effective Hamiltonian in edge-state subspace, which is block independent with the bulk and consistent with the numerical result. Actually, the effective edge Hamiltonian is exactly the transposition of  $H_{x}$ under OBC, which in momentum space reads
\begin{equation}
\label{exteff}
H_{eff}(k_{x})=2t_{x}\cos k_{x}\tau_{0}+2ig_{x}\sin k_{x}\tau_{z}.
\end{equation}

The complex energy spectrum of $H_{eff}(k_{x})$ under PBC~[orange loop in Fig.~\ref{fig: 4}(a)] surrounds the skin bulk complex spectrum under OBC~[black part in Fig.~\ref{fig: 4}(a)], which is identical with the second-order corner-localized modes under full OBC~[central part in Fig.~\ref{fig: 4}(b)]. The effective Hamiltonian $H_{eff}(k_{x})$ is actually two decoupled Hatano-Nelson models~\cite{hn} with opposite nonreciprocity. Hence, thanks to the left (right) localization behavior of the skin modes for the two Hatano-Nelson models, we can exactly deduce the low-left (up-right) corner modes~\cite{okugawa}. We emphasize that these corner modes in this model are categorized into hybrid $x$-skin and $y$-topological $ST$ modes~\cite{lee2019}. Nevertheless, the $TT$ zero modes localized at the same corners appear if the lattice site number is odd, since the zero edge state exists in the Hatano-Nelson model with odd number of lattice sites. The two $TT$ zero modes localized at the low-left and up-right corners read as
\begin{eqnarray}
&&\ket{\psi_{LL}}=\sum_{x=1}^{\frac{L_{x}+1}{2}}\sum_{y=1}^{\frac{L_{y}+1}{2}}\rho^{2x-1}\beta^{2y-1}\ket{\phi_{L}}\ket{2x-1}\ket{2y-1},
\nonumber\\
&&\ket{\psi_{RR}}=\sum_{x=1}^{\frac{L_{x}+1}{2}}\sum_{y=1}^{\frac{L_{y}+1}{2}}\rho^{L_{x}-2x}\beta^{L_{y}-2y}\ket{\phi_{R}}\ket{2x-1}\ket{2y-1},
\nonumber\\
\end{eqnarray}
where $\rho=\beta=i\sqrt{\frac{t_{x}-g_{x}}{t_{x}+g_{x}}}$ and $\ket{\phi_{L}}=\ket{2},\ket{\phi_{R}}=\ket{1}$ are the internal degrees of freedom. Similarly, the two numerical $TT$ corner modes are a linear combination of two analytical solutions. In addition, the extended Hermitian Hamiltonian of $H_{e}(\vec{k})$ is topologically characterized only by chiral symmetry~(for details in Ref.~\cite{okugawa}). This leads to the trivial topology of $H_{e}(\vec{k})$~(vanishing winding number of spectra under full PBC) and nontrivial topology of edge-state subspace effective Hamiltonian $H_{eff}(k_{x})$~(non-vanishing winding number of orange loop in Fig.~\ref{fig: 4}(a)), thus extrinsic feature of $ST$ corner modes. That is because the edges are topologically nontrivial while the bulk is trivial in the extrinsic higher-order topological phase~\cite{okugawa}.

\section{Conclusion and Discussion}
\label{CD}

In this paper, we construct the nested tight-binding formalism, within which we deduce the second-order corner-localized modes in non-Hermitian systems. Utilizing this formalism, we have strictly illustrated pure $SS$ modes for the simplest 2D model [Eq.~(\ref{2ds})], the $ST$ corner modes for the four-band model [Eq.~(\ref{2d4b})], and the extrinsic $ST$ corner modes for the 2D model [Eq.~(\ref{ext})]. Additionally, we have obtained the analytical solutions of zero-energy $TT$ corner modes for the four-band model [Eq.~(\ref{2d4b})] and the 2D model [Eq.~(\ref{ext})]. We also clarify the meaning and configuration of $TT$, $ST$, and $SS$ corner modes. Not distinguishing the zero- and nonzero-energy edge states, we conclude that the corner modes are classified into three types: (i) The pure second-order skin effect ($SS$) modes are the result of contribution from two directions with first-order skin effect. (ii) The pure second-order topological ($TT$) corner modes, inherited from Hermitian counterpart, are the result of contributions from two directions with topological edge states. Note that we should distinguish the topology of edge states from that of skin effect, in which the former is inherited from Hermitian counterpart and the latter is a pure non-Hermitian consequence. (iii) The most charming skin-topological ($ST$) modes are result of contribution from two directions with topological edge states and skin effect, respectively; in other words, the Hermitian ramification and pure non-Hermitian consequence for each of two directions, respectively.

The gapped edge-localized states of the Benalcazar-Bernevig-Hughes~(BBH) model~(Hermitian counterpart of the non-Hermitian four-band model) are protected by Wannier band topology instead of bulk-band topology, which reveals a new symmetry-protected topological~(SPT) phase in higher-order systems~\cite{dipole2017}. The non-Hermitian extension of this new SPT phase will be given in other work. Our nested tight-binding formalism for non-Hermitian higher-order topological insulators naturally applies to a Hermitian system, such as the BBH model, of which we can easily obtain the four second-order corner modes. In addition, a 2D non-Hermitian model given in a recent related work~\cite{kawabata2020}, can also be analytically studied by our nested tight-binding formalism. The study of a more general model utilizing our formalism is left for future work. We believe that there exists the mixture of bulk and edge in one direction due to the failure of block-diagonal of effective Hamiltonian $H_{eff}$ for a more general 2D model, and it will lead to more possible corner and edge states under full OBC.

\section*{Acknowledgments}
The authors thank Haoshu Li, Shuxuan Wang and Zhiwei Yin for helpful discussions. This work was supported by NSFC Grant No.11275180.

\begin{appendices}
\appendix
\section{The exact eigenstates of 1D tight-binding model}
\label{appendix: EEOTB}
Without loss of generality, any first-order tight-binding model can be ascribed to a 1D tight-binding model,
\begin{equation}
\hat{H}=\sum_{ij, \mu\nu}\hat{c}^{\dagger}_{i\mu}H_{ij, \mu\nu}(k_{2}, . . . k_{d}, {\lambda's})\hat{c}_{j\nu},
\end{equation}
where $k_{2}, . . . k_{d}, \lambda's$ are all parameters. The Hamiltonian of a 1D tight-binding model, with range of hopping $R$ and internal degrees of freedom $q$ per unit cell, is
\begin{equation}
\hat{H}=\sum_{n=1}^{L}\sum_{i=-R}^{R}\sum_{\mu, \nu=1}^{q}\hat{c}_{n+i}^{\mu\dagger}t_{i, \mu\nu}\hat{c}_{n}^{\nu}.
\end{equation}
We assume the solution as
\begin{equation}
\ket{\Phi}=\sum_{n=1}^{L}\ket{\phi_{n}}\ket{n}=\sum_{n=1}^{L}\sum_{\mu=1}^{q}\beta^{n}\phi_{\mu}\ket{\mu}\ket{n},
\end{equation}
and the Schrodinger equation is $\hat{H}\ket{\Phi}=E\ket{\Phi}$. We obtain the bulk equation
\begin{equation}
\sum_{\nu=1}^{q}H(\beta)_{\mu\nu}	\phi_{\nu}: =\sum_{\nu=1}^{q}\sum_{i=-R}^{R}t_{i, \mu\nu}\beta^{i}\phi_{\nu}=E\phi_{\mu}
\end{equation}
and the secular equation
\begin{equation}
\det(\sum_{i=-R}^{R}t_{i, \mu\nu}\beta^{i}-E)=0.
\end{equation}
From the above linear equation set of $\phi's$, we can linearly express the $(q-1)$ $\phi's$ by the remaining one
\begin{equation}
\label{39}
\phi_{\mu}=J_{\nu\mu}(\beta)\phi_{\nu},\, \, \, \, \, \mu=1, 2, \ldots\hat{\nu}, \ldots, q;\, \, \, \, \nu=1, 2, \ldots, q,
\end{equation}
The secular equation of the bulk equation can be solved, resulting in $2qR$ roots of $\beta$ in general. We briefly ignore the multiple roots case (it has been well studied in Ref.~\cite{alase2017}). Now the full solution is
\begin{equation}
\ket{\Phi}=\sum_{n=1}^{L}\sum_{\mu=1}^{q}\ket{\phi_{n\mu}}\ket{n}=\sum_{n=1}^{L}\sum_{\mu=1}^{q}\sum_{j=1}^{2qR}\beta^{n}_{j}\phi_{\mu}^{j}\ket{\mu}\ket{n}.
\end{equation}
Imposing the boundary condition both on the left and the right boundaries,
\begin{eqnarray}
&&\sum_{i=-s}^{R}t_{i}\ket{\phi_{s+i+1}}=E\ket{\phi_{1+s}},
\nonumber\\
&&\sum_{i=-R}^{s}t_{i}\ket{\phi_{L-s+i}}=E\ket{\phi_{L-s}},
\end{eqnarray}
where $s=0, 1, \ldots, (R-1)$. They can reduce to~\cite{kawabata20}
\begin{eqnarray}
&&\ket{\phi_{0}}=\ket{\phi_{-1}}=\ldots=\ket{\phi_{-R+1}}=0,
\nonumber\\
&&\ket{\phi_{L+1}}=\ket{\phi_{L+2}}=\ldots=\ket{\phi_{L+R}}=0.
\end{eqnarray}
We obtain
\begin{eqnarray}
&& \sum_{j=1}^{2qR}\beta_{j}^{-s}\phi_{\mu}^{j}=0;s=0, 1, \ldots, (R-1);\mu=1, 2, \ldots, q,
\nonumber\\
&&\sum_{j=1}^{2qR}\beta_{j}^{L+s}\phi_{\mu}^{j}=0; s=1, \ldots, R; \mu=1, 2, \ldots, q.
\end{eqnarray}
Using Eq.~(\ref{39}) for any $\nu$, we obtain
\begin{eqnarray}
&&\sum_{j=1}^{2qR}f_{s\mu}(\beta_{j}, E)\phi_{\nu}^{j}=0; s=0, 1, \ldots, (R-1); \mu=1, 2, \ldots, q,
\nonumber\\
&&\sum_{j=1}^{2qR}g_{s\mu}(\beta_{j}, E)\beta^{L}_{j}\phi_{\nu}^{j}=0; s=1, \ldots, R;  \mu=1, 2, \ldots, q,
\end{eqnarray}
where
\begin{eqnarray}
&&f_{s\mu}(\beta_{j}, E)=J_{\nu\mu}(\beta_{j})\beta_{j}^{-s},
\nonumber\\
&&g_{s\mu}(\beta_{j}, E)=J_{\nu\mu}(\beta_{j})\beta_{j}^{s}.
\end{eqnarray}
We can denote the $2qR$ functions $f_{s\mu}$ and $g_{s\mu}$ as $f_{j}, g_{j}, j=1, 2, \ldots, qR$ respectively. The boundary conditions require~\cite{yokomizo2019}
\begin{equation}
\label{bulkeq}
\det\left|\begin{matrix}
f_{1}(\beta_{1}, E)&\ldots &f_{1}(\beta_{2qR}, E)\\
\vdots&\ddots&\vdots\\
f_{qR}(\beta_{1}, E)&\ldots &f_{qR}(\beta_{2qR}, E)\\
g_{1}(\beta_{1}, E)\beta_{1}^{L}&\ldots &g_{1}(\beta_{2qR}, E)\beta_{2qR}^{L}\\
\vdots&\ddots&\vdots\\
g_{qR}(\beta_{1}, E)\beta_{1}^{L}&\ldots &g_{qR}(\beta_{2qR}, E)\beta_{2qR}^{L}
\end{matrix}\right|=0.
\end{equation}
We number the solutions satisfying $|\beta_{1}|\leqslant. . . \leqslant|\beta_{qR}|\leqslant|\beta_{qR+1}|\leqslant. . . \leqslant|\beta_{2qR}|$ and take limit $L\rightarrow \infty$. If $|\beta_{qR}|<|\beta_{qR+1}|$, only one leading term survives in Eq.~(\ref{bulkeq}),
\begin{eqnarray}
&&F(\beta_{i\in P_{1}}, \beta_{j\in Q_{1}}, E): =\det\left|\begin{matrix}
f_{1}(\beta_{1}, E)&\ldots&f_{1}(\beta_{qR}, E)\\
\vdots&\ddots&\vdots\\
f_{qR}(\beta_{1}, E)&\ldots&f_{qR}(\beta_{qR}, E)
\end{matrix}\right|
\nonumber\\
&&\qquad\times\det\left|\begin{matrix}
g_{1}(\beta_{qR+1}, E)&\ldots &g_{1}(\beta_{2qR}, E)\\
\vdots&\ddots&\vdots\\
g_{qR}(\beta_{qR+1}, E)&\ldots &g_{qR}(\beta_{2qR}, E)
\end{matrix}\right|=0,
\end{eqnarray}
where $P_{1}=\left\{\beta_{1}, \ldots, \beta_{qR}\right\}, Q_{1}=\left\{\beta_{qR+1}, \ldots, \beta_{2qR}\right\}$. The above equation gives discrete $\beta^{'}s$, deducing the edge states isolated from the continuous bulk states.

If $|\beta_{qR}|=|\beta_{qR+1}|$, two leading terms survive. Let $P_{0}=\left\{\beta_{1}, \ldots, \beta_{qR-1}, \beta_{qR+1}\right\}, Q_{0}=\left\{\beta_{qR}, \beta_{qR+2}\ldots, \beta_{2qR}\right\}$, then the continuous $\beta^{'}s$ are given~\cite{yokomizo2019}
\begin{equation}
-\frac{F(\beta_{i\in P_{1}}, \beta_{j\in Q_{1}}, E)}{F(\beta_{i\in P_{0}}, \beta_{j\in Q_{0}}, E)}=
\bigg(\frac{\beta_{qR}}{\beta_{qR+1}}\bigg)^{L}.
\end{equation}
We can obtain the bulk band spectra (or continuous band spectra) and generalized Brillouin zone (GBZ)  ~\cite{yang2019} as
\begin{eqnarray}
&&E_{bulk}=\left\{E\in\mathbb{C}:|\beta_{qR}(E)|=|\beta_{qR+1}(E)|\right\},
\nonumber\\
&&\mathcal{C}_{\beta}=\left\{\beta\in\mathbb{C}:\forall E\in E_{bulk}, |\beta_{qR}(E)|=|\beta_{qR+1}(E)|\right\}.
\nonumber\\
\end{eqnarray}
We emphasize that the GBZs depend on Riemann energy spectra sheets (i.e., complex energy bands) $E^{\mu}$ with $\mu=1, 2, \ldots, q$ in general. In other words, there are $q$ GBZs $\mathcal{C}_{\beta}^{\mu}$ one-to-one corresponding to $q$ Riemann energy spectra sheets $E^{\mu}$. However, the multiple GBZs are degenerate in some simple model, e.g., the non-Hermitian SSH model~\cite{yao2018}. In this paper, we only consider the degenerate GBZs or the single-band model and leave the multiple GBZs for numerical calculation in other work.

The above process to solve the eigenstates in non-Hermitian system is the non-Bloch band theory without any symmetry constraint proposed in Ref.~\cite{yokomizo2019}, which has been extended to symplectic class~\cite{yi2020,kawabata20} and $Z_{2}$ skin effect~\cite{origin2020} recently.
\section{Biorthogonal diagonalization of the single $y$-layer Hamiltonian}
\label{appendix: BD}

The $qL_{x}$ eigenvalue solutions can also be written as fermionic creation operators
\begin{equation}
\hat{\Phi}^{R, \mu\dagger}_{\alpha, y}=\sum_{x=1}^{L_{x}}\sum_{\nu=1}^{q}\tilde{\phi}^{R, \mu\nu}_{\alpha x}\hat{c}^{\nu\dagger}_{x, y},
\end{equation}
where $\tilde{\phi}^{R, \mu\nu}_{\alpha x}$ is the $x\nu$-th column component of $\alpha\mu$-th right eigenstate for generic non-Hermitian system. We define the right eigenstate matrix
\begin{equation}
U_{R}=\begin{bmatrix}
\tilde{\phi}_{11}^{R, 11}&\ldots&\tilde{\phi}_{L_{x}q}^{R, 11}\\
\vdots&\vdots&\vdots\\
\tilde{\phi}_{11}^{R, L_{x}q}&\ldots&\tilde{\phi}_{L_{x}q}^{R, L_{x}q}
\end{bmatrix}
\end{equation}
and
\begin{eqnarray}
&&\hat{c}_{y}=(\hat{c}_{1, y}^{1}, \ldots, \hat{c}_{1, y}^{q}, \ldots, \hat{c}_{L_{x}, y}^{1}, \ldots, \hat{c}_{L_{x}, y}^{q})^{T},
\nonumber\\
&&\hat{c}^{\dagger}_{y}=(\hat{c}^{1\dagger}_{1, y}, \ldots, \hat{c}^{q\dagger}_{1, y}, \ldots, \hat{c}^{1\dagger}_{L_{x}, y}, \ldots, \hat{c}^{q\dagger}_{L_{x}, y}),
\nonumber\\
&&\hat{\Phi}^{R\dagger}_{y}=(\hat{\Phi}^{R, 1\dagger}_{1, y}, \ldots, \hat{\Phi}^{R, q\dagger}_{1, y}, \ldots, \hat{\Phi}^{R, 1\dagger}_{L_{x}, y}, \ldots, \hat{\Phi}^{R, q\dagger}_{L_{x}, y}),
\nonumber\\
&&\hat{\Phi}^{L}_{y}=(\hat{\Phi}^{L, 1}_{1, y}, \ldots, \hat{\Phi}^{L, q}_{1, y}, \ldots, \hat{\Phi}^{L, 1}_{L_{x}, y}, \ldots, \hat{\Phi}^{L, q}_{L_{x}, y})^{T},
\end{eqnarray}
then
\begin{equation}
\hat{\Phi}^{R\dagger}_{y}=\hat{c}^{\dagger}_{y} U_{R}.
\end{equation}
From the eigenequation of $\hat{H}^{\dagger}$, we can obtain the left eigenstates with the equations
\begin{eqnarray}
&&\hat{\Phi}^{L\dagger}_{y}=\hat{c}^{\dagger}_{y} U_{L},
\nonumber\\
&&\hat{\Phi}^{L}_{y}=U_{L}^{\dagger} \hat{c}_{y}
\end{eqnarray}
and the biorthogonal relation
\begin{equation}
U_{R} U_{L}^{\dagger}=U_{L}U_{R}^{\dagger}=\hat{1}.
\end{equation}
The inverse relation between two fermionic operators is
\begin{eqnarray}
&&\hat{c}^{\dagger}_{y}=\hat{\Phi}^{R\dagger}_{y} U_{L}^{\dagger},
\nonumber\\
&&\hat{c}_{y}=U_{R}\hat{\Phi}_{y}^{L}.
\end{eqnarray}
The result transformed to the biorthogonal basis of the single $y$-layer Hamiltonian $\hat{H}_{y}$ is
\begin{eqnarray}
&&\epsilon=U_{L}^{\dagger} H_{y} U_{R},
\nonumber\\
&&\hat{H}_{y}=\hat{\Phi}_{y}^{R\dagger}\epsilon \hat{\Phi}_{y}^{L}.
\end{eqnarray}
where
\begin{eqnarray}
&&H_{y}=\begin{bmatrix}
T_{0}^{x}&\ldots&T_{R_{x}}^{x}&\dots&0\\
\vdots&\ddots&\vdots&\ddots&\vdots&\\
T_{-R_{x}}^{x}&\ldots&T_{0}^{x}&\ldots&T_{R_{x}}^{x}\\
\vdots&\ddots&\vdots&\ddots&\vdots\\
0&\ldots&T_{-R_{x}}^{x}&\ldots&T_{0}^{x}\\
\end{bmatrix},
\nonumber\\
&&\qquad\quad T_{i}^{x}=\begin{bmatrix}
T_{i, 11}^{x}&\ldots&T_{i, 1q}^{x}\\
\vdots&\ddots&\vdots\\
T_{i, q1}^{x}&\ldots&T_{i, qq}^{x}
\end{bmatrix},
\nonumber\\
&&\epsilon=\begin{bmatrix}
\epsilon^{1}(\beta_{1})&\ldots&0&\dots&0&\dots&0\\
\vdots&\ddots&\vdots&\ddots&\vdots&\ddots&\vdots\\
0&\ldots&\epsilon^{q}(\beta_{1})&\ldots&0&\ldots&0\\
\vdots&\ddots&\vdots&\ddots&\vdots&\ddots&\vdots\\
0&\ldots&0&\ldots&\epsilon^{1}(\beta_{L_{x}})&\ldots&0\\
\vdots&\ddots&\vdots&\ddots&\vdots&\ddots&\vdots\\
0&\ldots&0&\ldots&0&\ldots&\epsilon^{q}(\beta_{L_{x}})
\end{bmatrix}.
\end{eqnarray}
Note that the lowest $q$ eigenvalues are edge states energies which deduce the $ST$ and $TT$ corner modes. For Hermitian cases, the biorthogonal relation reduces to $U_{L}^{\dagger}=U_{R}^{-1}$ and the diagonalization process reduces to the standard one in linear algebra, $\epsilon=U^{-1}H_{y} U$.
\section{Other parameter choices for the four-band model}
\label{appendix: OPC}
Single nonreciprocity case: $\delta_{1}=\delta_{2}=-\delta_{3}=\delta_{4}=\gamma$. The net nonreciprocity only exists along the $y$ direction. The $M_{xy}$ is broken in this case. The corner modes contain: $ST$ modes and four $TT$ zero modes, while the $SS$ modes are absent. The forms of edge states read
\begin{eqnarray}
&&\ket{\phi_{L}^{\pm}}_{sn}=\ket{u_{1}}\pm r^{-1}\ket{u_{2}},
\nonumber\\
&&\ket{\phi_{R}^{\pm}}_{sn}=\ket{v_{1}}\pm r^{-1}\ket{v_{2}}.
\end{eqnarray}
The edge effective Hamiltonian is then deduced
\begin{equation}
t_{L}^{\pm}=\frac{\lambda}{2}r^{\mp}\begin{bmatrix}
1&\mp1\\
\pm1&-1
\end{bmatrix},
\end{equation}
\begin{equation}
t_{R}^{\pm}=\frac{\lambda}{2}r^{\pm}\begin{bmatrix}
1&\pm1\\
\mp1&-1
\end{bmatrix}.
\end{equation}

Double reciprocity case: $\delta_{1}=\delta_{2}=-\delta_{3}=-\delta_{4}=\gamma$. The $M_{xy}$ is also broken in this case. The numerical results are provided in Ref.~\cite{lee2019} for the $TT$ and $ST$ modes, while the $SS$ corner modes are absent.

Asymmetry case: $\delta_{1}=\delta_{2}=0$ or $\delta_{3}=\delta_{4}=0$ while the other direction is nonreciprocal. The mirror symmetry $M_{x}$ or $M_{y}$ is restored. The $TT$ and $ST$ corner modes are present while the $SS$ modes absent.

Hermitian case: $\delta_{1}=\delta_{2}=\delta_{3}=\delta_{4}=0$. Both $M_{x}$ and $M_{y}$ are restored as well as the fourfold rotation symmetry $C_{4}$; the only existent corner modes are the $TT$ zero modes.

Non-Hermitian case with on-site gain and loss~\cite{kawabaras}: $\delta_{1}=\delta_{2}=\delta_{3}=\delta_{4}=0$ but with the additional term $-iu \tau_{z}$. The $C_{4}$ is restored and the only existing corner modes are the in-gap $TT$ modes.
\section{Edge states for 2D extrinsic model }
\label{appendix:ESEM}
The zero edge modes have very simple forms for $H_{y}(k_{y})$ in the main text when the number of lattice sites is odd. The bulk equation for the Hamiltonian is
\begin{equation}
t_{y}^{+}\phi_{y+1}+t_{y}^{-}\phi_{y-1}=(t_{y}^{+}\beta+t_{y}^{-}\beta^{-1})\beta^{y}\phi=0,
\end{equation}
where
\begin{eqnarray}
&&t_{y}^{+}=\begin{bmatrix}
0&-t_{y}-g_{y}\\
t_{y}-g_{y}&0
\end{bmatrix},
\nonumber\\
&&t_{y}^{-}=\begin{bmatrix}
0&-t_{y}+g_{y}\\
t_{y}+g_{y}&0
\end{bmatrix}.
\end{eqnarray}
With the boundary conditions
\begin{eqnarray}
&& t_{y}^{+}\phi_{2}=0,
\nonumber\\
&& t_{y}^{-}\phi_{L_{y}-1}=0,
\end{eqnarray}
the amplitudes for exact zero edge states are destroyed on even lattice sites, which is consistent with the numerical results. A similar amplitude destruction is also found
in Ref.~\cite{kunst2018}. Utilizing the bulk equation, we obtain $\beta=i\sqrt{\frac{t_{y}-g_{y}}{t_{y}+g_{y}}}$ and two edge states given by Eq.~(\ref{extso}) in the main text.  In addition, the edge states form [Eq.~(\ref{extso})] is valid for the Hamiltonian $H_{x}(k_{x})$ along the $x$ direction under OBC for odd sites.

For an even number of lattice sites, the edge solutions of $H_{y}$ are the linear combination of two localized edge states, of which the left-~(right-)-localized edge state is destroyed on even~(odd) lattice sites. Consequently, we obtain the effective edge-state subspace Hamiltonian
\begin{equation}
H_{eff}(k_{x})=2t_{x}\cos k_{x}\tau_{0}+2ig_{x}\sin k_{x}\tau_{x}.
\end{equation}
From the above effective Hamiltonian, we find that the $ST$ modes are present, while the $TT$ zero modes are absent with an even number of lattice sites.
\end{appendices}

\bibliography{ref}

\begin{thebibliography}{92}%
\makeatletter
\providecommand \@ifxundefined [1]{%
 \@ifx{#1\undefined}
}%
\providecommand \@ifnum [1]{%
 \ifnum #1\expandafter \@firstoftwo
 \else \expandafter \@secondoftwo
 \fi
}%
\providecommand \@ifx [1]{%
 \ifx #1\expandafter \@firstoftwo
 \else \expandafter \@secondoftwo
 \fi
}%
\providecommand \natexlab [1]{#1}%
\providecommand \enquote  [1]{``#1''}%
\providecommand \bibnamefont  [1]{#1}%
\providecommand \bibfnamefont [1]{#1}%
\providecommand \citenamefont [1]{#1}%
\providecommand \href@noop [0]{\@secondoftwo}%
\providecommand \href [0]{\begingroup \@sanitize@url \@href}%
\providecommand \@href[1]{\@@startlink{#1}\@@href}%
\providecommand \@@href[1]{\endgroup#1\@@endlink}%
\providecommand \@sanitize@url [0]{\catcode `\\12\catcode `\$12\catcode
  `\&12\catcode `\#12\catcode `\^12\catcode `\_12\catcode `\%12\relax}%
\providecommand \@@startlink[1]{}%
\providecommand \@@endlink[0]{}%
\providecommand \url  [0]{\begingroup\@sanitize@url \@url }%
\providecommand \@url [1]{\endgroup\@href {#1}{\urlprefix }}%
\providecommand \urlprefix  [0]{URL }%
\providecommand \Eprint [0]{\href }%
\providecommand \doibase [0]{http://dx.doi.org/}%
\providecommand \selectlanguage [0]{\@gobble}%
\providecommand \bibinfo  [0]{\@secondoftwo}%
\providecommand \bibfield  [0]{\@secondoftwo}%
\providecommand \translation [1]{[#1]}%
\providecommand \BibitemOpen [0]{}%
\providecommand \bibitemStop [0]{}%
\providecommand \bibitemNoStop [0]{.\EOS\space}%
\providecommand \EOS [0]{\spacefactor3000\relax}%
\providecommand \BibitemShut  [1]{\csname bibitem#1\endcsname}%
\let\auto@bib@innerbib\@empty
\bibitem [{\citenamefont {Hasan}\ and\ \citenamefont {Kane}(2010)}]{hasan2010}%
  \BibitemOpen
  \bibfield  {author} {\bibinfo {author} {\bibfnamefont {M.~Z.}\ \bibnamefont
  {Hasan}}\ and\ \bibinfo {author} {\bibfnamefont {C.~L.}\ \bibnamefont
  {Kane}},\ }\bibfield  {title} {\enquote {\bibinfo {title} {Colloquium:
  Topological insulators},}\ }\href {\doibase 10.1103/RevModPhys.82.3045}
  {\bibfield  {journal} {\bibinfo  {journal} {Rev. Mod. Phys.}\ }\textbf
  {\bibinfo {volume} {82}},\ \bibinfo {pages} {3045--3067} (\bibinfo {year}
  {2010})}\BibitemShut {NoStop}%
\bibitem [{\citenamefont {Qi}\ and\ \citenamefont {Zhang}(2011)}]{qi2010}%
  \BibitemOpen
  \bibfield  {author} {\bibinfo {author} {\bibfnamefont {Xiao-Liang}\
  \bibnamefont {Qi}}\ and\ \bibinfo {author} {\bibfnamefont {Shou-Cheng}\
  \bibnamefont {Zhang}},\ }\bibfield  {title} {\enquote {\bibinfo {title}
  {Topological insulators and superconductors},}\ }\href {\doibase
  10.1103/RevModPhys.83.1057} {\bibfield  {journal} {\bibinfo  {journal} {Rev.
  Mod. Phys.}\ }\textbf {\bibinfo {volume} {83}},\ \bibinfo {pages}
  {1057--1110} (\bibinfo {year} {2011})}\BibitemShut {NoStop}%
\bibitem [{\citenamefont {Fu}\ and\ \citenamefont {Kane}(2007)}]{fu2007}%
  \BibitemOpen
  \bibfield  {author} {\bibinfo {author} {\bibfnamefont {Liang}\ \bibnamefont
  {Fu}}\ and\ \bibinfo {author} {\bibfnamefont {C.~L.}\ \bibnamefont {Kane}},\
  }\bibfield  {title} {\enquote {\bibinfo {title} {Topological insulators with
  inversion symmetry},}\ }\href {\doibase 10.1103/PhysRevB.76.045302}
  {\bibfield  {journal} {\bibinfo  {journal} {Phys. Rev. B}\ }\textbf {\bibinfo
  {volume} {76}},\ \bibinfo {pages} {045302} (\bibinfo {year}
  {2007})}\BibitemShut {NoStop}%
\bibitem [{\citenamefont {Qi}\ \emph {et~al.}(2008)\citenamefont {Qi},
  \citenamefont {Hughes},\ and\ \citenamefont {Zhang}}]{qi2008}%
  \BibitemOpen
  \bibfield  {author} {\bibinfo {author} {\bibfnamefont {Xiao-Liang}\
  \bibnamefont {Qi}}, \bibinfo {author} {\bibfnamefont {Taylor~L.}\
  \bibnamefont {Hughes}}, \ and\ \bibinfo {author} {\bibfnamefont {Shou-Cheng}\
  \bibnamefont {Zhang}},\ }\bibfield  {title} {\enquote {\bibinfo {title}
  {Topological field theory of time-reversal invariant insulators},}\ }\href
  {\doibase 10.1103/PhysRevB.78.195424} {\bibfield  {journal} {\bibinfo
  {journal} {Phys. Rev. B}\ }\textbf {\bibinfo {volume} {78}},\ \bibinfo
  {pages} {195424} (\bibinfo {year} {2008})}\BibitemShut {NoStop}%
\bibitem [{\citenamefont {Liu}\ \emph {et~al.}(2010)\citenamefont {Liu},
  \citenamefont {Qi}, \citenamefont {Zhang}, \citenamefont {Dai}, \citenamefont
  {Fang},\ and\ \citenamefont {Zhang}}]{liu2010}%
  \BibitemOpen
  \bibfield  {author} {\bibinfo {author} {\bibfnamefont {Chao-Xing}\
  \bibnamefont {Liu}}, \bibinfo {author} {\bibfnamefont {Xiao-Liang}\
  \bibnamefont {Qi}}, \bibinfo {author} {\bibfnamefont {HaiJun}\ \bibnamefont
  {Zhang}}, \bibinfo {author} {\bibfnamefont {Xi}~\bibnamefont {Dai}}, \bibinfo
  {author} {\bibfnamefont {Zhong}\ \bibnamefont {Fang}}, \ and\ \bibinfo
  {author} {\bibfnamefont {Shou-Cheng}\ \bibnamefont {Zhang}},\ }\bibfield
  {title} {\enquote {\bibinfo {title} {Model hamiltonian for topological
  insulators},}\ }\href {\doibase 10.1103/PhysRevB.82.045122} {\bibfield
  {journal} {\bibinfo  {journal} {Phys. Rev. B}\ }\textbf {\bibinfo {volume}
  {82}},\ \bibinfo {pages} {045122} (\bibinfo {year} {2010})}\BibitemShut
  {NoStop}%
\bibitem [{\citenamefont {Teo}\ and\ \citenamefont {Kane}(2010)}]{teo2010}%
  \BibitemOpen
  \bibfield  {author} {\bibinfo {author} {\bibfnamefont {Jeffrey C.~Y.}\
  \bibnamefont {Teo}}\ and\ \bibinfo {author} {\bibfnamefont {C.~L.}\
  \bibnamefont {Kane}},\ }\bibfield  {title} {\enquote {\bibinfo {title}
  {Topological defects and gapless modes in insulators and superconductors},}\
  }\href {\doibase 10.1103/PhysRevB.82.115120} {\bibfield  {journal} {\bibinfo
  {journal} {Phys. Rev. B}\ }\textbf {\bibinfo {volume} {82}},\ \bibinfo
  {pages} {115120} (\bibinfo {year} {2010})}\BibitemShut {NoStop}%
\bibitem [{\citenamefont {Mong}\ and\ \citenamefont
  {Shivamoggi}(2011)}]{mong2011}%
  \BibitemOpen
  \bibfield  {author} {\bibinfo {author} {\bibfnamefont {Roger S.~K.}\
  \bibnamefont {Mong}}\ and\ \bibinfo {author} {\bibfnamefont {Vasudha}\
  \bibnamefont {Shivamoggi}},\ }\bibfield  {title} {\enquote {\bibinfo {title}
  {Edge states and the bulk-boundary correspondence in dirac hamiltonians},}\
  }\href {\doibase 10.1103/PhysRevB.83.125109} {\bibfield  {journal} {\bibinfo
  {journal} {Phys. Rev. B}\ }\textbf {\bibinfo {volume} {83}},\ \bibinfo
  {pages} {125109} (\bibinfo {year} {2011})}\BibitemShut {NoStop}%
\bibitem [{\citenamefont {Hughes}\ \emph {et~al.}(2011)\citenamefont {Hughes},
  \citenamefont {Prodan},\ and\ \citenamefont {Bernevig}}]{hug2011}%
  \BibitemOpen
  \bibfield  {author} {\bibinfo {author} {\bibfnamefont {Taylor~L.}\
  \bibnamefont {Hughes}}, \bibinfo {author} {\bibfnamefont {Emil}\ \bibnamefont
  {Prodan}}, \ and\ \bibinfo {author} {\bibfnamefont {B.~Andrei}\ \bibnamefont
  {Bernevig}},\ }\bibfield  {title} {\enquote {\bibinfo {title}
  {Inversion-symmetric topological insulators},}\ }\href {\doibase
  10.1103/PhysRevB.83.245132} {\bibfield  {journal} {\bibinfo  {journal} {Phys.
  Rev. B}\ }\textbf {\bibinfo {volume} {83}},\ \bibinfo {pages} {245132}
  (\bibinfo {year} {2011})}\BibitemShut {NoStop}%
\bibitem [{\citenamefont {Schnyder}\ \emph {et~al.}(2008)\citenamefont
  {Schnyder}, \citenamefont {Ryu}, \citenamefont {Furusaki},\ and\
  \citenamefont {Ludwig}}]{ryu2008}%
  \BibitemOpen
  \bibfield  {author} {\bibinfo {author} {\bibfnamefont {Andreas~P.}\
  \bibnamefont {Schnyder}}, \bibinfo {author} {\bibfnamefont {Shinsei}\
  \bibnamefont {Ryu}}, \bibinfo {author} {\bibfnamefont {Akira}\ \bibnamefont
  {Furusaki}}, \ and\ \bibinfo {author} {\bibfnamefont {Andreas W.~W.}\
  \bibnamefont {Ludwig}},\ }\bibfield  {title} {\enquote {\bibinfo {title}
  {Classification of topological insulators and superconductors in three
  spatial dimensions},}\ }\href {\doibase 10.1103/PhysRevB.78.195125}
  {\bibfield  {journal} {\bibinfo  {journal} {Phys. Rev. B}\ }\textbf {\bibinfo
  {volume} {78}},\ \bibinfo {pages} {195125} (\bibinfo {year}
  {2008})}\BibitemShut {NoStop}%
\bibitem [{\citenamefont {Morimoto}\ and\ \citenamefont
  {Furusaki}(2013)}]{mor2013}%
  \BibitemOpen
  \bibfield  {author} {\bibinfo {author} {\bibfnamefont {Takahiro}\
  \bibnamefont {Morimoto}}\ and\ \bibinfo {author} {\bibfnamefont {Akira}\
  \bibnamefont {Furusaki}},\ }\bibfield  {title} {\enquote {\bibinfo {title}
  {Topological classification with additional symmetries from clifford
  algebras},}\ }\href {\doibase 10.1103/PhysRevB.88.125129} {\bibfield
  {journal} {\bibinfo  {journal} {Phys. Rev. B}\ }\textbf {\bibinfo {volume}
  {88}},\ \bibinfo {pages} {125129} (\bibinfo {year} {2013})}\BibitemShut
  {NoStop}%
\bibitem [{\citenamefont {Chiu}\ \emph {et~al.}(2013)\citenamefont {Chiu},
  \citenamefont {Yao},\ and\ \citenamefont {Ryu}}]{chiu2013}%
  \BibitemOpen
  \bibfield  {author} {\bibinfo {author} {\bibfnamefont {Ching-Kai}\
  \bibnamefont {Chiu}}, \bibinfo {author} {\bibfnamefont {Hong}\ \bibnamefont
  {Yao}}, \ and\ \bibinfo {author} {\bibfnamefont {Shinsei}\ \bibnamefont
  {Ryu}},\ }\bibfield  {title} {\enquote {\bibinfo {title} {Classification of
  topological insulators and superconductors in the presence of reflection
  symmetry},}\ }\href {\doibase 10.1103/PhysRevB.88.075142} {\bibfield
  {journal} {\bibinfo  {journal} {Phys. Rev. B}\ }\textbf {\bibinfo {volume}
  {88}},\ \bibinfo {pages} {075142} (\bibinfo {year} {2013})}\BibitemShut
  {NoStop}%
\bibitem [{\citenamefont {Shiozaki}\ and\ \citenamefont
  {Sato}(2014)}]{sato2014}%
  \BibitemOpen
  \bibfield  {author} {\bibinfo {author} {\bibfnamefont {Ken}\ \bibnamefont
  {Shiozaki}}\ and\ \bibinfo {author} {\bibfnamefont {Masatoshi}\ \bibnamefont
  {Sato}},\ }\bibfield  {title} {\enquote {\bibinfo {title} {Topology of
  crystalline insulators and superconductors},}\ }\href {\doibase
  10.1103/PhysRevB.90.165114} {\bibfield  {journal} {\bibinfo  {journal} {Phys.
  Rev. B}\ }\textbf {\bibinfo {volume} {90}},\ \bibinfo {pages} {165114}
  (\bibinfo {year} {2014})}\BibitemShut {NoStop}%
\bibitem [{\citenamefont {Chiu}\ and\ \citenamefont
  {Schnyder}(2014)}]{chiu2014}%
  \BibitemOpen
  \bibfield  {author} {\bibinfo {author} {\bibfnamefont {Ching-Kai}\
  \bibnamefont {Chiu}}\ and\ \bibinfo {author} {\bibfnamefont {Andreas~P.}\
  \bibnamefont {Schnyder}},\ }\bibfield  {title} {\enquote {\bibinfo {title}
  {Classification of reflection-symmetry-protected topological semimetals and
  nodal superconductors},}\ }\href {\doibase 10.1103/PhysRevB.90.205136}
  {\bibfield  {journal} {\bibinfo  {journal} {Phys. Rev. B}\ }\textbf {\bibinfo
  {volume} {90}},\ \bibinfo {pages} {205136} (\bibinfo {year}
  {2014})}\BibitemShut {NoStop}%
\bibitem [{\citenamefont {Chiu}\ \emph {et~al.}(2016)\citenamefont {Chiu},
  \citenamefont {Teo}, \citenamefont {Schnyder},\ and\ \citenamefont
  {Ryu}}]{qiu2015}%
  \BibitemOpen
  \bibfield  {author} {\bibinfo {author} {\bibfnamefont {Ching-Kai}\
  \bibnamefont {Chiu}}, \bibinfo {author} {\bibfnamefont {Jeffrey C.~Y.}\
  \bibnamefont {Teo}}, \bibinfo {author} {\bibfnamefont {Andreas~P.}\
  \bibnamefont {Schnyder}}, \ and\ \bibinfo {author} {\bibfnamefont {Shinsei}\
  \bibnamefont {Ryu}},\ }\bibfield  {title} {\enquote {\bibinfo {title}
  {Classification of topological quantum matter with symmetries},}\ }\href
  {\doibase 10.1103/RevModPhys.88.035005} {\bibfield  {journal} {\bibinfo
  {journal} {Rev. Mod. Phys.}\ }\textbf {\bibinfo {volume} {88}},\ \bibinfo
  {pages} {035005} (\bibinfo {year} {2016})}\BibitemShut {NoStop}%
\bibitem [{\citenamefont {Shiozaki}\ \emph {et~al.}(2016)\citenamefont
  {Shiozaki}, \citenamefont {Sato},\ and\ \citenamefont {Gomi}}]{sato2016}%
  \BibitemOpen
  \bibfield  {author} {\bibinfo {author} {\bibfnamefont {Ken}\ \bibnamefont
  {Shiozaki}}, \bibinfo {author} {\bibfnamefont {Masatoshi}\ \bibnamefont
  {Sato}}, \ and\ \bibinfo {author} {\bibfnamefont {Kiyonori}\ \bibnamefont
  {Gomi}},\ }\bibfield  {title} {\enquote {\bibinfo {title} {Topology of
  nonsymmorphic crystalline insulators and superconductors},}\ }\href {\doibase
  10.1103/PhysRevB.93.195413} {\bibfield  {journal} {\bibinfo  {journal} {Phys.
  Rev. B}\ }\textbf {\bibinfo {volume} {93}},\ \bibinfo {pages} {195413}
  (\bibinfo {year} {2016})}\BibitemShut {NoStop}%
\bibitem [{\citenamefont {Kruthoff}\ \emph {et~al.}(2017)\citenamefont
  {Kruthoff}, \citenamefont {de~Boer}, \citenamefont {van Wezel}, \citenamefont
  {Kane},\ and\ \citenamefont {Slager}}]{Kruthoff2017}%
  \BibitemOpen
  \bibfield  {author} {\bibinfo {author} {\bibfnamefont {Jorrit}\ \bibnamefont
  {Kruthoff}}, \bibinfo {author} {\bibfnamefont {Jan}\ \bibnamefont {de~Boer}},
  \bibinfo {author} {\bibfnamefont {Jasper}\ \bibnamefont {van Wezel}},
  \bibinfo {author} {\bibfnamefont {Charles~L.}\ \bibnamefont {Kane}}, \ and\
  \bibinfo {author} {\bibfnamefont {Robert-Jan}\ \bibnamefont {Slager}},\
  }\bibfield  {title} {\enquote {\bibinfo {title} {Topological classification
  of crystalline insulators through band structure combinatorics},}\ }\href
  {\doibase 10.1103/PhysRevX.7.041069} {\bibfield  {journal} {\bibinfo
  {journal} {Phys. Rev. X}\ }\textbf {\bibinfo {volume} {7}},\ \bibinfo {pages}
  {041069} (\bibinfo {year} {2017})}\BibitemShut {NoStop}%
\bibitem [{\citenamefont {Shiozaki}\ \emph {et~al.}(2017)\citenamefont
  {Shiozaki}, \citenamefont {Sato},\ and\ \citenamefont {Gomi}}]{sato2017}%
  \BibitemOpen
  \bibfield  {author} {\bibinfo {author} {\bibfnamefont {Ken}\ \bibnamefont
  {Shiozaki}}, \bibinfo {author} {\bibfnamefont {Masatoshi}\ \bibnamefont
  {Sato}}, \ and\ \bibinfo {author} {\bibfnamefont {Kiyonori}\ \bibnamefont
  {Gomi}},\ }\bibfield  {title} {\enquote {\bibinfo {title} {Topological
  crystalline materials: General formulation, module structure, and wallpaper
  groups},}\ }\href {\doibase 10.1103/PhysRevB.95.235425} {\bibfield  {journal}
  {\bibinfo  {journal} {Phys. Rev. B}\ }\textbf {\bibinfo {volume} {95}},\
  \bibinfo {pages} {235425} (\bibinfo {year} {2017})}\BibitemShut {NoStop}%
\bibitem [{\citenamefont {Cornfeld}\ and\ \citenamefont
  {Chapman}(2019)}]{con2019}%
  \BibitemOpen
  \bibfield  {author} {\bibinfo {author} {\bibfnamefont {Eyal}\ \bibnamefont
  {Cornfeld}}\ and\ \bibinfo {author} {\bibfnamefont {Adam}\ \bibnamefont
  {Chapman}},\ }\bibfield  {title} {\enquote {\bibinfo {title} {Classification
  of crystalline topological insulators and superconductors with point group
  symmetries},}\ }\href {\doibase 10.1103/PhysRevB.99.075105} {\bibfield
  {journal} {\bibinfo  {journal} {Phys. Rev. B}\ }\textbf {\bibinfo {volume}
  {99}},\ \bibinfo {pages} {075105} (\bibinfo {year} {2019})}\BibitemShut
  {NoStop}%
\bibitem [{\citenamefont {Slager}\ \emph {et~al.}(2015)\citenamefont {Slager},
  \citenamefont {Rademaker}, \citenamefont {Zaanen},\ and\ \citenamefont
  {Balents}}]{Slager2015}%
  \BibitemOpen
  \bibfield  {author} {\bibinfo {author} {\bibfnamefont {Robert-Jan}\
  \bibnamefont {Slager}}, \bibinfo {author} {\bibfnamefont {Louk}\ \bibnamefont
  {Rademaker}}, \bibinfo {author} {\bibfnamefont {Jan}\ \bibnamefont {Zaanen}},
  \ and\ \bibinfo {author} {\bibfnamefont {Leon}\ \bibnamefont {Balents}},\
  }\bibfield  {title} {\enquote {\bibinfo {title} {Impurity-bound states and
  green's function zeros as local signatures of topology},}\ }\href {\doibase
  10.1103/PhysRevB.92.085126} {\bibfield  {journal} {\bibinfo  {journal} {Phys.
  Rev. B}\ }\textbf {\bibinfo {volume} {92}},\ \bibinfo {pages} {085126}
  (\bibinfo {year} {2015})}\BibitemShut {NoStop}%
\bibitem [{\citenamefont {Schindler}\ \emph {et~al.}(2018)\citenamefont
  {Schindler}, \citenamefont {Cook}, \citenamefont {Vergniory}, \citenamefont
  {Wang}, \citenamefont {Parkin}, \citenamefont {Bernevig},\ and\ \citenamefont
  {Neupert}}]{schindler2018}%
  \BibitemOpen
  \bibfield  {author} {\bibinfo {author} {\bibfnamefont {Frank}\ \bibnamefont
  {Schindler}}, \bibinfo {author} {\bibfnamefont {Ashley~M.}\ \bibnamefont
  {Cook}}, \bibinfo {author} {\bibfnamefont {Maia~G.}\ \bibnamefont
  {Vergniory}}, \bibinfo {author} {\bibfnamefont {Zhijun}\ \bibnamefont
  {Wang}}, \bibinfo {author} {\bibfnamefont {Stuart S.~P.}\ \bibnamefont
  {Parkin}}, \bibinfo {author} {\bibfnamefont {B.~Andrei}\ \bibnamefont
  {Bernevig}}, \ and\ \bibinfo {author} {\bibfnamefont {Titus}\ \bibnamefont
  {Neupert}},\ }\bibfield  {title} {\enquote {\bibinfo {title} {Higher-order
  topological insulators},}\ }\href {\doibase 10.1126/sciadv.aat0346}
  {\bibfield  {journal} {\bibinfo  {journal} {Science Advances}\ }\textbf
  {\bibinfo {volume} {4}},\ \bibinfo {pages} {eaat0346} (\bibinfo {year}
  {2018})}\BibitemShut {NoStop}%
\bibitem [{\citenamefont {Benalcazar}\ \emph {et~al.}(2017)\citenamefont
  {Benalcazar}, \citenamefont {Bernevig},\ and\ \citenamefont
  {Hughes}}]{dipole2017}%
  \BibitemOpen
  \bibfield  {author} {\bibinfo {author} {\bibfnamefont {Wladimir~A.}\
  \bibnamefont {Benalcazar}}, \bibinfo {author} {\bibfnamefont {B.~Andrei}\
  \bibnamefont {Bernevig}}, \ and\ \bibinfo {author} {\bibfnamefont
  {Taylor~L.}\ \bibnamefont {Hughes}},\ }\bibfield  {title} {\enquote {\bibinfo
  {title} {Electric multipole moments, topological multipole moment pumping,
  and chiral hinge states in crystalline insulators},}\ }\href {\doibase
  10.1103/PhysRevB.96.245115} {\bibfield  {journal} {\bibinfo  {journal} {Phys.
  Rev. B}\ }\textbf {\bibinfo {volume} {96}},\ \bibinfo {pages} {245115}
  (\bibinfo {year} {2017})}\BibitemShut {NoStop}%
\bibitem [{\citenamefont {Huang}\ \emph {et~al.}(2017)\citenamefont {Huang},
  \citenamefont {Song}, \citenamefont {Huang},\ and\ \citenamefont
  {Hermele}}]{huang2017}%
  \BibitemOpen
  \bibfield  {author} {\bibinfo {author} {\bibfnamefont {Sheng-Jie}\
  \bibnamefont {Huang}}, \bibinfo {author} {\bibfnamefont {Hao}\ \bibnamefont
  {Song}}, \bibinfo {author} {\bibfnamefont {Yi-Ping}\ \bibnamefont {Huang}}, \
  and\ \bibinfo {author} {\bibfnamefont {Michael}\ \bibnamefont {Hermele}},\
  }\bibfield  {title} {\enquote {\bibinfo {title} {Building crystalline
  topological phases from lower-dimensional states},}\ }\href {\doibase
  10.1103/PhysRevB.96.205106} {\bibfield  {journal} {\bibinfo  {journal} {Phys.
  Rev. B}\ }\textbf {\bibinfo {volume} {96}},\ \bibinfo {pages} {205106}
  (\bibinfo {year} {2017})}\BibitemShut {NoStop}%
\bibitem [{\citenamefont {Shapourian}\ \emph {et~al.}(2018)\citenamefont
  {Shapourian}, \citenamefont {Wang},\ and\ \citenamefont {Ryu}}]{sha2018}%
  \BibitemOpen
  \bibfield  {author} {\bibinfo {author} {\bibfnamefont {Hassan}\ \bibnamefont
  {Shapourian}}, \bibinfo {author} {\bibfnamefont {Yuxuan}\ \bibnamefont
  {Wang}}, \ and\ \bibinfo {author} {\bibfnamefont {Shinsei}\ \bibnamefont
  {Ryu}},\ }\bibfield  {title} {\enquote {\bibinfo {title} {Topological
  crystalline superconductivity and second-order topological superconductivity
  in nodal-loop materials},}\ }\href {\doibase 10.1103/PhysRevB.97.094508}
  {\bibfield  {journal} {\bibinfo  {journal} {Phys. Rev. B}\ }\textbf {\bibinfo
  {volume} {97}},\ \bibinfo {pages} {094508} (\bibinfo {year}
  {2018})}\BibitemShut {NoStop}%
\bibitem [{\citenamefont {Ezawa}(2018)}]{ezawa2018}%
  \BibitemOpen
  \bibfield  {author} {\bibinfo {author} {\bibfnamefont {Motohiko}\
  \bibnamefont {Ezawa}},\ }\bibfield  {title} {\enquote {\bibinfo {title}
  {Magnetic second-order topological insulators and semimetals},}\ }\href
  {\doibase 10.1103/PhysRevB.97.155305} {\bibfield  {journal} {\bibinfo
  {journal} {Phys. Rev. B}\ }\textbf {\bibinfo {volume} {97}},\ \bibinfo
  {pages} {155305} (\bibinfo {year} {2018})}\BibitemShut {NoStop}%
\bibitem [{\citenamefont {Geier}\ \emph {et~al.}(2018)\citenamefont {Geier},
  \citenamefont {Trifunovic}, \citenamefont {Hoskam},\ and\ \citenamefont
  {Brouwer}}]{geier2018}%
  \BibitemOpen
  \bibfield  {author} {\bibinfo {author} {\bibfnamefont {Max}\ \bibnamefont
  {Geier}}, \bibinfo {author} {\bibfnamefont {Luka}\ \bibnamefont
  {Trifunovic}}, \bibinfo {author} {\bibfnamefont {Max}\ \bibnamefont
  {Hoskam}}, \ and\ \bibinfo {author} {\bibfnamefont {Piet~W.}\ \bibnamefont
  {Brouwer}},\ }\bibfield  {title} {\enquote {\bibinfo {title} {Second-order
  topological insulators and superconductors with an order-two crystalline
  symmetry},}\ }\href {\doibase 10.1103/PhysRevB.97.205135} {\bibfield
  {journal} {\bibinfo  {journal} {Phys. Rev. B}\ }\textbf {\bibinfo {volume}
  {97}},\ \bibinfo {pages} {205135} (\bibinfo {year} {2018})}\BibitemShut
  {NoStop}%
\bibitem [{\citenamefont {Khalaf}(2018)}]{khalaf2018}%
  \BibitemOpen
  \bibfield  {author} {\bibinfo {author} {\bibfnamefont {Eslam}\ \bibnamefont
  {Khalaf}},\ }\bibfield  {title} {\enquote {\bibinfo {title} {Higher-order
  topological insulators and superconductors protected by inversion
  symmetry},}\ }\href {\doibase 10.1103/PhysRevB.97.205136} {\bibfield
  {journal} {\bibinfo  {journal} {Phys. Rev. B}\ }\textbf {\bibinfo {volume}
  {97}},\ \bibinfo {pages} {205136} (\bibinfo {year} {2018})}\BibitemShut
  {NoStop}%
\bibitem [{\citenamefont {Kunst}\ \emph
  {et~al.}(2018{\natexlab{a}})\citenamefont {Kunst}, \citenamefont {van
  Miert},\ and\ \citenamefont {Bergholtz}}]{flore2018}%
  \BibitemOpen
  \bibfield  {author} {\bibinfo {author} {\bibfnamefont {Flore~K.}\
  \bibnamefont {Kunst}}, \bibinfo {author} {\bibfnamefont {Guido}\ \bibnamefont
  {van Miert}}, \ and\ \bibinfo {author} {\bibfnamefont {Emil~J.}\ \bibnamefont
  {Bergholtz}},\ }\bibfield  {title} {\enquote {\bibinfo {title} {Lattice
  models with exactly solvable topological hinge and corner states},}\ }\href
  {\doibase 10.1103/PhysRevB.97.241405} {\bibfield  {journal} {\bibinfo
  {journal} {Phys. Rev. B}\ }\textbf {\bibinfo {volume} {97}},\ \bibinfo
  {pages} {241405} (\bibinfo {year} {2018}{\natexlab{a}})}\BibitemShut
  {NoStop}%
\bibitem [{\citenamefont {Matsugatani}\ and\ \citenamefont
  {Watanabe}(2018)}]{akishi2018}%
  \BibitemOpen
  \bibfield  {author} {\bibinfo {author} {\bibfnamefont {Akishi}\ \bibnamefont
  {Matsugatani}}\ and\ \bibinfo {author} {\bibfnamefont {Haruki}\ \bibnamefont
  {Watanabe}},\ }\bibfield  {title} {\enquote {\bibinfo {title} {Connecting
  higher-order topological insulators to lower-dimensional topological
  insulators},}\ }\href {\doibase 10.1103/PhysRevB.98.205129} {\bibfield
  {journal} {\bibinfo  {journal} {Phys. Rev. B}\ }\textbf {\bibinfo {volume}
  {98}},\ \bibinfo {pages} {205129} (\bibinfo {year} {2018})}\BibitemShut
  {NoStop}%
\bibitem [{\citenamefont {Li}\ \emph {et~al.}(2018)\citenamefont {Li},
  \citenamefont {Umer},\ and\ \citenamefont {Gong}}]{li2018}%
  \BibitemOpen
  \bibfield  {author} {\bibinfo {author} {\bibfnamefont {Linhu}\ \bibnamefont
  {Li}}, \bibinfo {author} {\bibfnamefont {Muhammad}\ \bibnamefont {Umer}}, \
  and\ \bibinfo {author} {\bibfnamefont {Jiangbin}\ \bibnamefont {Gong}},\
  }\bibfield  {title} {\enquote {\bibinfo {title} {Direct prediction of corner
  state configurations from edge winding numbers in two- and three-dimensional
  chiral-symmetric lattice systems},}\ }\href {\doibase
  10.1103/PhysRevB.98.205422} {\bibfield  {journal} {\bibinfo  {journal} {Phys.
  Rev. B}\ }\textbf {\bibinfo {volume} {98}},\ \bibinfo {pages} {205422}
  (\bibinfo {year} {2018})}\BibitemShut {NoStop}%
\bibitem [{\citenamefont {Lin}\ and\ \citenamefont {Hughes}(2018)}]{lin2018}%
  \BibitemOpen
  \bibfield  {author} {\bibinfo {author} {\bibfnamefont {Mao}\ \bibnamefont
  {Lin}}\ and\ \bibinfo {author} {\bibfnamefont {Taylor~L.}\ \bibnamefont
  {Hughes}},\ }\bibfield  {title} {\enquote {\bibinfo {title} {Topological
  quadrupolar semimetals},}\ }\href {\doibase 10.1103/PhysRevB.98.241103}
  {\bibfield  {journal} {\bibinfo  {journal} {Phys. Rev. B}\ }\textbf {\bibinfo
  {volume} {98}},\ \bibinfo {pages} {241103} (\bibinfo {year}
  {2018})}\BibitemShut {NoStop}%
\bibitem [{\citenamefont {Okugawa}\ \emph {et~al.}(2019)\citenamefont
  {Okugawa}, \citenamefont {Hayashi},\ and\ \citenamefont
  {Nakanishi}}]{ryo2019}%
  \BibitemOpen
  \bibfield  {author} {\bibinfo {author} {\bibfnamefont {Ryo}\ \bibnamefont
  {Okugawa}}, \bibinfo {author} {\bibfnamefont {Shin}\ \bibnamefont {Hayashi}},
  \ and\ \bibinfo {author} {\bibfnamefont {Takeshi}\ \bibnamefont
  {Nakanishi}},\ }\bibfield  {title} {\enquote {\bibinfo {title} {Second-order
  topological phases protected by chiral symmetry},}\ }\href {\doibase
  10.1103/PhysRevB.100.235302} {\bibfield  {journal} {\bibinfo  {journal}
  {Phys. Rev. B}\ }\textbf {\bibinfo {volume} {100}},\ \bibinfo {pages}
  {235302} (\bibinfo {year} {2019})}\BibitemShut {NoStop}%
\bibitem [{\citenamefont {Tanaka}\ \emph {et~al.}(2020)\citenamefont {Tanaka},
  \citenamefont {Takahashi},\ and\ \citenamefont {Murakami}}]{tanaka2020}%
  \BibitemOpen
  \bibfield  {author} {\bibinfo {author} {\bibfnamefont {Yutaro}\ \bibnamefont
  {Tanaka}}, \bibinfo {author} {\bibfnamefont {Ryo}\ \bibnamefont {Takahashi}},
  \ and\ \bibinfo {author} {\bibfnamefont {Shuichi}\ \bibnamefont {Murakami}},\
  }\bibfield  {title} {\enquote {\bibinfo {title} {Appearance of hinge states
  in second-order topological insulators via the cutting procedure},}\ }\href
  {\doibase 10.1103/PhysRevB.101.115120} {\bibfield  {journal} {\bibinfo
  {journal} {Phys. Rev. B}\ }\textbf {\bibinfo {volume} {101}},\ \bibinfo
  {pages} {115120} (\bibinfo {year} {2020})}\BibitemShut {NoStop}%
\bibitem [{\citenamefont {Langbehn}\ \emph {et~al.}(2017)\citenamefont
  {Langbehn}, \citenamefont {Peng}, \citenamefont {Trifunovic}, \citenamefont
  {von Oppen},\ and\ \citenamefont {Brouwer}}]{reflection}%
  \BibitemOpen
  \bibfield  {author} {\bibinfo {author} {\bibfnamefont {Josias}\ \bibnamefont
  {Langbehn}}, \bibinfo {author} {\bibfnamefont {Yang}\ \bibnamefont {Peng}},
  \bibinfo {author} {\bibfnamefont {Luka}\ \bibnamefont {Trifunovic}}, \bibinfo
  {author} {\bibfnamefont {Felix}\ \bibnamefont {von Oppen}}, \ and\ \bibinfo
  {author} {\bibfnamefont {Piet~W.}\ \bibnamefont {Brouwer}},\ }\bibfield
  {title} {\enquote {\bibinfo {title} {Reflection-symmetric second-order
  topological insulators and superconductors},}\ }\href {\doibase
  10.1103/PhysRevLett.119.246401} {\bibfield  {journal} {\bibinfo  {journal}
  {Phys. Rev. Lett.}\ }\textbf {\bibinfo {volume} {119}},\ \bibinfo {pages}
  {246401} (\bibinfo {year} {2017})}\BibitemShut {NoStop}%
\bibitem [{\citenamefont {Fu}(2011)}]{fu2011}%
  \BibitemOpen
  \bibfield  {author} {\bibinfo {author} {\bibfnamefont {Liang}\ \bibnamefont
  {Fu}},\ }\bibfield  {title} {\enquote {\bibinfo {title} {Topological
  crystalline insulators},}\ }\href {\doibase 10.1103/PhysRevLett.106.106802}
  {\bibfield  {journal} {\bibinfo  {journal} {Phys. Rev. Lett.}\ }\textbf
  {\bibinfo {volume} {106}},\ \bibinfo {pages} {106802} (\bibinfo {year}
  {2011})}\BibitemShut {NoStop}%
\bibitem [{\citenamefont {Song}\ \emph {et~al.}(2017)\citenamefont {Song},
  \citenamefont {Fang},\ and\ \citenamefont {Fang}}]{song2017}%
  \BibitemOpen
  \bibfield  {author} {\bibinfo {author} {\bibfnamefont {Zhida}\ \bibnamefont
  {Song}}, \bibinfo {author} {\bibfnamefont {Zhong}\ \bibnamefont {Fang}}, \
  and\ \bibinfo {author} {\bibfnamefont {Chen}\ \bibnamefont {Fang}},\
  }\bibfield  {title} {\enquote {\bibinfo {title}
  {$(d\ensuremath{-}2)$-dimensional edge states of rotation symmetry protected
  topological states},}\ }\href {\doibase 10.1103/PhysRevLett.119.246402}
  {\bibfield  {journal} {\bibinfo  {journal} {Phys. Rev. Lett.}\ }\textbf
  {\bibinfo {volume} {119}},\ \bibinfo {pages} {246402} (\bibinfo {year}
  {2017})}\BibitemShut {NoStop}%
\bibitem [{\citenamefont {Park}\ \emph {et~al.}(2019)\citenamefont {Park},
  \citenamefont {Kim}, \citenamefont {Cho},\ and\ \citenamefont
  {Lee}}]{park2019}%
  \BibitemOpen
  \bibfield  {author} {\bibinfo {author} {\bibfnamefont {Moon~Jip}\
  \bibnamefont {Park}}, \bibinfo {author} {\bibfnamefont {Youngkuk}\
  \bibnamefont {Kim}}, \bibinfo {author} {\bibfnamefont {Gil~Young}\
  \bibnamefont {Cho}}, \ and\ \bibinfo {author} {\bibfnamefont {SungBin}\
  \bibnamefont {Lee}},\ }\bibfield  {title} {\enquote {\bibinfo {title}
  {Higher-order topological insulator in twisted bilayer graphene},}\ }\href
  {\doibase 10.1103/PhysRevLett.123.216803} {\bibfield  {journal} {\bibinfo
  {journal} {Phys. Rev. Lett.}\ }\textbf {\bibinfo {volume} {123}},\ \bibinfo
  {pages} {216803} (\bibinfo {year} {2019})}\BibitemShut {NoStop}%
\bibitem [{\citenamefont {Yan}\ \emph {et~al.}(2018)\citenamefont {Yan},
  \citenamefont {Song},\ and\ \citenamefont {Wang}}]{yan2018}%
  \BibitemOpen
  \bibfield  {author} {\bibinfo {author} {\bibfnamefont {Zhongbo}\ \bibnamefont
  {Yan}}, \bibinfo {author} {\bibfnamefont {Fei}\ \bibnamefont {Song}}, \ and\
  \bibinfo {author} {\bibfnamefont {Zhong}\ \bibnamefont {Wang}},\ }\bibfield
  {title} {\enquote {\bibinfo {title} {Majorana corner modes in a
  high-temperature platform},}\ }\href {\doibase
  10.1103/PhysRevLett.121.096803} {\bibfield  {journal} {\bibinfo  {journal}
  {Phys. Rev. Lett.}\ }\textbf {\bibinfo {volume} {121}},\ \bibinfo {pages}
  {096803} (\bibinfo {year} {2018})}\BibitemShut {NoStop}%
\bibitem [{\citenamefont {Wang}\ \emph {et~al.}(2018)\citenamefont {Wang},
  \citenamefont {Liu}, \citenamefont {Lu},\ and\ \citenamefont
  {Zhang}}]{yan22018}%
  \BibitemOpen
  \bibfield  {author} {\bibinfo {author} {\bibfnamefont {Qiyue}\ \bibnamefont
  {Wang}}, \bibinfo {author} {\bibfnamefont {Cheng-Cheng}\ \bibnamefont {Liu}},
  \bibinfo {author} {\bibfnamefont {Yuan-Ming}\ \bibnamefont {Lu}}, \ and\
  \bibinfo {author} {\bibfnamefont {Fan}\ \bibnamefont {Zhang}},\ }\bibfield
  {title} {\enquote {\bibinfo {title} {High-temperature majorana corner
  states},}\ }\href {\doibase 10.1103/PhysRevLett.121.186801} {\bibfield
  {journal} {\bibinfo  {journal} {Phys. Rev. Lett.}\ }\textbf {\bibinfo
  {volume} {121}},\ \bibinfo {pages} {186801} (\bibinfo {year}
  {2018})}\BibitemShut {NoStop}%
\bibitem [{\citenamefont {Esaki}\ \emph {et~al.}(2011)\citenamefont {Esaki},
  \citenamefont {Sato}, \citenamefont {Hasebe},\ and\ \citenamefont
  {Kohmoto}}]{sato2011}%
  \BibitemOpen
  \bibfield  {author} {\bibinfo {author} {\bibfnamefont {Kenta}\ \bibnamefont
  {Esaki}}, \bibinfo {author} {\bibfnamefont {Masatoshi}\ \bibnamefont {Sato}},
  \bibinfo {author} {\bibfnamefont {Kazuki}\ \bibnamefont {Hasebe}}, \ and\
  \bibinfo {author} {\bibfnamefont {Mahito}\ \bibnamefont {Kohmoto}},\
  }\bibfield  {title} {\enquote {\bibinfo {title} {Edge states and topological
  phases in non-hermitian systems},}\ }\href {\doibase
  10.1103/PhysRevB.84.205128} {\bibfield  {journal} {\bibinfo  {journal} {Phys.
  Rev. B}\ }\textbf {\bibinfo {volume} {84}},\ \bibinfo {pages} {205128}
  (\bibinfo {year} {2011})}\BibitemShut {NoStop}%
\bibitem [{\citenamefont {Kawabata}\ \emph {et~al.}(2018)\citenamefont
  {Kawabata}, \citenamefont {Shiozaki},\ and\ \citenamefont
  {Ueda}}]{kawabata2018}%
  \BibitemOpen
  \bibfield  {author} {\bibinfo {author} {\bibfnamefont {Kohei}\ \bibnamefont
  {Kawabata}}, \bibinfo {author} {\bibfnamefont {Ken}\ \bibnamefont
  {Shiozaki}}, \ and\ \bibinfo {author} {\bibfnamefont {Masahito}\ \bibnamefont
  {Ueda}},\ }\bibfield  {title} {\enquote {\bibinfo {title} {Anomalous helical
  edge states in a non-hermitian chern insulator},}\ }\href {\doibase
  10.1103/PhysRevB.98.165148} {\bibfield  {journal} {\bibinfo  {journal} {Phys.
  Rev. B}\ }\textbf {\bibinfo {volume} {98}},\ \bibinfo {pages} {165148}
  (\bibinfo {year} {2018})}\BibitemShut {NoStop}%
\bibitem [{\citenamefont {Martinez~Alvarez}\ \emph {et~al.}(2018)\citenamefont
  {Martinez~Alvarez}, \citenamefont {Barrios~Vargas},\ and\ \citenamefont
  {Foa~Torres}}]{Alvarez2018}%
  \BibitemOpen
  \bibfield  {author} {\bibinfo {author} {\bibfnamefont {V.~M.}\ \bibnamefont
  {Martinez~Alvarez}}, \bibinfo {author} {\bibfnamefont {J.~E.}\ \bibnamefont
  {Barrios~Vargas}}, \ and\ \bibinfo {author} {\bibfnamefont {L.~E.~F.}\
  \bibnamefont {Foa~Torres}},\ }\bibfield  {title} {\enquote {\bibinfo {title}
  {Non-hermitian robust edge states in one dimension: Anomalous localization
  and eigenspace condensation at exceptional points},}\ }\href {\doibase
  10.1103/PhysRevB.97.121401} {\bibfield  {journal} {\bibinfo  {journal} {Phys.
  Rev. B}\ }\textbf {\bibinfo {volume} {97}},\ \bibinfo {pages} {121401}
  (\bibinfo {year} {2018})}\BibitemShut {NoStop}%
\bibitem [{\citenamefont {Lee}(2016)}]{lee2016}%
  \BibitemOpen
  \bibfield  {author} {\bibinfo {author} {\bibfnamefont {Tony~E.}\ \bibnamefont
  {Lee}},\ }\bibfield  {title} {\enquote {\bibinfo {title} {Anomalous edge
  state in a non-hermitian lattice},}\ }\href {\doibase
  10.1103/PhysRevLett.116.133903} {\bibfield  {journal} {\bibinfo  {journal}
  {Phys. Rev. Lett.}\ }\textbf {\bibinfo {volume} {116}},\ \bibinfo {pages}
  {133903} (\bibinfo {year} {2016})}\BibitemShut {NoStop}%
\bibitem [{\citenamefont {Xiong}(2018)}]{xiong2017}%
  \BibitemOpen
  \bibfield  {author} {\bibinfo {author} {\bibfnamefont {Ye}~\bibnamefont
  {Xiong}},\ }\bibfield  {title} {\enquote {\bibinfo {title} {Why does bulk
  boundary correspondence fail in some non-hermitian topological models},}\
  }\href {\doibase 10.1088/2399-6528/aab64a} {\bibfield  {journal} {\bibinfo
  {journal} {Journal of Physics Communications}\ }\textbf {\bibinfo {volume}
  {2}},\ \bibinfo {pages} {035043} (\bibinfo {year} {2018})}\BibitemShut
  {NoStop}%
\bibitem [{\citenamefont {Leykam}\ \emph {et~al.}(2017)\citenamefont {Leykam},
  \citenamefont {Bliokh}, \citenamefont {Huang}, \citenamefont {Chong},\ and\
  \citenamefont {Nori}}]{leykam2018}%
  \BibitemOpen
  \bibfield  {author} {\bibinfo {author} {\bibfnamefont {Daniel}\ \bibnamefont
  {Leykam}}, \bibinfo {author} {\bibfnamefont {Konstantin~Y.}\ \bibnamefont
  {Bliokh}}, \bibinfo {author} {\bibfnamefont {Chunli}\ \bibnamefont {Huang}},
  \bibinfo {author} {\bibfnamefont {Y.~D.}\ \bibnamefont {Chong}}, \ and\
  \bibinfo {author} {\bibfnamefont {Franco}\ \bibnamefont {Nori}},\ }\bibfield
  {title} {\enquote {\bibinfo {title} {Edge modes, degeneracies, and
  topological numbers in non-hermitian systems},}\ }\href {\doibase
  10.1103/PhysRevLett.118.040401} {\bibfield  {journal} {\bibinfo  {journal}
  {Phys. Rev. Lett.}\ }\textbf {\bibinfo {volume} {118}},\ \bibinfo {pages}
  {040401} (\bibinfo {year} {2017})}\BibitemShut {NoStop}%
\bibitem [{\citenamefont {Shen}\ \emph {et~al.}(2018)\citenamefont {Shen},
  \citenamefont {Zhen},\ and\ \citenamefont {Fu}}]{shen2018}%
  \BibitemOpen
  \bibfield  {author} {\bibinfo {author} {\bibfnamefont {Huitao}\ \bibnamefont
  {Shen}}, \bibinfo {author} {\bibfnamefont {Bo}~\bibnamefont {Zhen}}, \ and\
  \bibinfo {author} {\bibfnamefont {Liang}\ \bibnamefont {Fu}},\ }\bibfield
  {title} {\enquote {\bibinfo {title} {Topological band theory for
  non-hermitian hamiltonians},}\ }\href {\doibase
  10.1103/PhysRevLett.120.146402} {\bibfield  {journal} {\bibinfo  {journal}
  {Phys. Rev. Lett.}\ }\textbf {\bibinfo {volume} {120}},\ \bibinfo {pages}
  {146402} (\bibinfo {year} {2018})}\BibitemShut {NoStop}%
\bibitem [{\citenamefont {Kunst}\ \emph
  {et~al.}(2018{\natexlab{b}})\citenamefont {Kunst}, \citenamefont
  {Edvardsson}, \citenamefont {Budich},\ and\ \citenamefont
  {Bergholtz}}]{kunst2018}%
  \BibitemOpen
  \bibfield  {author} {\bibinfo {author} {\bibfnamefont {Flore~K.}\
  \bibnamefont {Kunst}}, \bibinfo {author} {\bibfnamefont {Elisabet}\
  \bibnamefont {Edvardsson}}, \bibinfo {author} {\bibfnamefont {Jan~Carl}\
  \bibnamefont {Budich}}, \ and\ \bibinfo {author} {\bibfnamefont {Emil~J.}\
  \bibnamefont {Bergholtz}},\ }\bibfield  {title} {\enquote {\bibinfo {title}
  {Biorthogonal bulk-boundary correspondence in non-hermitian systems},}\
  }\href {\doibase 10.1103/PhysRevLett.121.026808} {\bibfield  {journal}
  {\bibinfo  {journal} {Phys. Rev. Lett.}\ }\textbf {\bibinfo {volume} {121}},\
  \bibinfo {pages} {026808} (\bibinfo {year} {2018}{\natexlab{b}})}\BibitemShut
  {NoStop}%
\bibitem [{\citenamefont {Yao}\ and\ \citenamefont {Wang}(2018)}]{yao2018}%
  \BibitemOpen
  \bibfield  {author} {\bibinfo {author} {\bibfnamefont {Shunyu}\ \bibnamefont
  {Yao}}\ and\ \bibinfo {author} {\bibfnamefont {Zhong}\ \bibnamefont {Wang}},\
  }\bibfield  {title} {\enquote {\bibinfo {title} {Edge states and topological
  invariants of non-hermitian systems},}\ }\href {\doibase
  10.1103/PhysRevLett.121.086803} {\bibfield  {journal} {\bibinfo  {journal}
  {Phys. Rev. Lett.}\ }\textbf {\bibinfo {volume} {121}},\ \bibinfo {pages}
  {086803} (\bibinfo {year} {2018})}\BibitemShut {NoStop}%
\bibitem [{\citenamefont {Yao}\ \emph {et~al.}(2018)\citenamefont {Yao},
  \citenamefont {Song},\ and\ \citenamefont {Wang}}]{yao20182}%
  \BibitemOpen
  \bibfield  {author} {\bibinfo {author} {\bibfnamefont {Shunyu}\ \bibnamefont
  {Yao}}, \bibinfo {author} {\bibfnamefont {Fei}\ \bibnamefont {Song}}, \ and\
  \bibinfo {author} {\bibfnamefont {Zhong}\ \bibnamefont {Wang}},\ }\bibfield
  {title} {\enquote {\bibinfo {title} {Non-hermitian chern bands},}\ }\href
  {\doibase 10.1103/PhysRevLett.121.136802} {\bibfield  {journal} {\bibinfo
  {journal} {Phys. Rev. Lett.}\ }\textbf {\bibinfo {volume} {121}},\ \bibinfo
  {pages} {136802} (\bibinfo {year} {2018})}\BibitemShut {NoStop}%
\bibitem [{\citenamefont {Lee}\ and\ \citenamefont
  {Thomale}(2019)}]{Thomale2019}%
  \BibitemOpen
  \bibfield  {author} {\bibinfo {author} {\bibfnamefont {Ching~Hua}\
  \bibnamefont {Lee}}\ and\ \bibinfo {author} {\bibfnamefont {Ronny}\
  \bibnamefont {Thomale}},\ }\bibfield  {title} {\enquote {\bibinfo {title}
  {Anatomy of skin modes and topology in non-hermitian systems},}\ }\href
  {\doibase 10.1103/PhysRevB.99.201103} {\bibfield  {journal} {\bibinfo
  {journal} {Phys. Rev. B}\ }\textbf {\bibinfo {volume} {99}},\ \bibinfo
  {pages} {201103} (\bibinfo {year} {2019})}\BibitemShut {NoStop}%
\bibitem [{\citenamefont {Longhi}(2019)}]{londhi2019}%
  \BibitemOpen
  \bibfield  {author} {\bibinfo {author} {\bibfnamefont {S.}~\bibnamefont
  {Longhi}},\ }\bibfield  {title} {\enquote {\bibinfo {title} {Topological
  phase transition in non-hermitian quasicrystals},}\ }\href {\doibase
  10.1103/PhysRevLett.122.237601} {\bibfield  {journal} {\bibinfo  {journal}
  {Phys. Rev. Lett.}\ }\textbf {\bibinfo {volume} {122}},\ \bibinfo {pages}
  {237601} (\bibinfo {year} {2019})}\BibitemShut {NoStop}%
\bibitem [{\citenamefont {Yokomizo}\ and\ \citenamefont
  {Murakami}(2019)}]{yokomizo2019}%
  \BibitemOpen
  \bibfield  {author} {\bibinfo {author} {\bibfnamefont {Kazuki}\ \bibnamefont
  {Yokomizo}}\ and\ \bibinfo {author} {\bibfnamefont {Shuichi}\ \bibnamefont
  {Murakami}},\ }\bibfield  {title} {\enquote {\bibinfo {title} {Non-bloch band
  theory of non-hermitian systems},}\ }\href {\doibase
  10.1103/PhysRevLett.123.066404} {\bibfield  {journal} {\bibinfo  {journal}
  {Phys. Rev. Lett.}\ }\textbf {\bibinfo {volume} {123}},\ \bibinfo {pages}
  {066404} (\bibinfo {year} {2019})}\BibitemShut {NoStop}%
\bibitem [{\citenamefont {Kawabata}\ \emph
  {et~al.}(2019{\natexlab{a}})\citenamefont {Kawabata}, \citenamefont
  {Bessho},\ and\ \citenamefont {Sato}}]{kawabata2019}%
  \BibitemOpen
  \bibfield  {author} {\bibinfo {author} {\bibfnamefont {Kohei}\ \bibnamefont
  {Kawabata}}, \bibinfo {author} {\bibfnamefont {Takumi}\ \bibnamefont
  {Bessho}}, \ and\ \bibinfo {author} {\bibfnamefont {Masatoshi}\ \bibnamefont
  {Sato}},\ }\bibfield  {title} {\enquote {\bibinfo {title} {Classification of
  exceptional points and non-hermitian topological semimetals},}\ }\href
  {\doibase 10.1103/PhysRevLett.123.066405} {\bibfield  {journal} {\bibinfo
  {journal} {Phys. Rev. Lett.}\ }\textbf {\bibinfo {volume} {123}},\ \bibinfo
  {pages} {066405} (\bibinfo {year} {2019}{\natexlab{a}})}\BibitemShut
  {NoStop}%
\bibitem [{\citenamefont {Song}\ \emph {et~al.}(2019)\citenamefont {Song},
  \citenamefont {Yao},\ and\ \citenamefont {Wang}}]{song2019}%
  \BibitemOpen
  \bibfield  {author} {\bibinfo {author} {\bibfnamefont {Fei}\ \bibnamefont
  {Song}}, \bibinfo {author} {\bibfnamefont {Shunyu}\ \bibnamefont {Yao}}, \
  and\ \bibinfo {author} {\bibfnamefont {Zhong}\ \bibnamefont {Wang}},\
  }\bibfield  {title} {\enquote {\bibinfo {title} {Non-hermitian skin effect
  and chiral damping in open quantum systems},}\ }\href {\doibase
  10.1103/PhysRevLett.123.170401} {\bibfield  {journal} {\bibinfo  {journal}
  {Phys. Rev. Lett.}\ }\textbf {\bibinfo {volume} {123}},\ \bibinfo {pages}
  {170401} (\bibinfo {year} {2019})}\BibitemShut {NoStop}%
\bibitem [{\citenamefont {Imura}\ and\ \citenamefont
  {Takane}(2019)}]{imura2019}%
  \BibitemOpen
  \bibfield  {author} {\bibinfo {author} {\bibfnamefont {Ken-Ichiro}\
  \bibnamefont {Imura}}\ and\ \bibinfo {author} {\bibfnamefont {Yositake}\
  \bibnamefont {Takane}},\ }\bibfield  {title} {\enquote {\bibinfo {title}
  {Generalized bulk-edge correspondence for non-hermitian topological
  systems},}\ }\href {\doibase 10.1103/PhysRevB.100.165430} {\bibfield
  {journal} {\bibinfo  {journal} {Phys. Rev. B}\ }\textbf {\bibinfo {volume}
  {100}},\ \bibinfo {pages} {165430} (\bibinfo {year} {2019})}\BibitemShut
  {NoStop}%
\bibitem [{\citenamefont {Okuma}\ and\ \citenamefont {Sato}(2019)}]{okuma2019}%
  \BibitemOpen
  \bibfield  {author} {\bibinfo {author} {\bibfnamefont {Nobuyuki}\
  \bibnamefont {Okuma}}\ and\ \bibinfo {author} {\bibfnamefont {Masatoshi}\
  \bibnamefont {Sato}},\ }\bibfield  {title} {\enquote {\bibinfo {title}
  {Topological phase transition driven by infinitesimal instability: Majorana
  fermions in non-hermitian spintronics},}\ }\href {\doibase
  10.1103/PhysRevLett.123.097701} {\bibfield  {journal} {\bibinfo  {journal}
  {Phys. Rev. Lett.}\ }\textbf {\bibinfo {volume} {123}},\ \bibinfo {pages}
  {097701} (\bibinfo {year} {2019})}\BibitemShut {NoStop}%
\bibitem [{\citenamefont {Borgnia}\ \emph {et~al.}(2020)\citenamefont
  {Borgnia}, \citenamefont {Kruchkov},\ and\ \citenamefont
  {Slager}}]{Borgnia2020}%
  \BibitemOpen
  \bibfield  {author} {\bibinfo {author} {\bibfnamefont {Dan~S.}\ \bibnamefont
  {Borgnia}}, \bibinfo {author} {\bibfnamefont {Alex~Jura}\ \bibnamefont
  {Kruchkov}}, \ and\ \bibinfo {author} {\bibfnamefont {Robert-Jan}\
  \bibnamefont {Slager}},\ }\bibfield  {title} {\enquote {\bibinfo {title}
  {Non-hermitian boundary modes and topology},}\ }\href {\doibase
  10.1103/PhysRevLett.124.056802} {\bibfield  {journal} {\bibinfo  {journal}
  {Phys. Rev. Lett.}\ }\textbf {\bibinfo {volume} {124}},\ \bibinfo {pages}
  {056802} (\bibinfo {year} {2020})}\BibitemShut {NoStop}%
\bibitem [{\citenamefont {Okuma}\ \emph {et~al.}(2020)\citenamefont {Okuma},
  \citenamefont {Kawabata}, \citenamefont {Shiozaki},\ and\ \citenamefont
  {Sato}}]{origin2020}%
  \BibitemOpen
  \bibfield  {author} {\bibinfo {author} {\bibfnamefont {Nobuyuki}\
  \bibnamefont {Okuma}}, \bibinfo {author} {\bibfnamefont {Kohei}\ \bibnamefont
  {Kawabata}}, \bibinfo {author} {\bibfnamefont {Ken}\ \bibnamefont
  {Shiozaki}}, \ and\ \bibinfo {author} {\bibfnamefont {Masatoshi}\
  \bibnamefont {Sato}},\ }\bibfield  {title} {\enquote {\bibinfo {title}
  {Topological origin of non-hermitian skin effects},}\ }\href {\doibase
  10.1103/PhysRevLett.124.086801} {\bibfield  {journal} {\bibinfo  {journal}
  {Phys. Rev. Lett.}\ }\textbf {\bibinfo {volume} {124}},\ \bibinfo {pages}
  {086801} (\bibinfo {year} {2020})}\BibitemShut {NoStop}%
\bibitem [{\citenamefont {Xue}\ \emph {et~al.}(2020)\citenamefont {Xue},
  \citenamefont {Wang}, \citenamefont {Zhang},\ and\ \citenamefont
  {Chong}}]{xue2020}%
  \BibitemOpen
  \bibfield  {author} {\bibinfo {author} {\bibfnamefont {Haoran}\ \bibnamefont
  {Xue}}, \bibinfo {author} {\bibfnamefont {Qiang}\ \bibnamefont {Wang}},
  \bibinfo {author} {\bibfnamefont {Baile}\ \bibnamefont {Zhang}}, \ and\
  \bibinfo {author} {\bibfnamefont {Y.~D.}\ \bibnamefont {Chong}},\ }\bibfield
  {title} {\enquote {\bibinfo {title} {Non-hermitian dirac cones},}\ }\href
  {\doibase 10.1103/PhysRevLett.124.236403} {\bibfield  {journal} {\bibinfo
  {journal} {Phys. Rev. Lett.}\ }\textbf {\bibinfo {volume} {124}},\ \bibinfo
  {pages} {236403} (\bibinfo {year} {2020})}\BibitemShut {NoStop}%
\bibitem [{\citenamefont {Gong}\ \emph {et~al.}(2018)\citenamefont {Gong},
  \citenamefont {Ashida}, \citenamefont {Kawabata}, \citenamefont {Takasan},
  \citenamefont {Higashikawa},\ and\ \citenamefont {Ueda}}]{gong2018}%
  \BibitemOpen
  \bibfield  {author} {\bibinfo {author} {\bibfnamefont {Zongping}\
  \bibnamefont {Gong}}, \bibinfo {author} {\bibfnamefont {Yuto}\ \bibnamefont
  {Ashida}}, \bibinfo {author} {\bibfnamefont {Kohei}\ \bibnamefont
  {Kawabata}}, \bibinfo {author} {\bibfnamefont {Kazuaki}\ \bibnamefont
  {Takasan}}, \bibinfo {author} {\bibfnamefont {Sho}\ \bibnamefont
  {Higashikawa}}, \ and\ \bibinfo {author} {\bibfnamefont {Masahito}\
  \bibnamefont {Ueda}},\ }\bibfield  {title} {\enquote {\bibinfo {title}
  {Topological phases of non-hermitian systems},}\ }\href {\doibase
  10.1103/PhysRevX.8.031079} {\bibfield  {journal} {\bibinfo  {journal} {Phys.
  Rev. X}\ }\textbf {\bibinfo {volume} {8}},\ \bibinfo {pages} {031079}
  (\bibinfo {year} {2018})}\BibitemShut {NoStop}%
\bibitem [{\citenamefont {Kawabata}\ \emph
  {et~al.}(2019{\natexlab{b}})\citenamefont {Kawabata}, \citenamefont
  {Shiozaki}, \citenamefont {Ueda},\ and\ \citenamefont {Sato}}]{kawabataprx}%
  \BibitemOpen
  \bibfield  {author} {\bibinfo {author} {\bibfnamefont {Kohei}\ \bibnamefont
  {Kawabata}}, \bibinfo {author} {\bibfnamefont {Ken}\ \bibnamefont
  {Shiozaki}}, \bibinfo {author} {\bibfnamefont {Masahito}\ \bibnamefont
  {Ueda}}, \ and\ \bibinfo {author} {\bibfnamefont {Masatoshi}\ \bibnamefont
  {Sato}},\ }\bibfield  {title} {\enquote {\bibinfo {title} {Symmetry and
  topology in non-hermitian physics},}\ }\href {\doibase
  10.1103/PhysRevX.9.041015} {\bibfield  {journal} {\bibinfo  {journal} {Phys.
  Rev. X}\ }\textbf {\bibinfo {volume} {9}},\ \bibinfo {pages} {041015}
  (\bibinfo {year} {2019}{\natexlab{b}})}\BibitemShut {NoStop}%
\bibitem [{\citenamefont {Malzard}\ \emph {et~al.}(2015)\citenamefont
  {Malzard}, \citenamefont {Poli},\ and\ \citenamefont
  {Schomerus}}]{malzard2015}%
  \BibitemOpen
  \bibfield  {author} {\bibinfo {author} {\bibfnamefont {Simon}\ \bibnamefont
  {Malzard}}, \bibinfo {author} {\bibfnamefont {Charles}\ \bibnamefont {Poli}},
  \ and\ \bibinfo {author} {\bibfnamefont {Henning}\ \bibnamefont
  {Schomerus}},\ }\bibfield  {title} {\enquote {\bibinfo {title} {Topologically
  protected defect states in open photonic systems with non-hermitian
  charge-conjugation and parity-time symmetry},}\ }\href {\doibase
  10.1103/PhysRevLett.115.200402} {\bibfield  {journal} {\bibinfo  {journal}
  {Phys. Rev. Lett.}\ }\textbf {\bibinfo {volume} {115}},\ \bibinfo {pages}
  {200402} (\bibinfo {year} {2015})}\BibitemShut {NoStop}%
\bibitem [{\citenamefont {Carmichael}(1993)}]{open1}%
  \BibitemOpen
  \bibfield  {author} {\bibinfo {author} {\bibfnamefont {H.~J.}\ \bibnamefont
  {Carmichael}},\ }\bibfield  {title} {\enquote {\bibinfo {title} {Quantum
  trajectory theory for cascaded open systems},}\ }\href {\doibase
  10.1103/PhysRevLett.70.2273} {\bibfield  {journal} {\bibinfo  {journal}
  {Phys. Rev. Lett.}\ }\textbf {\bibinfo {volume} {70}},\ \bibinfo {pages}
  {2273--2276} (\bibinfo {year} {1993})}\BibitemShut {NoStop}%
\bibitem [{\citenamefont {Cao}\ and\ \citenamefont {Wiersig}(2015)}]{open2}%
  \BibitemOpen
  \bibfield  {author} {\bibinfo {author} {\bibfnamefont {Hui}\ \bibnamefont
  {Cao}}\ and\ \bibinfo {author} {\bibfnamefont {Jan}\ \bibnamefont
  {Wiersig}},\ }\bibfield  {title} {\enquote {\bibinfo {title} {Dielectric
  microcavities: Model systems for wave chaos and non-hermitian physics},}\
  }\href {\doibase 10.1103/RevModPhys.87.61} {\bibfield  {journal} {\bibinfo
  {journal} {Rev. Mod. Phys.}\ }\textbf {\bibinfo {volume} {87}},\ \bibinfo
  {pages} {61--111} (\bibinfo {year} {2015})}\BibitemShut {NoStop}%
\bibitem [{\citenamefont {Lee}\ and\ \citenamefont {Chan}(2014)}]{open3}%
  \BibitemOpen
  \bibfield  {author} {\bibinfo {author} {\bibfnamefont {Tony~E.}\ \bibnamefont
  {Lee}}\ and\ \bibinfo {author} {\bibfnamefont {Ching-Kit}\ \bibnamefont
  {Chan}},\ }\bibfield  {title} {\enquote {\bibinfo {title} {Heralded magnetism
  in non-hermitian atomic systems},}\ }\href {\doibase
  10.1103/PhysRevX.4.041001} {\bibfield  {journal} {\bibinfo  {journal} {Phys.
  Rev. X}\ }\textbf {\bibinfo {volume} {4}},\ \bibinfo {pages} {041001}
  (\bibinfo {year} {2014})}\BibitemShut {NoStop}%
\bibitem [{\citenamefont {Choi}\ \emph {et~al.}(2010)\citenamefont {Choi},
  \citenamefont {Kang}, \citenamefont {Lim}, \citenamefont {Kim}, \citenamefont
  {Kim}, \citenamefont {Lee},\ and\ \citenamefont {An}}]{open4}%
  \BibitemOpen
  \bibfield  {author} {\bibinfo {author} {\bibfnamefont {Youngwoon}\
  \bibnamefont {Choi}}, \bibinfo {author} {\bibfnamefont {Sungsam}\
  \bibnamefont {Kang}}, \bibinfo {author} {\bibfnamefont {Sooin}\ \bibnamefont
  {Lim}}, \bibinfo {author} {\bibfnamefont {Wookrae}\ \bibnamefont {Kim}},
  \bibinfo {author} {\bibfnamefont {Jung-Ryul}\ \bibnamefont {Kim}}, \bibinfo
  {author} {\bibfnamefont {Jai-Hyung}\ \bibnamefont {Lee}}, \ and\ \bibinfo
  {author} {\bibfnamefont {Kyungwon}\ \bibnamefont {An}},\ }\bibfield  {title}
  {\enquote {\bibinfo {title} {Quasieigenstate coalescence in an atom-cavity
  quantum composite},}\ }\href {\doibase 10.1103/PhysRevLett.104.153601}
  {\bibfield  {journal} {\bibinfo  {journal} {Phys. Rev. Lett.}\ }\textbf
  {\bibinfo {volume} {104}},\ \bibinfo {pages} {153601} (\bibinfo {year}
  {2010})}\BibitemShut {NoStop}%
\bibitem [{\citenamefont {Lee}\ \emph {et~al.}(2014)\citenamefont {Lee},
  \citenamefont {Reiter},\ and\ \citenamefont {Moiseyev}}]{open5}%
  \BibitemOpen
  \bibfield  {author} {\bibinfo {author} {\bibfnamefont {Tony~E.}\ \bibnamefont
  {Lee}}, \bibinfo {author} {\bibfnamefont {Florentin}\ \bibnamefont {Reiter}},
  \ and\ \bibinfo {author} {\bibfnamefont {Nimrod}\ \bibnamefont {Moiseyev}},\
  }\bibfield  {title} {\enquote {\bibinfo {title} {Entanglement and spin
  squeezing in non-hermitian phase transitions},}\ }\href {\doibase
  10.1103/PhysRevLett.113.250401} {\bibfield  {journal} {\bibinfo  {journal}
  {Phys. Rev. Lett.}\ }\textbf {\bibinfo {volume} {113}},\ \bibinfo {pages}
  {250401} (\bibinfo {year} {2014})}\BibitemShut {NoStop}%
\bibitem [{\citenamefont {Makris}\ \emph {et~al.}(2008)\citenamefont {Makris},
  \citenamefont {El-Ganainy}, \citenamefont {Christodoulides},\ and\
  \citenamefont {Musslimani}}]{gain1}%
  \BibitemOpen
  \bibfield  {author} {\bibinfo {author} {\bibfnamefont {K.~G.}\ \bibnamefont
  {Makris}}, \bibinfo {author} {\bibfnamefont {R.}~\bibnamefont {El-Ganainy}},
  \bibinfo {author} {\bibfnamefont {D.~N.}\ \bibnamefont {Christodoulides}}, \
  and\ \bibinfo {author} {\bibfnamefont {Z.~H.}\ \bibnamefont {Musslimani}},\
  }\bibfield  {title} {\enquote {\bibinfo {title} {Beam dynamics in
  $\mathcal{P}\mathcal{T}$ symmetric optical lattices},}\ }\href {\doibase
  10.1103/PhysRevLett.100.103904} {\bibfield  {journal} {\bibinfo  {journal}
  {Phys. Rev. Lett.}\ }\textbf {\bibinfo {volume} {100}},\ \bibinfo {pages}
  {103904} (\bibinfo {year} {2008})}\BibitemShut {NoStop}%
\bibitem [{\citenamefont {Longhi}(2009)}]{gain2}%
  \BibitemOpen
  \bibfield  {author} {\bibinfo {author} {\bibfnamefont {S.}~\bibnamefont
  {Longhi}},\ }\bibfield  {title} {\enquote {\bibinfo {title} {Bloch
  oscillations in complex crystals with $\mathcal{P}\mathcal{T}$ symmetry},}\
  }\href {\doibase 10.1103/PhysRevLett.103.123601} {\bibfield  {journal}
  {\bibinfo  {journal} {Phys. Rev. Lett.}\ }\textbf {\bibinfo {volume} {103}},\
  \bibinfo {pages} {123601} (\bibinfo {year} {2009})}\BibitemShut {NoStop}%
\bibitem [{\citenamefont {Klaiman}\ \emph {et~al.}(2008)\citenamefont
  {Klaiman}, \citenamefont {G\"unther},\ and\ \citenamefont
  {Moiseyev}}]{gain3}%
  \BibitemOpen
  \bibfield  {author} {\bibinfo {author} {\bibfnamefont {Shachar}\ \bibnamefont
  {Klaiman}}, \bibinfo {author} {\bibfnamefont {Uwe}\ \bibnamefont
  {G\"unther}}, \ and\ \bibinfo {author} {\bibfnamefont {Nimrod}\ \bibnamefont
  {Moiseyev}},\ }\bibfield  {title} {\enquote {\bibinfo {title} {Visualization
  of branch points in $\mathcal{P}\mathcal{T}$-symmetric waveguides},}\ }\href
  {\doibase 10.1103/PhysRevLett.101.080402} {\bibfield  {journal} {\bibinfo
  {journal} {Phys. Rev. Lett.}\ }\textbf {\bibinfo {volume} {101}},\ \bibinfo
  {pages} {080402} (\bibinfo {year} {2008})}\BibitemShut {NoStop}%
\bibitem [{\citenamefont {Bittner}\ \emph {et~al.}(2012)\citenamefont
  {Bittner}, \citenamefont {Dietz}, \citenamefont {G\"unther}, \citenamefont
  {Harney}, \citenamefont {Miski-Oglu}, \citenamefont {Richter},\ and\
  \citenamefont {Sch\"afer}}]{gain4}%
  \BibitemOpen
  \bibfield  {author} {\bibinfo {author} {\bibfnamefont {S.}~\bibnamefont
  {Bittner}}, \bibinfo {author} {\bibfnamefont {B.}~\bibnamefont {Dietz}},
  \bibinfo {author} {\bibfnamefont {U.}~\bibnamefont {G\"unther}}, \bibinfo
  {author} {\bibfnamefont {H.~L.}\ \bibnamefont {Harney}}, \bibinfo {author}
  {\bibfnamefont {M.}~\bibnamefont {Miski-Oglu}}, \bibinfo {author}
  {\bibfnamefont {A.}~\bibnamefont {Richter}}, \ and\ \bibinfo {author}
  {\bibfnamefont {F.}~\bibnamefont {Sch\"afer}},\ }\bibfield  {title} {\enquote
  {\bibinfo {title} {$\mathcal{P}\mathcal{T}$ symmetry and spontaneous symmetry
  breaking in a microwave billiard},}\ }\href {\doibase
  10.1103/PhysRevLett.108.024101} {\bibfield  {journal} {\bibinfo  {journal}
  {Phys. Rev. Lett.}\ }\textbf {\bibinfo {volume} {108}},\ \bibinfo {pages}
  {024101} (\bibinfo {year} {2012})}\BibitemShut {NoStop}%
\bibitem [{\citenamefont {Guo}\ \emph {et~al.}(2009)\citenamefont {Guo},
  \citenamefont {Salamo}, \citenamefont {Duchesne}, \citenamefont {Morandotti},
  \citenamefont {Volatier-Ravat}, \citenamefont {Aimez}, \citenamefont
  {Siviloglou},\ and\ \citenamefont {Christodoulides}}]{gain5}%
  \BibitemOpen
  \bibfield  {author} {\bibinfo {author} {\bibfnamefont {A.}~\bibnamefont
  {Guo}}, \bibinfo {author} {\bibfnamefont {G.~J.}\ \bibnamefont {Salamo}},
  \bibinfo {author} {\bibfnamefont {D.}~\bibnamefont {Duchesne}}, \bibinfo
  {author} {\bibfnamefont {R.}~\bibnamefont {Morandotti}}, \bibinfo {author}
  {\bibfnamefont {M.}~\bibnamefont {Volatier-Ravat}}, \bibinfo {author}
  {\bibfnamefont {V.}~\bibnamefont {Aimez}}, \bibinfo {author} {\bibfnamefont
  {G.~A.}\ \bibnamefont {Siviloglou}}, \ and\ \bibinfo {author} {\bibfnamefont
  {D.~N.}\ \bibnamefont {Christodoulides}},\ }\bibfield  {title} {\enquote
  {\bibinfo {title} {Observation of $\mathcal{P}\mathcal{T}$-symmetry breaking
  in complex optical potentials},}\ }\href {\doibase
  10.1103/PhysRevLett.103.093902} {\bibfield  {journal} {\bibinfo  {journal}
  {Phys. Rev. Lett.}\ }\textbf {\bibinfo {volume} {103}},\ \bibinfo {pages}
  {093902} (\bibinfo {year} {2009})}\BibitemShut {NoStop}%
\bibitem [{\citenamefont {Liertzer}\ \emph {et~al.}(2012)\citenamefont
  {Liertzer}, \citenamefont {Ge}, \citenamefont {Cerjan}, \citenamefont
  {Stone}, \citenamefont {T\"ureci},\ and\ \citenamefont {Rotter}}]{gain6}%
  \BibitemOpen
  \bibfield  {author} {\bibinfo {author} {\bibfnamefont {M.}~\bibnamefont
  {Liertzer}}, \bibinfo {author} {\bibfnamefont {Li}~\bibnamefont {Ge}},
  \bibinfo {author} {\bibfnamefont {A.}~\bibnamefont {Cerjan}}, \bibinfo
  {author} {\bibfnamefont {A.~D.}\ \bibnamefont {Stone}}, \bibinfo {author}
  {\bibfnamefont {H.~E.}\ \bibnamefont {T\"ureci}}, \ and\ \bibinfo {author}
  {\bibfnamefont {S.}~\bibnamefont {Rotter}},\ }\bibfield  {title} {\enquote
  {\bibinfo {title} {Pump-induced exceptional points in lasers},}\ }\href
  {\doibase 10.1103/PhysRevLett.108.173901} {\bibfield  {journal} {\bibinfo
  {journal} {Phys. Rev. Lett.}\ }\textbf {\bibinfo {volume} {108}},\ \bibinfo
  {pages} {173901} (\bibinfo {year} {2012})}\BibitemShut {NoStop}%
\bibitem [{\citenamefont {Lin}\ \emph {et~al.}(2011)\citenamefont {Lin},
  \citenamefont {Ramezani}, \citenamefont {Eichelkraut}, \citenamefont
  {Kottos}, \citenamefont {Cao},\ and\ \citenamefont
  {Christodoulides}}]{gain7}%
  \BibitemOpen
  \bibfield  {author} {\bibinfo {author} {\bibfnamefont {Zin}\ \bibnamefont
  {Lin}}, \bibinfo {author} {\bibfnamefont {Hamidreza}\ \bibnamefont
  {Ramezani}}, \bibinfo {author} {\bibfnamefont {Toni}\ \bibnamefont
  {Eichelkraut}}, \bibinfo {author} {\bibfnamefont {Tsampikos}\ \bibnamefont
  {Kottos}}, \bibinfo {author} {\bibfnamefont {Hui}\ \bibnamefont {Cao}}, \
  and\ \bibinfo {author} {\bibfnamefont {Demetrios~N.}\ \bibnamefont
  {Christodoulides}},\ }\bibfield  {title} {\enquote {\bibinfo {title}
  {Unidirectional invisibility induced by $\mathcal{P}\mathcal{T}$-symmetric
  periodic structures},}\ }\href {\doibase 10.1103/PhysRevLett.106.213901}
  {\bibfield  {journal} {\bibinfo  {journal} {Phys. Rev. Lett.}\ }\textbf
  {\bibinfo {volume} {106}},\ \bibinfo {pages} {213901} (\bibinfo {year}
  {2011})}\BibitemShut {NoStop}%
\bibitem [{\citenamefont {Peng}\ \emph {et~al.}(2014)\citenamefont {Peng},
  \citenamefont {{\"O}zdemir}, \citenamefont {Rotter}, \citenamefont {Yilmaz},
  \citenamefont {Liertzer}, \citenamefont {Monifi}, \citenamefont {Bender},
  \citenamefont {Nori},\ and\ \citenamefont {Yang}}]{gain8}%
  \BibitemOpen
  \bibfield  {author} {\bibinfo {author} {\bibfnamefont {B.}~\bibnamefont
  {Peng}}, \bibinfo {author} {\bibfnamefont {{\c S}.~K.}\ \bibnamefont
  {{\"O}zdemir}}, \bibinfo {author} {\bibfnamefont {S.}~\bibnamefont {Rotter}},
  \bibinfo {author} {\bibfnamefont {H.}~\bibnamefont {Yilmaz}}, \bibinfo
  {author} {\bibfnamefont {M.}~\bibnamefont {Liertzer}}, \bibinfo {author}
  {\bibfnamefont {F.}~\bibnamefont {Monifi}}, \bibinfo {author} {\bibfnamefont
  {C.~M.}\ \bibnamefont {Bender}}, \bibinfo {author} {\bibfnamefont
  {F.}~\bibnamefont {Nori}}, \ and\ \bibinfo {author} {\bibfnamefont
  {L.}~\bibnamefont {Yang}},\ }\bibfield  {title} {\enquote {\bibinfo {title}
  {Loss-induced suppression and revival of lasing},}\ }\href {\doibase
  10.1126/science.1258004} {\bibfield  {journal} {\bibinfo  {journal}
  {Science}\ }\textbf {\bibinfo {volume} {346}},\ \bibinfo {pages} {328--332}
  (\bibinfo {year} {2014})},\ \Eprint
  {http://arxiv.org/abs/https://science.sciencemag.org/content/346/6207/328.full.pdf}
  {https://science.sciencemag.org/content/346/6207/328.full.pdf} \BibitemShut
  {NoStop}%
\bibitem [{\citenamefont {Feng}\ \emph {et~al.}(2014)\citenamefont {Feng},
  \citenamefont {Wong}, \citenamefont {Ma}, \citenamefont {Wang},\ and\
  \citenamefont {Zhang}}]{gain9}%
  \BibitemOpen
  \bibfield  {author} {\bibinfo {author} {\bibfnamefont {Liang}\ \bibnamefont
  {Feng}}, \bibinfo {author} {\bibfnamefont {Zi~Jing}\ \bibnamefont {Wong}},
  \bibinfo {author} {\bibfnamefont {Ren-Min}\ \bibnamefont {Ma}}, \bibinfo
  {author} {\bibfnamefont {Yuan}\ \bibnamefont {Wang}}, \ and\ \bibinfo
  {author} {\bibfnamefont {Xiang}\ \bibnamefont {Zhang}},\ }\bibfield  {title}
  {\enquote {\bibinfo {title} {Single-mode laser by parity-time symmetry
  breaking},}\ }\href {\doibase 10.1126/science.1258479} {\bibfield  {journal}
  {\bibinfo  {journal} {Science}\ }\textbf {\bibinfo {volume} {346}},\ \bibinfo
  {pages} {972--975} (\bibinfo {year} {2014})},\ \Eprint
  {http://arxiv.org/abs/https://science.sciencemag.org/content/346/6212/972.full.pdf}
  {https://science.sciencemag.org/content/346/6212/972.full.pdf} \BibitemShut
  {NoStop}%
\bibitem [{\citenamefont {Kawabata}\ \emph {et~al.}(2017)\citenamefont
  {Kawabata}, \citenamefont {Ashida},\ and\ \citenamefont {Ueda}}]{gain10}%
  \BibitemOpen
  \bibfield  {author} {\bibinfo {author} {\bibfnamefont {Kohei}\ \bibnamefont
  {Kawabata}}, \bibinfo {author} {\bibfnamefont {Yuto}\ \bibnamefont {Ashida}},
  \ and\ \bibinfo {author} {\bibfnamefont {Masahito}\ \bibnamefont {Ueda}},\
  }\bibfield  {title} {\enquote {\bibinfo {title} {Information retrieval and
  criticality in parity-time-symmetric systems},}\ }\href {\doibase
  10.1103/PhysRevLett.119.190401} {\bibfield  {journal} {\bibinfo  {journal}
  {Phys. Rev. Lett.}\ }\textbf {\bibinfo {volume} {119}},\ \bibinfo {pages}
  {190401} (\bibinfo {year} {2017})}\BibitemShut {NoStop}%
\bibitem [{\citenamefont {Ozawa}\ \emph {et~al.}(2019)\citenamefont {Ozawa},
  \citenamefont {Price}, \citenamefont {Amo}, \citenamefont {Goldman},
  \citenamefont {Hafezi}, \citenamefont {Lu}, \citenamefont {Rechtsman},
  \citenamefont {Schuster}, \citenamefont {Simon}, \citenamefont {Zilberberg},\
  and\ \citenamefont {Carusotto}}]{gain11}%
  \BibitemOpen
  \bibfield  {author} {\bibinfo {author} {\bibfnamefont {Tomoki}\ \bibnamefont
  {Ozawa}}, \bibinfo {author} {\bibfnamefont {Hannah~M.}\ \bibnamefont
  {Price}}, \bibinfo {author} {\bibfnamefont {Alberto}\ \bibnamefont {Amo}},
  \bibinfo {author} {\bibfnamefont {Nathan}\ \bibnamefont {Goldman}}, \bibinfo
  {author} {\bibfnamefont {Mohammad}\ \bibnamefont {Hafezi}}, \bibinfo {author}
  {\bibfnamefont {Ling}\ \bibnamefont {Lu}}, \bibinfo {author} {\bibfnamefont
  {Mikael~C.}\ \bibnamefont {Rechtsman}}, \bibinfo {author} {\bibfnamefont
  {David}\ \bibnamefont {Schuster}}, \bibinfo {author} {\bibfnamefont
  {Jonathan}\ \bibnamefont {Simon}}, \bibinfo {author} {\bibfnamefont {Oded}\
  \bibnamefont {Zilberberg}}, \ and\ \bibinfo {author} {\bibfnamefont {Iacopo}\
  \bibnamefont {Carusotto}},\ }\bibfield  {title} {\enquote {\bibinfo {title}
  {Topological photonics},}\ }\href {\doibase 10.1103/RevModPhys.91.015006}
  {\bibfield  {journal} {\bibinfo  {journal} {Rev. Mod. Phys.}\ }\textbf
  {\bibinfo {volume} {91}},\ \bibinfo {pages} {015006} (\bibinfo {year}
  {2019})}\BibitemShut {NoStop}%
\bibitem [{\citenamefont {Longhi}(2017)}]{gain12}%
  \BibitemOpen
  \bibfield  {author} {\bibinfo {author} {\bibfnamefont {Stefano}\ \bibnamefont
  {Longhi}},\ }\bibfield  {title} {\enquote {\bibinfo {title} {Parity-time
  symmetry meets photonics: A new twist in non-hermitian optics},}\ }\href
  {\doibase 10.1209/0295-5075/120/64001} {\bibfield  {journal} {\bibinfo
  {journal} {{EPL} (Europhysics Letters)}\ }\textbf {\bibinfo {volume} {120}},\
  \bibinfo {pages} {64001} (\bibinfo {year} {2017})}\BibitemShut {NoStop}%
\bibitem [{\citenamefont {Heiss}(2012)}]{heiss2012}%
  \BibitemOpen
  \bibfield  {author} {\bibinfo {author} {\bibfnamefont {W~D}\ \bibnamefont
  {Heiss}},\ }\bibfield  {title} {\enquote {\bibinfo {title} {The physics of
  exceptional points},}\ }\href {\doibase 10.1088/1751-8113/45/44/444016}
  {\bibfield  {journal} {\bibinfo  {journal} {Journal of Physics A:
  Mathematical and Theoretical}\ }\textbf {\bibinfo {volume} {45}},\ \bibinfo
  {pages} {444016} (\bibinfo {year} {2012})}\BibitemShut {NoStop}%
\bibitem [{\citenamefont {Edvardsson}\ \emph {et~al.}(2019)\citenamefont
  {Edvardsson}, \citenamefont {Kunst},\ and\ \citenamefont
  {Bergholtz}}]{Edvardsson2019}%
  \BibitemOpen
  \bibfield  {author} {\bibinfo {author} {\bibfnamefont {Elisabet}\
  \bibnamefont {Edvardsson}}, \bibinfo {author} {\bibfnamefont {Flore~K.}\
  \bibnamefont {Kunst}}, \ and\ \bibinfo {author} {\bibfnamefont {Emil~J.}\
  \bibnamefont {Bergholtz}},\ }\bibfield  {title} {\enquote {\bibinfo {title}
  {Non-hermitian extensions of higher-order topological phases and their
  biorthogonal bulk-boundary correspondence},}\ }\href {\doibase
  10.1103/PhysRevB.99.081302} {\bibfield  {journal} {\bibinfo  {journal} {Phys.
  Rev. B}\ }\textbf {\bibinfo {volume} {99}},\ \bibinfo {pages} {081302}
  (\bibinfo {year} {2019})}\BibitemShut {NoStop}%
\bibitem [{\citenamefont {Ezawa}(2019)}]{Ezawa2019}%
  \BibitemOpen
  \bibfield  {author} {\bibinfo {author} {\bibfnamefont {Motohiko}\
  \bibnamefont {Ezawa}},\ }\bibfield  {title} {\enquote {\bibinfo {title}
  {Non-hermitian boundary and interface states in nonreciprocal higher-order
  topological metals and electrical circuits},}\ }\href {\doibase
  10.1103/PhysRevB.99.121411} {\bibfield  {journal} {\bibinfo  {journal} {Phys.
  Rev. B}\ }\textbf {\bibinfo {volume} {99}},\ \bibinfo {pages} {121411}
  (\bibinfo {year} {2019})}\BibitemShut {NoStop}%
\bibitem [{\citenamefont {Liu}\ \emph {et~al.}(2019)\citenamefont {Liu},
  \citenamefont {Zhang}, \citenamefont {Ai}, \citenamefont {Gong},
  \citenamefont {Kawabata}, \citenamefont {Ueda},\ and\ \citenamefont
  {Nori}}]{kawabaras}%
  \BibitemOpen
  \bibfield  {author} {\bibinfo {author} {\bibfnamefont {Tao}\ \bibnamefont
  {Liu}}, \bibinfo {author} {\bibfnamefont {Yu-Ran}\ \bibnamefont {Zhang}},
  \bibinfo {author} {\bibfnamefont {Qing}\ \bibnamefont {Ai}}, \bibinfo
  {author} {\bibfnamefont {Zongping}\ \bibnamefont {Gong}}, \bibinfo {author}
  {\bibfnamefont {Kohei}\ \bibnamefont {Kawabata}}, \bibinfo {author}
  {\bibfnamefont {Masahito}\ \bibnamefont {Ueda}}, \ and\ \bibinfo {author}
  {\bibfnamefont {Franco}\ \bibnamefont {Nori}},\ }\bibfield  {title} {\enquote
  {\bibinfo {title} {Second-order topological phases in non-hermitian
  systems},}\ }\href {\doibase 10.1103/PhysRevLett.122.076801} {\bibfield
  {journal} {\bibinfo  {journal} {Phys. Rev. Lett.}\ }\textbf {\bibinfo
  {volume} {122}},\ \bibinfo {pages} {076801} (\bibinfo {year}
  {2019})}\BibitemShut {NoStop}%
\bibitem [{\citenamefont {Zhang}\ \emph {et~al.}(2019)\citenamefont {Zhang},
  \citenamefont {Rosendo~L\'opez}, \citenamefont {Cheng}, \citenamefont {Liu},\
  and\ \citenamefont {Christensen}}]{zhang2019}%
  \BibitemOpen
  \bibfield  {author} {\bibinfo {author} {\bibfnamefont {Zhiwang}\ \bibnamefont
  {Zhang}}, \bibinfo {author} {\bibfnamefont {Mar\'{\i}a}\ \bibnamefont
  {Rosendo~L\'opez}}, \bibinfo {author} {\bibfnamefont {Ying}\ \bibnamefont
  {Cheng}}, \bibinfo {author} {\bibfnamefont {Xiaojun}\ \bibnamefont {Liu}}, \
  and\ \bibinfo {author} {\bibfnamefont {Johan}\ \bibnamefont {Christensen}},\
  }\bibfield  {title} {\enquote {\bibinfo {title} {Non-hermitian sonic
  second-order topological insulator},}\ }\href {\doibase
  10.1103/PhysRevLett.122.195501} {\bibfield  {journal} {\bibinfo  {journal}
  {Phys. Rev. Lett.}\ }\textbf {\bibinfo {volume} {122}},\ \bibinfo {pages}
  {195501} (\bibinfo {year} {2019})}\BibitemShut {NoStop}%
\bibitem [{\citenamefont {Lee}\ \emph {et~al.}(2019)\citenamefont {Lee},
  \citenamefont {Li},\ and\ \citenamefont {Gong}}]{lee2019}%
  \BibitemOpen
  \bibfield  {author} {\bibinfo {author} {\bibfnamefont {Ching~Hua}\
  \bibnamefont {Lee}}, \bibinfo {author} {\bibfnamefont {Linhu}\ \bibnamefont
  {Li}}, \ and\ \bibinfo {author} {\bibfnamefont {Jiangbin}\ \bibnamefont
  {Gong}},\ }\bibfield  {title} {\enquote {\bibinfo {title} {Hybrid
  higher-order skin-topological modes in nonreciprocal systems},}\ }\href
  {\doibase 10.1103/PhysRevLett.123.016805} {\bibfield  {journal} {\bibinfo
  {journal} {Phys. Rev. Lett.}\ }\textbf {\bibinfo {volume} {123}},\ \bibinfo
  {pages} {016805} (\bibinfo {year} {2019})}\BibitemShut {NoStop}%
\bibitem [{\citenamefont {Denner}\ \emph {et~al.}(2020)\citenamefont {Denner},
  \citenamefont {Skurativska}, \citenamefont {Schindler}, \citenamefont
  {Fischer}, \citenamefont {Thomale}, \citenamefont {Bzdušek},\ and\
  \citenamefont {Neupert}}]{denner2020}%
  \BibitemOpen
  \bibfield  {author} {\bibinfo {author} {\bibfnamefont {M.~Michael}\
  \bibnamefont {Denner}}, \bibinfo {author} {\bibfnamefont {Anastasiia}\
  \bibnamefont {Skurativska}}, \bibinfo {author} {\bibfnamefont {Frank}\
  \bibnamefont {Schindler}}, \bibinfo {author} {\bibfnamefont {Mark~H.}\
  \bibnamefont {Fischer}}, \bibinfo {author} {\bibfnamefont {Ronny}\
  \bibnamefont {Thomale}}, \bibinfo {author} {\bibfnamefont {Tomáš}\
  \bibnamefont {Bzdušek}}, \ and\ \bibinfo {author} {\bibfnamefont {Titus}\
  \bibnamefont {Neupert}},\ }\href@noop {} {\enquote {\bibinfo {title}
  {Exceptional topological insulators},}\ } (\bibinfo {year} {2020}),\ \Eprint
  {http://arxiv.org/abs/2008.01090} {arXiv:2008.01090 [cond-mat.mes-hall]}
  \BibitemShut {NoStop}%
\bibitem [{\citenamefont {Okugawa}\ \emph {et~al.}(2020)\citenamefont
  {Okugawa}, \citenamefont {Takahashi},\ and\ \citenamefont
  {Yokomizo}}]{okugawa}%
  \BibitemOpen
  \bibfield  {author} {\bibinfo {author} {\bibfnamefont {Ryo}\ \bibnamefont
  {Okugawa}}, \bibinfo {author} {\bibfnamefont {Ryo}\ \bibnamefont
  {Takahashi}}, \ and\ \bibinfo {author} {\bibfnamefont {Kazuki}\ \bibnamefont
  {Yokomizo}},\ }\bibfield  {title} {\enquote {\bibinfo {title} {Second-order
  topological non-hermitian skin effects},}\ }\href {\doibase
  10.1103/PhysRevB.102.241202} {\bibfield  {journal} {\bibinfo  {journal}
  {Phys. Rev. B}\ }\textbf {\bibinfo {volume} {102}},\ \bibinfo {pages}
  {241202} (\bibinfo {year} {2020})}\BibitemShut {NoStop}%
\bibitem [{\citenamefont {Yang}\ \emph {et~al.}(2020)\citenamefont {Yang},
  \citenamefont {Zhang}, \citenamefont {Fang},\ and\ \citenamefont
  {Hu}}]{yang2019}%
  \BibitemOpen
  \bibfield  {author} {\bibinfo {author} {\bibfnamefont {Zhesen}\ \bibnamefont
  {Yang}}, \bibinfo {author} {\bibfnamefont {Kai}\ \bibnamefont {Zhang}},
  \bibinfo {author} {\bibfnamefont {Chen}\ \bibnamefont {Fang}}, \ and\
  \bibinfo {author} {\bibfnamefont {Jiangping}\ \bibnamefont {Hu}},\ }\bibfield
   {title} {\enquote {\bibinfo {title} {Non-hermitian bulk-boundary
  correspondence and auxiliary generalized brillouin zone theory},}\ }\href
  {\doibase 10.1103/PhysRevLett.125.226402} {\bibfield  {journal} {\bibinfo
  {journal} {Phys. Rev. Lett.}\ }\textbf {\bibinfo {volume} {125}},\ \bibinfo
  {pages} {226402} (\bibinfo {year} {2020})}\BibitemShut {NoStop}%
\bibitem [{\citenamefont {Hatano}\ and\ \citenamefont {Nelson}(1997)}]{hn}%
  \BibitemOpen
  \bibfield  {author} {\bibinfo {author} {\bibfnamefont {Naomichi}\
  \bibnamefont {Hatano}}\ and\ \bibinfo {author} {\bibfnamefont {David~R.}\
  \bibnamefont {Nelson}},\ }\bibfield  {title} {\enquote {\bibinfo {title}
  {Vortex pinning and non-hermitian quantum mechanics},}\ }\href {\doibase
  10.1103/PhysRevB.56.8651} {\bibfield  {journal} {\bibinfo  {journal} {Phys.
  Rev. B}\ }\textbf {\bibinfo {volume} {56}},\ \bibinfo {pages} {8651--8673}
  (\bibinfo {year} {1997})}\BibitemShut {NoStop}%
\bibitem [{\citenamefont {Alase}\ \emph {et~al.}(2017)\citenamefont {Alase},
  \citenamefont {Cobanera}, \citenamefont {Ortiz},\ and\ \citenamefont
  {Viola}}]{alase2017}%
  \BibitemOpen
  \bibfield  {author} {\bibinfo {author} {\bibfnamefont {Abhijeet}\
  \bibnamefont {Alase}}, \bibinfo {author} {\bibfnamefont {Emilio}\
  \bibnamefont {Cobanera}}, \bibinfo {author} {\bibfnamefont {Gerardo}\
  \bibnamefont {Ortiz}}, \ and\ \bibinfo {author} {\bibfnamefont {Lorenza}\
  \bibnamefont {Viola}},\ }\bibfield  {title} {\enquote {\bibinfo {title}
  {Generalization of bloch's theorem for arbitrary boundary conditions:
  Theory},}\ }\href {\doibase 10.1103/PhysRevB.96.195133} {\bibfield  {journal}
  {\bibinfo  {journal} {Phys. Rev. B}\ }\textbf {\bibinfo {volume} {96}},\
  \bibinfo {pages} {195133} (\bibinfo {year} {2017})}\BibitemShut {NoStop}%
\bibitem [{\citenamefont {Kawabata}\ \emph
  {et~al.}(2020{\natexlab{a}})\citenamefont {Kawabata}, \citenamefont {Sato},\
  and\ \citenamefont {Shiozaki}}]{kawabata2020}%
  \BibitemOpen
  \bibfield  {author} {\bibinfo {author} {\bibfnamefont {Kohei}\ \bibnamefont
  {Kawabata}}, \bibinfo {author} {\bibfnamefont {Masatoshi}\ \bibnamefont
  {Sato}}, \ and\ \bibinfo {author} {\bibfnamefont {Ken}\ \bibnamefont
  {Shiozaki}},\ }\bibfield  {title} {\enquote {\bibinfo {title} {Higher-order
  non-hermitian skin effect},}\ }\href {\doibase 10.1103/PhysRevB.102.205118}
  {\bibfield  {journal} {\bibinfo  {journal} {Phys. Rev. B}\ }\textbf {\bibinfo
  {volume} {102}},\ \bibinfo {pages} {205118} (\bibinfo {year}
  {2020}{\natexlab{a}})}\BibitemShut {NoStop}%
\bibitem [{\citenamefont {Kawabata}\ \emph
  {et~al.}(2020{\natexlab{b}})\citenamefont {Kawabata}, \citenamefont {Okuma},\
  and\ \citenamefont {Sato}}]{kawabata20}%
  \BibitemOpen
  \bibfield  {author} {\bibinfo {author} {\bibfnamefont {Kohei}\ \bibnamefont
  {Kawabata}}, \bibinfo {author} {\bibfnamefont {Nobuyuki}\ \bibnamefont
  {Okuma}}, \ and\ \bibinfo {author} {\bibfnamefont {Masatoshi}\ \bibnamefont
  {Sato}},\ }\bibfield  {title} {\enquote {\bibinfo {title} {Non-bloch band
  theory of non-hermitian hamiltonians in the symplectic class},}\ }\href
  {\doibase 10.1103/PhysRevB.101.195147} {\bibfield  {journal} {\bibinfo
  {journal} {Phys. Rev. B}\ }\textbf {\bibinfo {volume} {101}},\ \bibinfo
  {pages} {195147} (\bibinfo {year} {2020}{\natexlab{b}})}\BibitemShut
  {NoStop}%
\bibitem [{\citenamefont {Yi}\ and\ \citenamefont {Yang}(2020)}]{yi2020}%
  \BibitemOpen
  \bibfield  {author} {\bibinfo {author} {\bibfnamefont {Yifei}\ \bibnamefont
  {Yi}}\ and\ \bibinfo {author} {\bibfnamefont {Zhesen}\ \bibnamefont {Yang}},\
  }\bibfield  {title} {\enquote {\bibinfo {title} {Non-hermitian skin modes
  induced by on-site dissipations and chiral tunneling effect},}\ }\href
  {\doibase 10.1103/PhysRevLett.125.186802} {\bibfield  {journal} {\bibinfo
  {journal} {Phys. Rev. Lett.}\ }\textbf {\bibinfo {volume} {125}},\ \bibinfo
  {pages} {186802} (\bibinfo {year} {2020})}\BibitemShut {NoStop}%
\end{thebibliography}%

\end{document}